\documentclass[useAMS,usenatbib,usedcolumn]{mn2e}
\usepackage[figuresright]{rotating}
\usepackage{lscape}
\usepackage{graphics}
\usepackage{epsfig}
\usepackage{multirow}
\usepackage{bigdelim}
\usepackage{bigstrut}
\usepackage{amsmath}
\usepackage{amssymb}
\usepackage{lscape}
\usepackage{supertabular}
\usepackage{gensymb}
\usepackage{times}
\usepackage[T1]{fontenc}
\usepackage{aecompl}
\usepackage{subcaption}
\usepackage{afterpage}
\usepackage{verbatim}

\newcommand\weeFig{0.24}
\newcommand\bigFig{0.48}

\DeclareMathOperator{\sech}{sech}
\title[External field effect on galaxies in galaxy clusters]{Consequences of the external field effect for MOND disk galaxies in galaxy clusters}
\author[G.~N. Candlish et al.]{G.~N. Candlish$^{1}$\thanks{E-mail: graeme.candlish@ifa.uv.cl} R. Smith$^{2}$, Y. Jaff\'e$^{1}$, A. Cortesi$^{3}$\\
$^{1}$Instituto de F\'isica y Astronom\'ia, Universidad de Valpara\'iso, Gran Breta\~na 1111, Valpara\'iso, Chile\\
$^{2}$Korea Astronomy and Space Science Institute, 766, Daedeokdae-ro, Yuseon-gu, Daejon, 34055, Korea\\
$^{3}$Departamento de Astronomia, Instituto de Astronomia, Geofisica e Ciencias Atmosfericas da USP, Cidade Universitaria, CEP:05508900, Sao Paulo, SP, Brazil}
\begin{document}

\maketitle

\begin{abstract}
Galaxies within galaxy clusters are known to be subject to a wide variety of environmental effects, both gravitational and hydrodynamical. In this study, we examine the purely gravitational interaction of idealised galaxy models falling into a galaxy cluster in the context of Modified Newtonian Dynamics (MOND). This modification of gravity gives rise to an external field effect (EFE), where the internal dynamics of a system are affected by the presence of external gravitational fields. We examine the consequences of the EFE on low and high mass disk galaxies in time-evolving analytic background cluster potentials, considering orbits with weak and strong tidal fields. By varying the orbital plane of the galaxies we also test the effect of having the tidal interaction orthogonal or parallel to the disk. Furthermore, we consider as a control sample models where the EFE has been removed and they are only affected by tides. Our results suggest that MOND cluster galaxies should exhibit clear asymmetries in their isophotes, suffer increased mass loss and a reduction in their rotation curves due to the combined effect of cluster tides and the external field. In particular, low mass galaxies are hit hard by the EFE, becoming dominated by dispersion rather than rotation even in the absence of tides.
\end{abstract}


\section{Introduction}
The ``missing mass'' problem is one of the classic problems of astronomy. In the context of individual disk galaxies, this missing mass, referred to as \textit{dark matter}, manifests itself in the non-Keplerian behaviour of their rotation curves. This additional unseen mass is also required to explain observations of the velocity dispersions of galaxies within galaxy clusters as well as the X-ray temperature profile of the cluster gas. At the level of cosmology, dark matter is required to explain the power spectrum of temperature fluctuations in the CMB, and the process of structure formation more generally.

Direct and indirect detection searches for dark matter particles have made impressive improvements in sensitivity over the years, yet an unambiguous signal has so far proven elusive (see \citealp{DMpaper} for a recent review). Furthermore, from an aesthetic point-of-view, it may seem unsatisfactory to introduce a new particle species, with properties constructed precisely to elude immediate detection. Combined with the ongoing mystery of dark energy, the current standard cosmological model may be considered as a model with a (small) number of free parameters which can be adjusted to match observations extremely well, particularly at high redshift. The freedom within this model represents the entirely unknown nature of the dark sector: effectively a parameterisation of our ignorance. Furthermore, well into the non-linear regime, there are potential small-scale problems with the $\Lambda$CDM model \citep{smallscale}, although these may eventually be resolved by improved baryonic physics in simulations \citep{baryonicPhysSim}. In addition, the recently discovered planes of satellites around the Milky Way, Andromeda and now Centaurus A \citep{centA}, present a potentially significant challenge to the standard picture of cold dark matter (for a review, see \citealp{pawlowskiPlanes}).

Given these issues and the fact that the evidence for the existence of dark matter is (so far) entirely derived from gravitational dynamics, it remains an intriguing possibility to consider a modification of gravity as being responsible for the missing mass phenomenology. A very well-known and studied modification is that of Milgrom's Modified Newtonian Dynamics \citep{milgrom1,milgrom2,milgrom3}, referred to as MOND (for a thorough review see \citealp{mondreview}). The most common formulation of this paradigm is that of a modified gravitational potential, arising from an adjustment of the standard Poisson equation of Newtonian gravity. This adjustment is a function of the local accelerations, such that, in circumstances of weak accelerations below an empirically determined scale, the gradient of the gravitational potential is increased, and thus gravity is ``strengthened.'' Such a modification of gravity has interesting consequences throughout all of stellar and galactic dynamics \citep{kroupa}, suggesting the need to study numerical N-body simulations within this regime.

Several such codes are now available, such as PoR \citep{por}, N-MODY \citep{nmody} and a MOND version of the cosmological code AMIGA \citep{amiga_mond}. In this work we will use the MOND N-body/hydrodynamics code known as RAyMOND \citep{raymond}, which is based upon the RAMSES code \citep{ramses}. This code has previously been used to study possible signals of MOND gravity in cosmological structure formation \citep{raymondCosmo}. In this study we work at the smaller scale of idealised galaxy models within a galaxy cluster.

One of the most spectacular examples of the effects of gravity on galactic dynamics is the tidal disruption of galaxies when they encounter the external gravitational fields of other galaxies or galaxy clusters. These gravitational interactions give rise to a wide range of environmental effects such as mergers, harassment (high-speed encounters between galaxies) and tidal stripping. A huge number of examples of such systems have been observed over the years, making clear that these effects are a crucial ingredient in the process of galaxy evolution, and can even lead to their formation, in the case of tidal dwarf galaxies. Furthermore, the tidal disruption of galaxies by the cluster potential itself is thought to be a crucial ingredient in the build-up of the intracluster light, which may require substantial stripping of massive galaxies \citep{ICLbuildup}.

The central importance of these gravitational encounters for galaxy evolution compels us to consider them in the context of the MOND paradigm, possibly as a means to distinguish between MOND and dark matter models, as well as to explore the process of galaxy evolution in such modified gravity theories.

The main aim of our study is to examine the consequences of the external field effect on disk galaxies as they fall into a cluster. Previous analytical studies of the EFE have focussed on warping of galactic disks \citep{brada_warps} and the distortions of isophotes in giant ellipticals well within galaxy clusters \citep{elliptical_efeI}. A numerical approach (without full N-body dynamics) was taken to analyse the consequences of the EFE on the velocity dispersions of satellite dwarfs of the Milky Way \citep{efe_mw}. Beyond this, there have been a small number of studies of the dynamical effect of the EFE using N-body simulations. The study of \cite{elliptical_efeII} searched for equilibrium N-body models of elliptical galaxies within an external field, while the studies of \cite{tidal_streamsI,tidal_streamsII} analysed the evolution of globular clusters in the external field of a host galaxy, finding that the EFE both distorts the shape of the globular clusters and induces asymmetries in the tidal streams. Recent analytical work has discussed the implications of ultra-diffuse galaxies within clusters for MOND \citep{milgromUDG}. In this paper, however, we will analyse the effect of the external field on isolated galaxies that fall into a galaxy cluster, thus accounting for a time-varying EFE by including the infall process. One would expect the strength of the EFE in a low mass galaxy as it falls into a massive cluster to be considerable, given the large mass ratio and associated large difference in external and internal accelerations, and therefore potentially observable.

We begin with a brief description of the formulation of MOND used in our study in Section~\ref{sec:theory}, before the simulations are discussed in Section~\ref{sec:sims}. Our results are presented in Section~\ref{sec:results} and finally our conclusions are in Section~\ref{sec:conclusions}.

\section{Theoretical background}
\label{sec:theory}
The underlying concept of MOND has been formulated in a wide variety of theoretical contexts. While the origins of the paradigm lie in modifying Newtonian gravity, there are now multiple relativistic completions, including Generalised Einstein-Aether \citep{GEA}, TeVeS \citep{teves} and Bi-metric MOND (or BiMOND, \citealp{bimetricmond}). These formulations may be considered as ``pure'' modifications of gravity by the introduction of additional degrees of freedom in the gravitational sector (a single vector field, a scalar and a vector field, and a rank-2 symmetric tensor field, respectively). Other works have proposed new forms of matter that would give rise to the MOND phenomenology in the weak field regime, such as dipolar dark matter \citep{dipolarDM1,dipolarDM2}, dissipative dark matter \citep{dissipativeDM}, superfluid dark matter \citep{superfluidDM} and the dark fluid formalism \citep{darkfluid}, which attempts to unify the dark sector within one framework.

In the weak-field limit, the most common non-relativistic formulations of MOND (to which most of the above theories reduce) are referred to as AQUAL and QUMOND. Both formulations are included in the RAyMOND code. In this study we will consider only the AQUAL formulation. This is so-called because the modified Poisson equation, that may be taken as the definition of this formulation, is derivable from an aquadratic (non-relativistic) Lagrangian. The modified Poisson equation is
\begin{equation}
\label{aqual}
\nabla \cdot \left( \mu\left(\frac{|\nabla \Phi|}{a_0}\right) \nabla \Phi \right) = 4\pi G \rho
\end{equation}
where $\Phi$ is the gravitational potential, $a_0$ is the MOND acceleration scale (empirically determined to be $a_0 \approx 1.2 \times 10^{-10} m/s^2$ which is the value chosen for this study), and $\rho$ is the matter density distribution. The non-linear nature of MOND is clear from the presence of the gradient of the potential $\Phi$ within the function $\mu$. This function, known as the MOND interpolation function, is constrained only in its limiting behaviour: $\mu(x) \to 1$ when $x \to \infty$ and $\mu(x) \to 0$ when $x \to 0$. As such, an infinite number of functional forms may be provided for $\mu$. We choose the following form:
\begin{equation}
\label{mondTrans}
\mu(x) = \frac{x}{(1+x^n)^{1/n}}
\end{equation}
usually with $n=1$, the so-called 'simple' interpolation function\footnote{It should be noted that this form for the MOND interpolation function is excluded on observational grounds as described in \cite{MONDsolar}, where a more rapid transition from Newtonian gravity to MONDian gravity appears to be required. For our purposes the simple form suffices to analyse the EFE, and we include an $n=5$ model for comparison.}. For one model we will use a more rapid transition function where $n=5$ to examine the consequences. The behaviour of MOND may be inferred from a comparison with the standard Newtonian Poisson equation, and the behaviour of the function $\mu$. The standard Poisson equation is
\begin{equation}
\label{poisson}
\nabla^2 \Phi_N = 4\pi G \rho.
\end{equation}
Comparing with equation (\ref{aqual}) and considering the behaviour of $\mu$, we see that, for a given density distribution, the MONDian potential will have steeper gradients if the magnitude of the local accelerations $|\nabla \Phi|$ is comparable to, or less than, the MOND acceleration scale. This is because $\mu \to 0$ in this case, requiring a `scaling' of $\Phi$ to satisfy equation (\ref{aqual}) for the same density distribution. Conversely, for large accelerations, $\mu \to 1$, and equation (\ref{aqual}) reduces to equation (\ref{poisson}).

\subsection{The external field effect}
\label{subsec:efe}
The presence of the quantity $|\nabla \Phi|$ in equation (\ref{aqual}) is crucial to ensure that the MOND gravitational enhancement applies only whenever the accelerations are sufficiently weak, and does not affect the known gravitational physics in the Solar System or on Earth. The introduction of this quantity to the definition of the gravitational potential, however, leads to an interesting physical consequence known as the \textit{external field effect} (EFE): if system $A$ moves within the gravitational field of system $B$, the internal gravitational dynamics of system $A$ will be affected by the field generated by system $B$ (and vice versa).

One can understand how the EFE arises by considering equations (\ref{aqual}) and (\ref{poisson}). In the case of Newtonian gravity the linearity of the Laplacian in equation (\ref{poisson}) ensures that the internal and external potentials of a system embedded in an external field can be treated separately. That is, we can write $\Phi_t = \Phi_i + \Phi_e$ where $\Phi_t$ is the total potential, $\Phi_i$ is the internal potential (e.g. the potential sourced by the galaxy) and $\Phi_e$ is the external potential (e.g. the potential sourced by the galaxy cluster). The two potentials are ``decoupled'' in the sense that they are each separately sourced by their respective density distributions, and the total potential is simply the two added together: $\Phi_i$ does not depend on $\Phi_e$. Therefore the internal dynamics have no awareness of the external field, apart from tidal forces. In the MOND case, the modified Poisson equation which is solved to determine the MOND potential depends on the magnitude of the total local acceleration, which depends on \emph{both} internal and external potentials. Therefore the internal potential \emph{does} depend on the external potential (and vice versa) in MOND gravity.

Given that the `strength' of the MOND gravitational potential is controlled by the $\mu$ function, the presence of large background accelerations (even in a uniform gravitational field) leads to a value of $|\nabla \Phi|$ well above the MOND scale, and so the internal gravitational potential of the system will become more Newtonian. This is a violation of the Strong Equivalence Principle, since it implies that \emph{gravitational} experiments within the system (if it is within the MOND regime) can detect the presence of the background gravitational field.

In this study we are primarily interested in exploring the consequences of the EFE on disk galaxies within cluster environments. One would expect such consequences because of the large background accelerations in the potential of a galaxy cluster, particularly near the cluster centre. In order to study this, we must use numerical N-body simulations within the context of MOND gravity, to which we now turn.

\section{Simulations}
\label{sec:sims}
The simulations used in this study are idealised galaxy models, in the sense that we consider only single galaxies falling into single isolated clusters. In particular, our models are purely gravitational, without any hydrodynamics and the background galaxy cluster is taken to be a time-evolving analytic density distribution.

\subsection{Initial setup}
\label{subsec:initialSetup}
The disk galaxies that we will use are initially set up using the (Newtonian gravity) DICE code\footnote{Available at bitbucket.org/vperret/dice} \citep{DICE}. In doing so, we include a dark matter halo and allow DICE to determine the initial velocities of the particles as standard, i.e. using a Newtonian gravitational potential in the standard Jeans equations, before then removing the dark matter halo. Given that we have no \textit{a priori} reason to expect such models to begin in equilibrium, we then evolve the disks (and bulge in the case of the high mass model) in isolation in MOND gravity, until they reach an equilibrium configuration. Generating the initial particle velocities with a dark matter halo is merely a simple way to generate a model that is likely to be not too far from equilibrium in MOND gravity. The subsequent evolution in isolation then ensures our models are indeed equilibrated.

The total system (halo, disk and bulge) is generated with $M_{200} = 3.3 \times 10^{12} M_{\odot}$, and mass fractions of $0.965$, $0.028$ and $0.007$ for the halo, disk and bulge respectively for the high mass model. The low mass model has $M_{200} = 1.0 \times 10^{11} M_{\odot}$, and mass fractions of $0.99$ and $0.01$ for the halo and disk respectively (no bulge). Therefore the halo masses are $1.0 \times 10^{11} M_{\odot}$ and $3.2 \times 10^{12} M_{\odot}$ for the low and high mass models.

The disks have $10^5$ particles, and a total mass of $1.0 \times 10^{9} M_{\odot}$ and $9.2 \times 10^{10} M_{\odot}$ for the low and high mass models respectively. The density profiles used are those of an exponential disk radially, with a $\sech(z)$ profile vertically. The radial scale lengths are set to $h_R = 1$~kpc and $h_R = 3$~kpc (low and high mass), with the vertical scale height set to $h_z = 0.15$~kpc and $h_z = 0.25$~kpc (low and high mass). The radial cut-offs are $16$~kpc and $50$~kpc.

For the high mass model the bulge component is represented by a Hernquist profile with $5 \times 10^4$ particles, and has a total mass of $2.3 \times 10^{10} M_{\odot}$. The bulge scale radius is $1$~kpc and the cut-off is $5$~kpc. Note that the low mass model does not include a bulge component.

\subsection{Evolution in isolation}
\label{subsec:evoIso}
To ensure that our tidal interaction simulations use galaxies in equilibrium (at least when outside the cluster) we first evolve the galaxy models described above in isolation, in MONDian gravity.

We use the AQUAL formulation of MOND included in the RAyMOND code, with the standard MOND acceleration scale of $a_0 = 1.2 \times 10^{-10}$~m/s$^2$, and the simple MOND interpolation function of Eq.~(\ref{mondTrans}) with $n=1$, except for the model with a more rapid transition function where $n=5$. The dark matter halos used to generate the disks are removed, and only the stellar components are included. Thus the halos are used simply to ensure that the galaxies begin with reasonable rotation curves.

The models are evolved for approximately $9$~Gyr, using a box length of $1000$~kpc for the low mass models and $2000$~kpc for the high mass models. Due to adaptive mesh refinement the maximum spatial resolution achieved is $0.061$~kpc for the low mass models (refinement level 14) and $0.244$~kpc (refinement level 13) for the high mass models. The coarsest resolutions are $7.8125$~kpc and $15.625$~kpc (refinement level 7) for the low and high mass models respectively. The refinement criterion is such that a grid cell is split into $8$ new cells on the next level of refinement whenever there are $8$ or more particles in a cell. Note that the vertical scale heights of the disks correspond to only a few grid cells even at the highest refinement level, so we would expect some disk thickening as the isolated model evolves.

\begin{figure*}
\centering
\begin{tabular}{cc}
\includegraphics[width=\bigFig\textwidth]{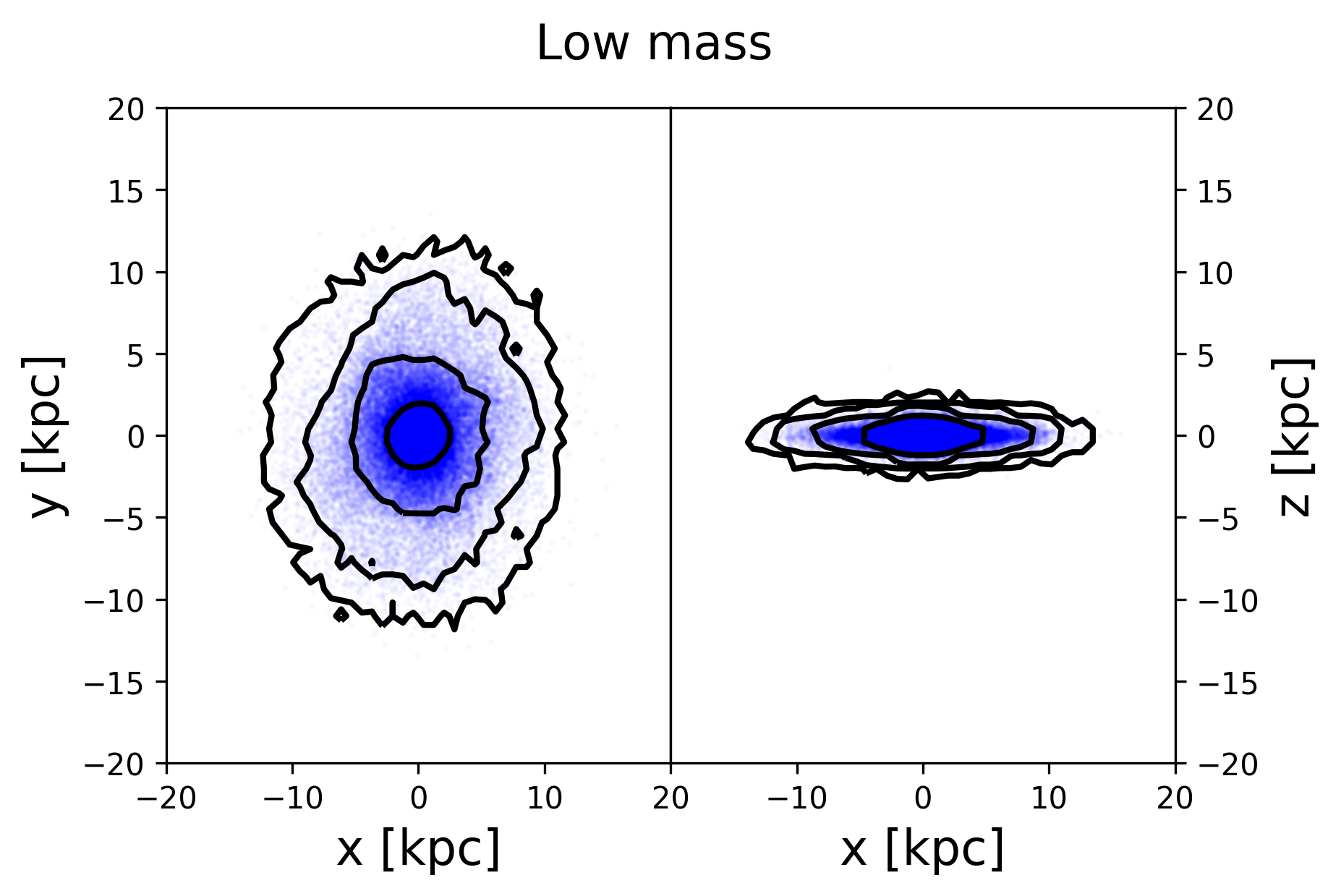} & \includegraphics[width=\bigFig\textwidth]{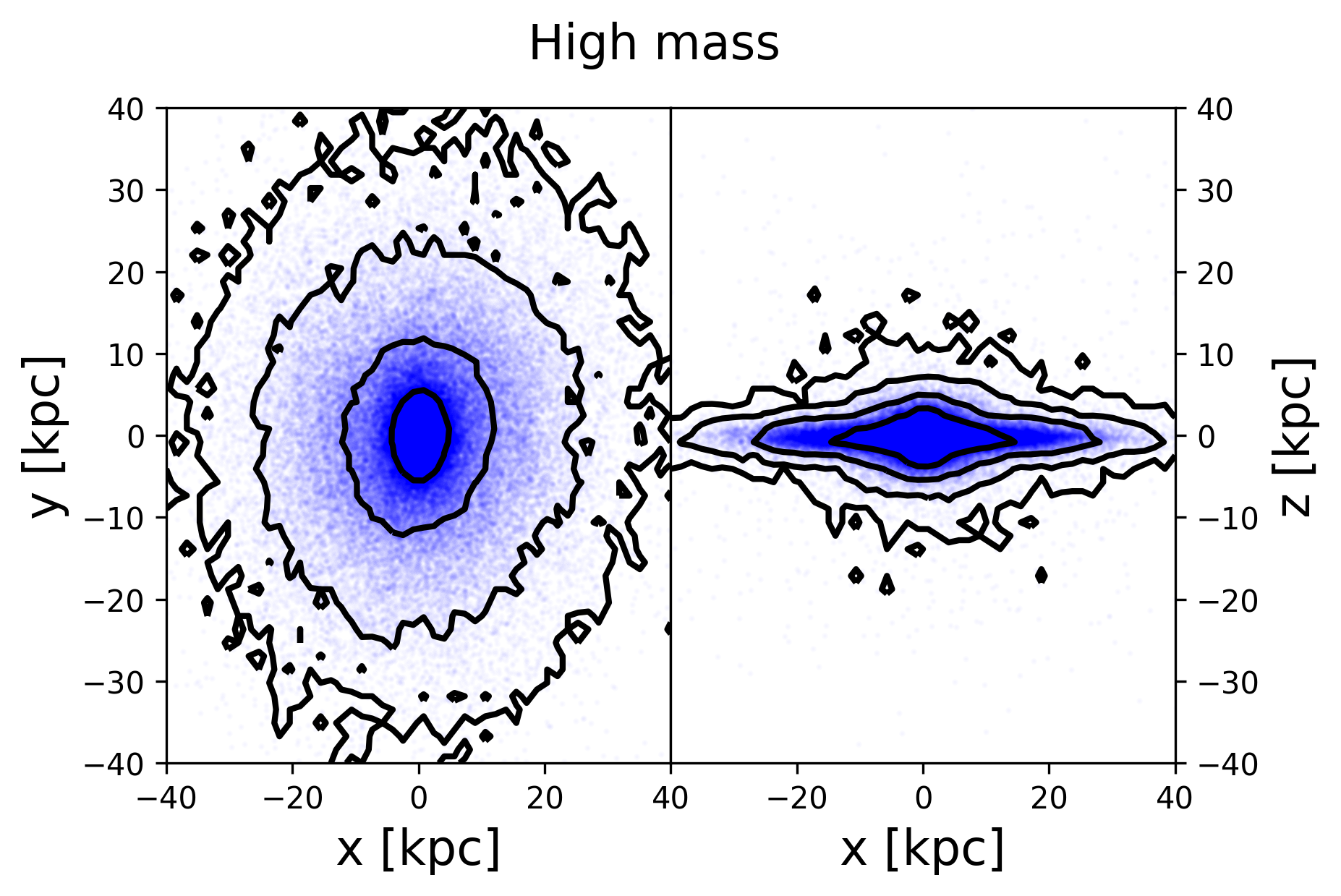}
\end{tabular}
\caption{Simulated disks after evolution in isolation for $9.4$~Gyr.}
\label{fig:evo_isolated_visual}
\end{figure*}

The final isodensity contours for the low and high mass disks are shown in Fig.~\ref{fig:evo_isolated_visual}. While there is some degree of disk thickening, both models maintain clear disk structure. We can quantify the evolution in isolation by comparing the initial and final radial and vertical scale lengths (assuming exponential distributions in both directions). The low mass model evolves from $h_R^{initial} = 1.04$ to $h_R^{final} = 1.72$ in the radial direction, and from $h_z^{initial} = 0.15$ to $h_z^{final} = 0.25$ in the vertical. The high mass model evolves from $h_R^{initial} = 2.99$ to $h_R^{final} = 4.01$ in the radial direction, and from $h_z^{initial} = 0.24$ to $h_z^{final} = 0.75$ in the vertical. Thus both models ``spread'' somewhat in both the radial and vertical directions. It should be borne in mind that neither model is constructed to begin evolution in equilibrium: we have simply set up two Newtonian models, removed their dark matter halos, and then evolved them in MONDian gravity. Moreover, we merely require rotationally-supported stable disk galaxies for our study: the precise structural parameters are of lesser importance.

\subsection{Evolution within the cluster}
The background cluster is analytic (not live) therefore there is no dynamical friction in these models. We do include, however, a time-evolving cluster mass, which also ensures that our infalling galaxies remain within the cluster. We do this in a simple way by linearly interpolating the cluster mass from an initial value at the beginning of the simulation to a final value after $\sim 8$~Gyr, which corresponds to evolution from $z \approx 1$ to $z = 0$ (assuming a standard background cosmology). During this time a cosmological cluster mass may be expected to grow by approximately a factor of $2$ through merging \citep{gill2004,laporte} as chosen here.

Specifically, we consider two background cluster models:
\begin{itemize}
\item Our analysis concentrates on a model of a more massive cluster with an initial mass of $4.5 \times 10^{14}$ M$_{\odot}$ and a final mass of $1 \times 10^{15}$ M$_{\odot}$, roughly corresponding to the Coma cluster. 
\item We also consider a lower mass cluster with an initial mass of $1.5 \times 10^{14}$ M$_{\odot}$ and a final mass of $3 \times 10^{14}$ M$_{\odot}$, roughly corresponding to the Virgo cluster.
\end{itemize}

For realistic galaxy dynamics within the clusters, we use an NFW model with a virial radius of approximately $R_{200} = 2$~Mpc and concentration of $c=3$ for the high mass cluster, and $R_{200} = 1$~Mpc and concentration of $c=8$ for the low mass cluster. The higher concentration for the lower mass cluster is motivated by cosmological simulations \citep{massConc} which show a trend of increasing concentration for decreasing cluster mass. The amount of tidal disruption experienced by the live galaxy models will be partly controlled by the choice of orbits, as explained in Section~\ref{sec:trajectories}.

\subsubsection{Generating the background potentials}
\label{subsubsec:backPot}
The velocity dispersions of galaxy clusters are well known to require some additional unseen mass in MOND, although significantly less than in $\Lambda$CDM \citep{mondgalcl}. For our purposes, the manner in which the cluster has formed is of no importance. We simply require a background gravitational potential that gives rise to reasonable galaxy dynamics within the cluster. This is observationally motivated, and therefore such a potential must somehow arise in the MOND context as well.

Thus we must consider how to include a static background cluster potential within our simulations. RAMSES (and by extension RAyMOND) allows the user to include an analytic \textit{density} field within the simulation, which is then added to the density field reconstructed from the particles, before being used as the source in the Poisson equation. We will determine a density distribution that corresponds to the gravitational potential of a cluster with an NFW profile, given by
\begin{equation}
\rho_N = \frac{M_{200}}{4\pi R_{200}^3} \frac{g(c) c^2}{s(1+cs)^2}
\end{equation}
where
\begin{equation}
g(c) = \left(\ln(1+c) - c/(1+c) \right)^{-1}
\end{equation}
and $s = r/R_{200}$.

We must modify this density profile in order to obtain the required NFW potential given by
\begin{equation}
\phi = -g(c)\frac{GM_{200}}{R_{200}}\frac{\ln(1+cs)}{s}.
\end{equation}
To achieve this, we substitute the above potential into the modified Poisson equation of the AQUAL formulation of MOND:
\begin{equation}
\label{aqual_eq}
\nabla \cdot \left( \mu\left(\frac{|\nabla \phi |}{a_0} \right) \nabla \phi \right) = 4\pi G \rho.
\end{equation}
Thus the density used for the background cluster in the simulations is given by
\begin{equation}
\label{density}
\rho_M = \frac{1}{4\pi G} \mu' \left(\frac{|\nabla \phi|}{a_0}\right) \frac{1}{a_0} \nabla | \nabla \phi | \cdot \nabla \phi + \mu \left( \frac{|\nabla \phi|}{a_0} \right) \rho_N,
\end{equation}
where $\mu'(x)$ is the derivative of the MOND interpolation function with respect to its argument. The first term on the right hand side of the above equation has negative values at all radii, in order to reproduce a (shallower) Newtonian NFW potential within MOND gravity. Given that the NFW potential is asymptotically $\ln(r)/r$, as compared to $\ln(r)$ for the asymptotic deep-MOND potential, the effective dark matter density we are using to reproduce an NFW potential for the cluster in MOND has negative values at large radii. Note that within approximately the virial radius the densities are positive.

It is worth pointing out that RAMSES/RAyMOND also provides the possibility for specifying the gravitational \textit{accelerations} experienced by the particles, i.e. a background analytic gravitational force field. This is typically used without the self-gravity of the particles being included, although it is a relatively simple matter to modify the code to add these accelerations as well. The reason for using a background \textit{density} rather than specifying a background gravitational force field is to allow us to model the EFE. In Section~\ref{subsec:switchingOffEFE} we will discuss the use of an analytic \emph{acceleration} field to enable us to deactivate the EFE, allowing us to quantify its impact on our models. The ability to use the analytic NFW potential to determine the exact tides in the model is one of the primary motivations for using the effective DM density described above, despite the negative densities at large radii.

\subsubsection{Trajectories and numerical setup}
\label{sec:trajectories}
The galaxies are placed on orbits within the Coma-mass cluster that correspond to either weak or strong tidal disruption, depending on the closeness of the pericentric passage. This is controlled by changing the initial velocity of the galaxy. We choose two different initial velocity vectors, which lead to two different orbital trajectories. Furthermore, we change the plane of orbit within the cluster to explore the effect of the tidal interaction acting within the disk plane, or perpendicular to the disk. Thus we have orbits that lie within the $xz$ plane, referred to as ``xz'', and two orbits that lie within the $xy$ plane, referred to as ``xyRet'' and ``xyPro''. The former corresponds to retrograde motion of the galaxy around the cluster centre (the galaxies rotate counter-clockwise, and the retrograde orbit is clockwise) and the latter corresponds to prograde motion. For the Virgo-mass cluster we concentrate only on the prograde orbits (for which the tidal interaction is most disruptive), with three initial velocity vectors corresponding to weak, intermediate and strong tides.

\begin{figure*}
\centering
\begin{tabular}{cc}
\includegraphics[width=\bigFig\textwidth]{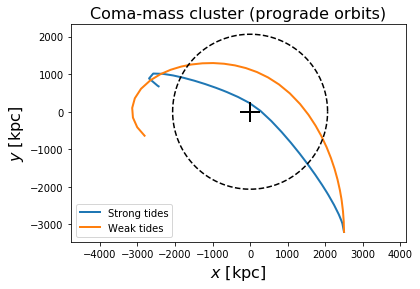} & \includegraphics[width=\bigFig\textwidth]{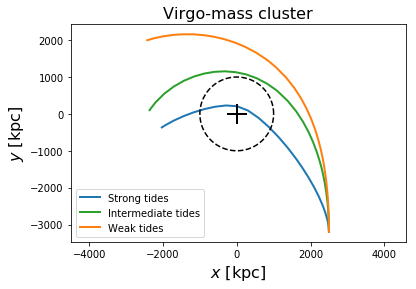} \\
\end{tabular}
\caption{Orbital trajectories in the \textit{Left:} Coma-mass cluster; \textit{Right:} Virgo-mass cluster. These trajectories are determined using the centre-of-mass location of the galaxy. The black dashed circle indicates $R_{200}$ of the cluster, and the cross indicates the cluster centre. We only show the prograde orbits for the Coma-mass cluster. All the other orbital trajectories are very similar: they are just rotations of these orbits around the $x$ and $y$ axes.}
\label{fig:trajectories}
\end{figure*}

The model names and parameters are specified in Table \ref{modParamsTable}. We use a naming convention whereby letters indicate the galaxy mass, tidal strength and orbit type. The orbital trajectories of the models in both cluster backgrounds are given in Fig.~\ref{fig:trajectories}. In the Coma-mass cluster the pericentric distance for the tidally weak orbit is $\sim 1090$~kpc and $\sim 130$~kpc for the tidally strong orbit. These correspond to $0.5$ $R_{200}$ and $\sim 0.06$ $R_{200}$ respectively. Note that we only show the prograde orbits in Fig.~\ref{fig:trajectories}. The retrograde orbits are the same, but rotated $180^{\degree}$ in the $y$-axis. The perpendicular orbit is given by rotating the prograde orbit $90^{\degree}$ around the $x$-axis. Note that the disks of the galaxies always lie in the $xy$ plane and always rotate in the same direction. In the Virgo-mass cluster the pericentre distances for the weak, intermediate and strong tides are $\sim 1650$~kpc, $\sim 1100$~kpc and $\sim 208$~kpc respectively.

\begin{table*}
\centering
\begin{tabular}{c c c c c c c} 
 \hline
 Background cluster & Model name & Galaxy mass [M$_{\odot}$] & Tide strength & Orbit plane & Initial position [kpc] & Initial velocity [km/s] \\
 \hline\hline
 Coma-mass & mlwxz & $10^9$ (Low) & Weak & xz & $(2500,0,3200)$ & $(0,0,-831)$ \\ 
           & mlwxyPro & $10^9$ (Low) & Weak & xy & $(2500,-3200,0)$ & $(0,831,0)$ \\
           & mlwxyRet & $10^9$ (Low) & Weak & xy & $(-2500,-3200,0)$ & $(0,831,0)$ \\
           & mlsxz & $10^9$ (Low) & Strong & xz & $(2500,0,3200)$ & $(0,0,-209)$ \\
           & mlsxyPro & $10^9$ (Low) & Strong & xy & $(2500,-3200,0)$ & $(0,209,0)$ \\
           & mlsxyRet & $10^9$ (Low) & Strong & xy & $(-2500,-3200,0)$ & $(0,209,0)$ \\
           & mhwxz & $10^{11}$ (High) & Weak & xz & $(2500,0,3200)$ & $(0,0,-831)$ \\ 
           & mhwxyPro & $10^{11}$ (High) & Weak & xy & $(2500,-3200,0)$ & $(0,831,0)$ \\
           & mhwxyRet & $10^{11}$ (High) & Weak & xy & $(-2500,-3200,0)$ & $(0,831,0)$ \\
           & mhsxz & $10^{11}$ (High) & Strong & xz & $(2500,0,3200)$ & $(0,0,-209)$ \\
           & mhsxyPro & $10^{11}$ (High) & Strong & xy & $(2500,-3200,0)$ & $(0,209,0)$ \\
           & mhsxyRet & $10^{11}$ (High) & Strong & xy & $(-2500,-3200,0)$ & $(0,209,0)$ \\
           & mlinfall & $10^9$ (Low) & Infall & xz & $(5500,-6200,0)$ & $(0,0,0)$ \\
           & mlwxyPro\_n5 & $10^9$ (Low) & Weak & xy & $(2500,-3200,0)$ & $(0,209,0)$ \\
           & mlwxz\_noEFE & $10^9$ (Low) & Weak & xz & $(2500,0,3200)$ & $(0,0,-831)$ \\
           & mlsxyPro\_noEFE & $10^9$ (Low) & Strong & xy & $(2500,-3200,0)$ & $(0,209,0)$ \\
           & mhwxz\_noEFE & $10^{11}$ (High) & Weak & xz & $(2500,0,3200)$ & $(0,0,-831)$ \\
           & mhsxyPro\_noEFE & $10^{11}$ (High) & Strong & xy & $(2500,-3200,0)$ & $(0,209,0)$ \\
\hline
 Virgo-mass & mlwxyPro & $10^9$ (Low) & Weak & xy & $(2500,-3200,0)$ & $(0,831,0)$ \\
            & mlixyPro & $10^9$ (Low) & Intermediate & xy & $(2500,-3200,0)$ & $(0,622,0)$ \\
            & mlsxyPro & $10^9$ (Low) & Strong & xy & $(2500,-3200,0)$ & $(0,209,0)$ \\
            & mhwxyPro & $10^{11}$ (High) & Weak & xy & $(2500,-3200,0)$ & $(0,831,0)$ \\
            & mhixyPro & $10^{11}$ (High) & Intermediate & xy & $(2500,-3200,0)$ & $(0,622,0)$ \\
            & mhsxyPro & $10^{11}$ (High) & Strong & xy & $(2500,-3200,0)$ & $(0,209,0)$ \\
            & mlsxyPro\_noEFE & $10^9$ (Low) & Strong & xy & $(2500,-3200,0)$ & $(0,209,0)$ \\
            & mhsxyPro\_noEFE & $10^{11}$ (High) & Strong & xy & $(2500,-3200,0)$ & $(0,209,0)$ \\
 \hline
\end{tabular}
\caption{Names and parameters of all models considered in this study. The letters in the model names correspond to model properties: `l' is low mass; `h' is high mass; `w' is weak tides; `i' is intermediate tides; `s' is strong tides. The orbit types are `xz' (orbit plane is perpendicular to the galactic plane), `xyRet' and `xyPro' (retrograde and prograde orbits with orbit plane parallel to the disk plane). All the Coma-mass models begin the simulation with a clustercentric radius of $\sim 2R_{200}$, except for the ``mlinfall'' model which begins at $\sim 4R_{200}$. The Virgo-mass models all begin with a clustercentric radius of $\sim 4R_{200}$, given the smaller size of this cluster. All models use the simple MOND transition function with $n=1$ except for the ``mlwxyPro\_n5'' model which uses $n=5$. Models that do not include an EFE have ``\_noEFE'' appended to the model name.}
\label{modParamsTable}
\end{table*}

\begin{figure*}
\centering
\begin{tabular}{cc}
\includegraphics[width=\bigFig\textwidth]{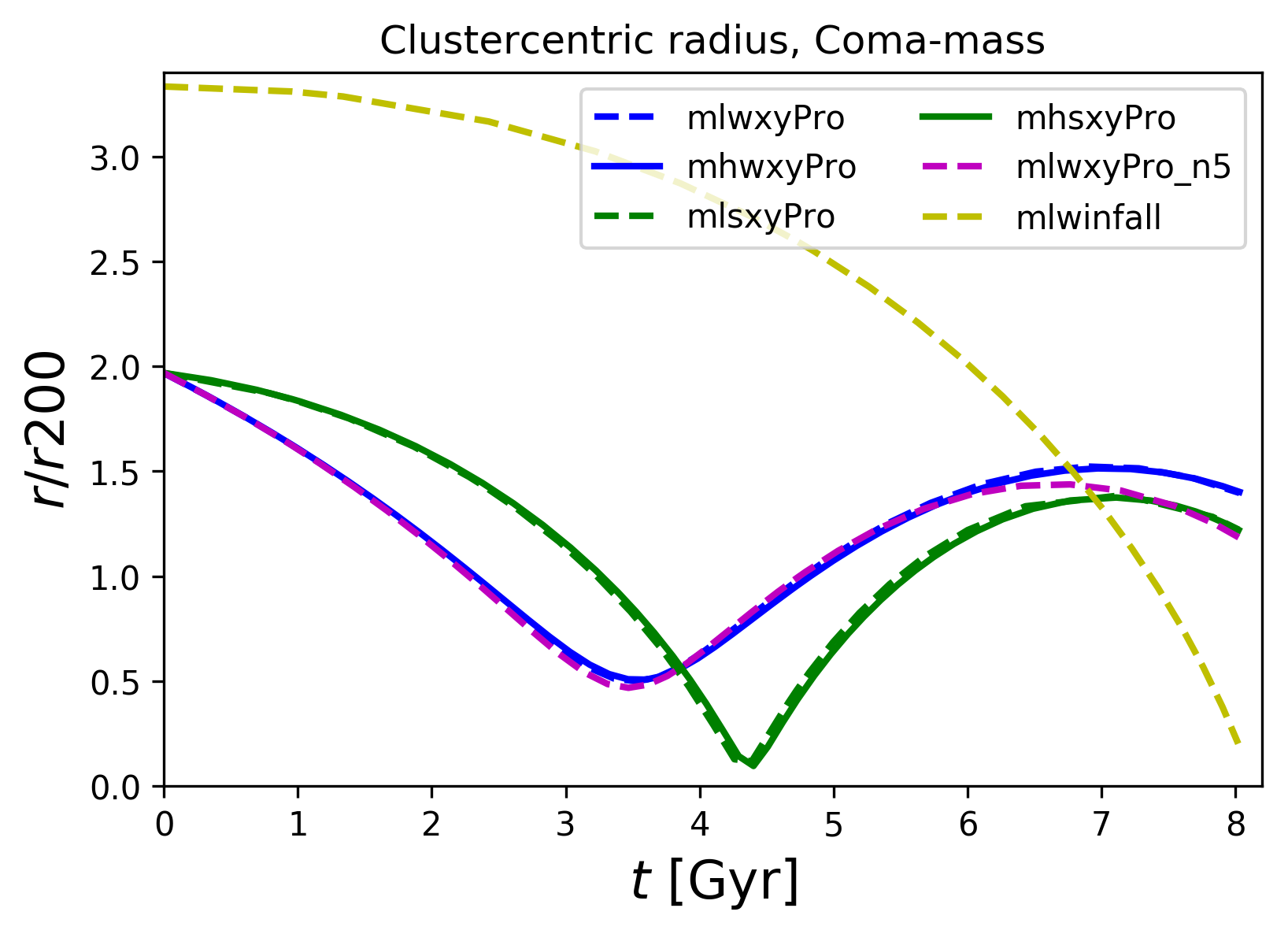} & \includegraphics[width=\bigFig\textwidth]{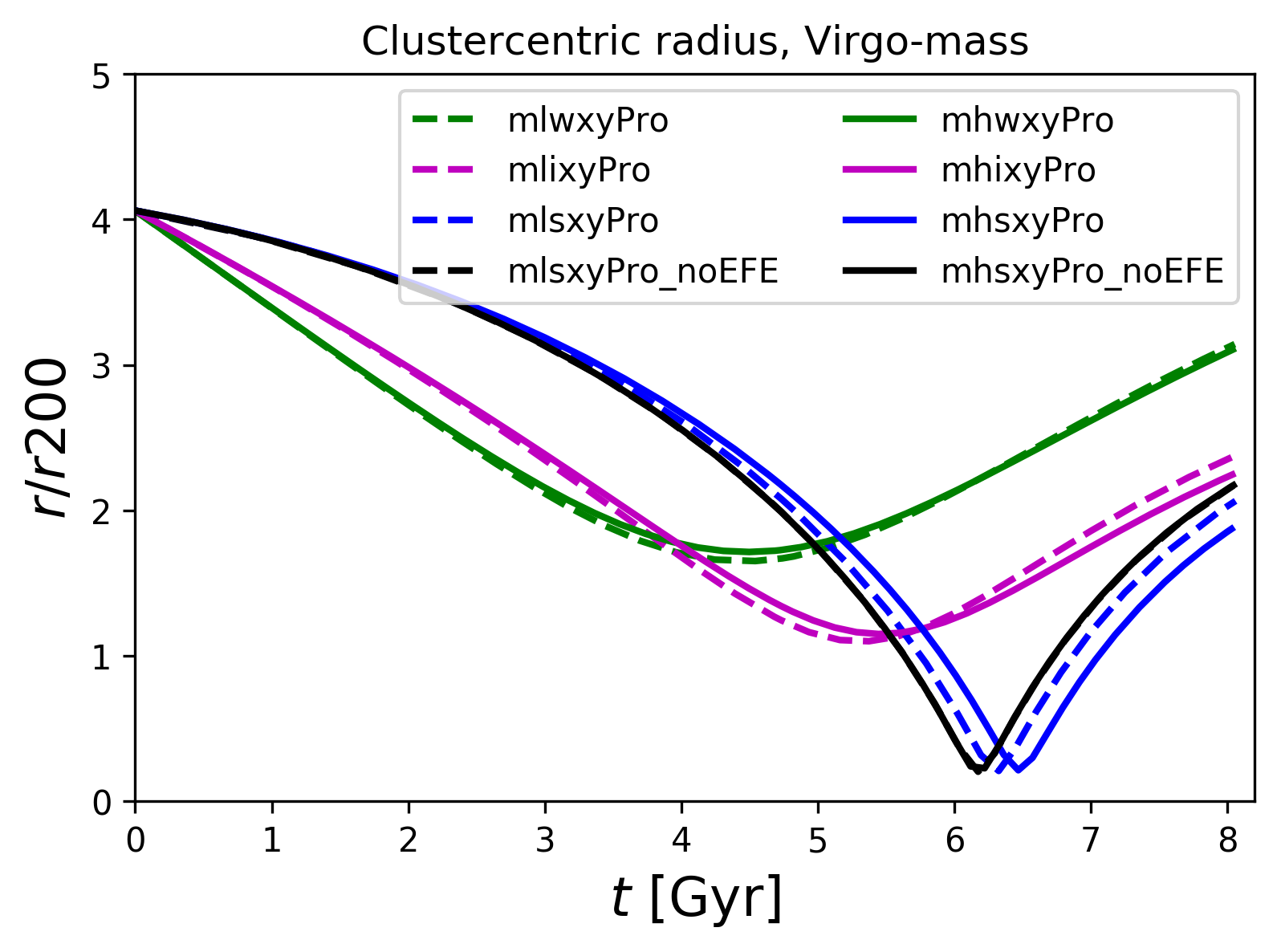}
\end{tabular}
\caption{Clustercentric radii (normalised by $R_{200}$ of the background cluster) of prograde orbits and the infall model as a function of time. \textit{Left}: Coma-mass cluster; \textit{Right}: Virgo-mass cluster.}
\label{setup:orbitVels}
\end{figure*}

The box length in the cluster simulations is increased to $8000$~kpc to accommodate the orbital trajectories. The maximum refinement level is set to level 15 for the low mass models and 14 for the high mass models, meaning the maximum resolution achieved is $0.244$~kpc and $0.488$~kpc respectively. The minimum resolution is now $31.25$~kpc.

\subsubsection{Switching off the EFE}
\label{subsec:switchingOffEFE}
It would be invaluable for our analysis to be able to somehow deactivate the EFE within our simulations. Fortunately, RAyMOND provides a convenient means for doing so. Our strategy is the following. Firstly, we include the use of a background analytical \emph{acceleration} field within the simulation, as opposed to the analytic density discussed in Section~\ref{subsubsec:backPot}, in the normal manner provided by RAMSES. This background acceleration field is easily calculated for all models, as we are fixing the background \textit{potential} to be that of an NFW profile with the parameters given earlier. Typically the inclusion of an analytic external acceleration field in RAMSES takes the place of the internal self-gravity of the N-body particles. Therefore, we slightly modify the code to ensure that the gravitational solver is used to determine the self-gravity of the galaxy as normal, with the background accelerations then added on before the particle positions are updated.

The important point is that the background potential now makes \textit{no appearance} in the gravitational solver. The internal self-gravity of the galaxy is calculated without considering the background cluster, and therefore there is no external field effect. The overall motion of the galaxy (and therefore the motion of its constituent particles) within the cluster is accounted for by adding the analytically determined acceleration vectors after the gravitational solver has concluded. Therefore the \textit{tidal disruption} of the background cluster is still included in the simulation.

We run several models in which the EFE has been switched off in the manner just described, as listed in Table~\ref{modParamsTable}.

\subsection{Newtonian models}
We have also run models in Newtonian gravity, maintaining the dark matter halos generated by the DICE code, as a comparison sample. Upon analysis of these models we found that the galaxies undergo essentially no evolution due to interaction with the background cluster potential. The galaxies exhibit very little tidal disruption, even for strong tides, essentially no mass loss (although their halos are gradually stripped by the cluster tides), no kinematical evolution and no morphological evolution. These results are to be expected for Newtonian galaxies deeply embedded in dark matter halos on first infall trajectories, as we are considering here. Multiple studies (see e.g. \citealp{roryMassLoss}) have demonstrated that the dark matter halo must be stripped by as much as $80-90\%$ before the bayronic material begins to be affected. This implies that the cluster potential can only begin to have an impact on the galaxy on subsequent pericentric passages. The fact that our MOND galaxies show clear effects of the interaction with the cluster even on first infall is therefore already an important difference to the picture seen in the $\Lambda$CDM cosmological model.

In summary, given the unsurprising results of the Newtonian models in DM halos, we deliberately do not include the Newtonian models in this paper, and instead focus entirely on the MOND galaxies, with and without an external field effect.

\section{Results}
\label{sec:results}

\subsection{External field induced asymmetries}
\label{subsec:EFEasymm}
The central focus of our study is the consequence of the EFE on MOND galaxies as they fall into a galaxy cluster, an effect that does not exist in Newtonian gravity. In Sections~\ref{subsec:massLoss} and \ref{subsec:rotCurves} we will examine in more detail the mass loss and the evolution of the dynamical state of all models due to both the EFE and tidal disruption. In this Section we will analyse the impact of the EFE on the appearance of our infalling galaxies.

Firstly, as will be demonstrated later in Sections~\ref{subsec:massLoss} and \ref{subsec:rotCurves}, the MOND models without an EFE in weak tides are essentially undisturbed by the presence of the cluster potential. Therefore, for now, we will concentrate on the models without an EFE in \textit{strong} tides, and the models \textit{with} an EFE in both strong and weak tides. Furthermore, while the specific nature of the deformation induced in the galaxies depends on the disk orientation, we can demonstrate the effect very clearly by choosing the prograde orbit models, which are typically more affected by the tidal interaction than either of the other orbits, as is well known from galaxy-galaxy interaction simulations.

\begin{figure*}
\centering
\begin{tabular}{cccc}
\includegraphics[width=\weeFig\textwidth,trim={1cm 1cm 1cm 1cm},clip]{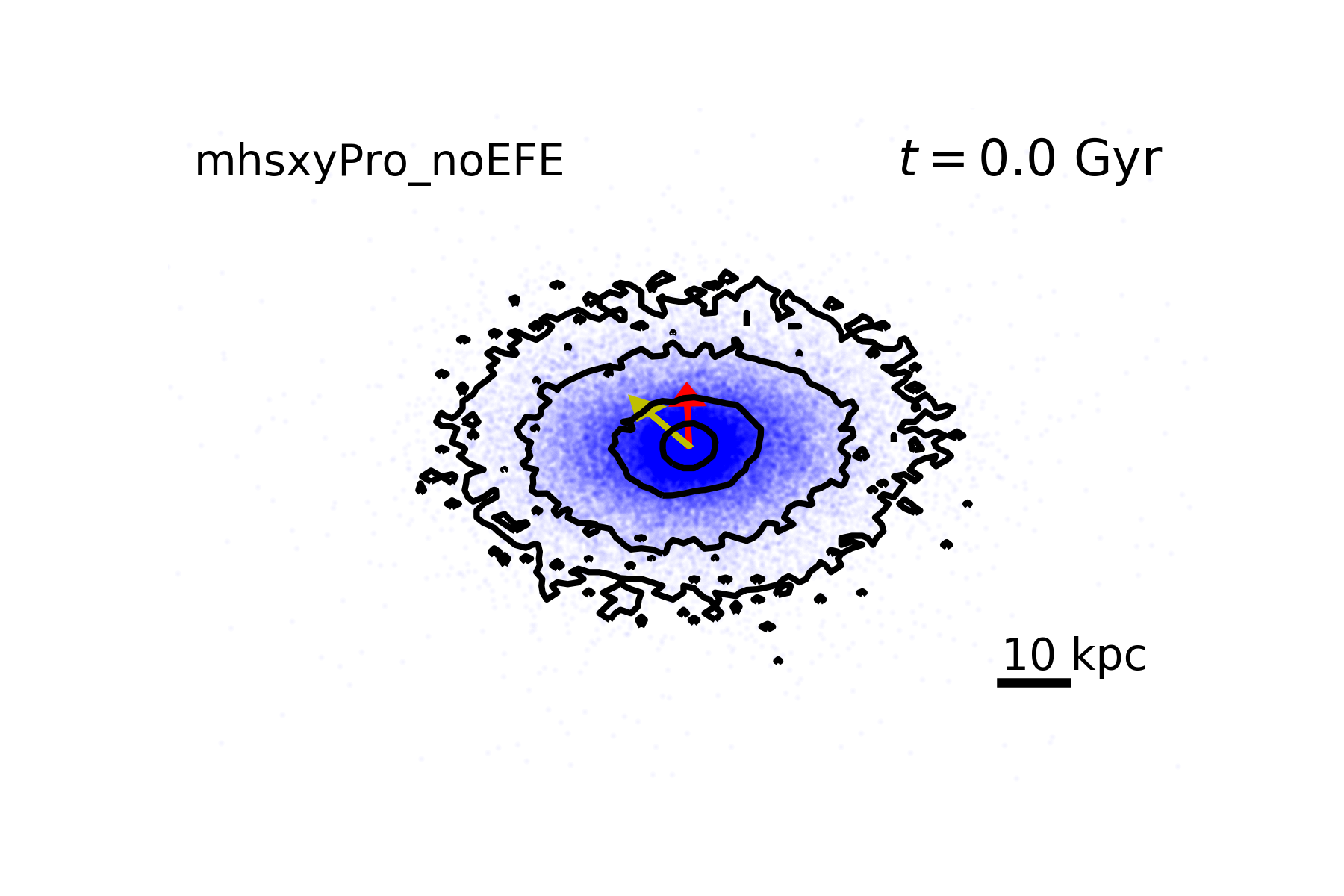} &
\includegraphics[width=\weeFig\textwidth,trim={1cm 1cm 1cm 1cm},clip]{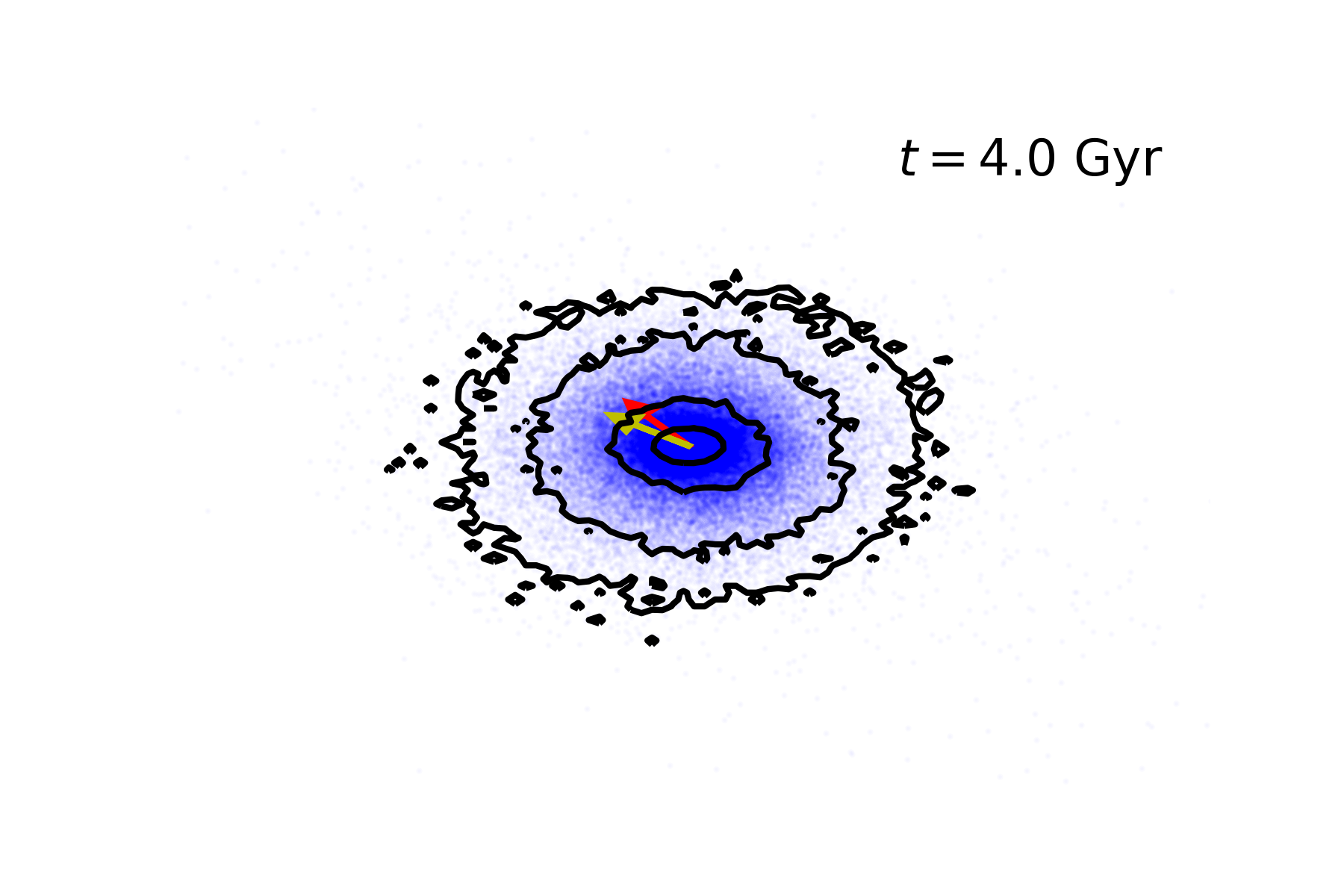} &
\includegraphics[width=\weeFig\textwidth,trim={1cm 1cm 1cm 1cm},clip]{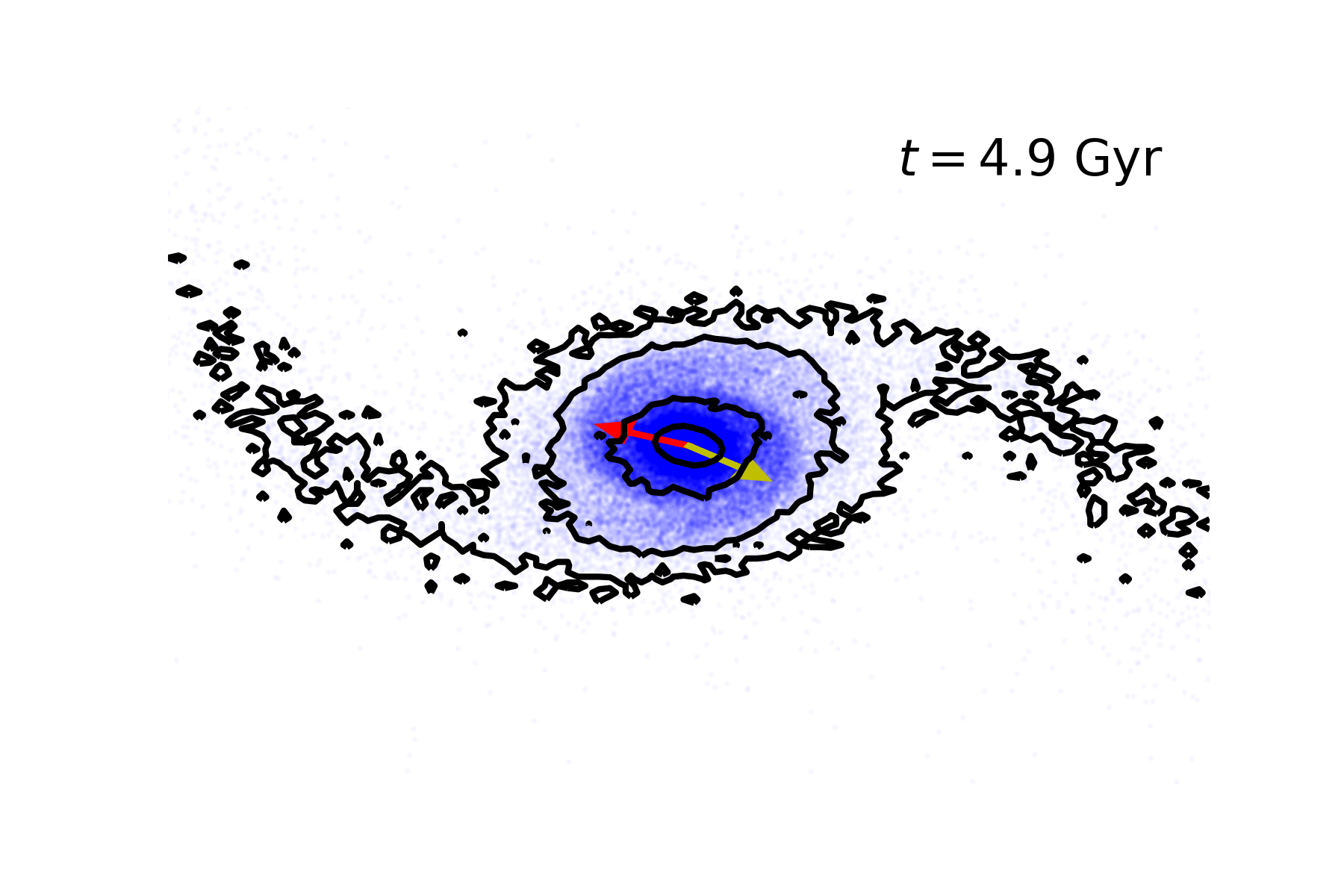} &
\includegraphics[width=\weeFig\textwidth,trim={1cm 1cm 1cm 1cm},clip]{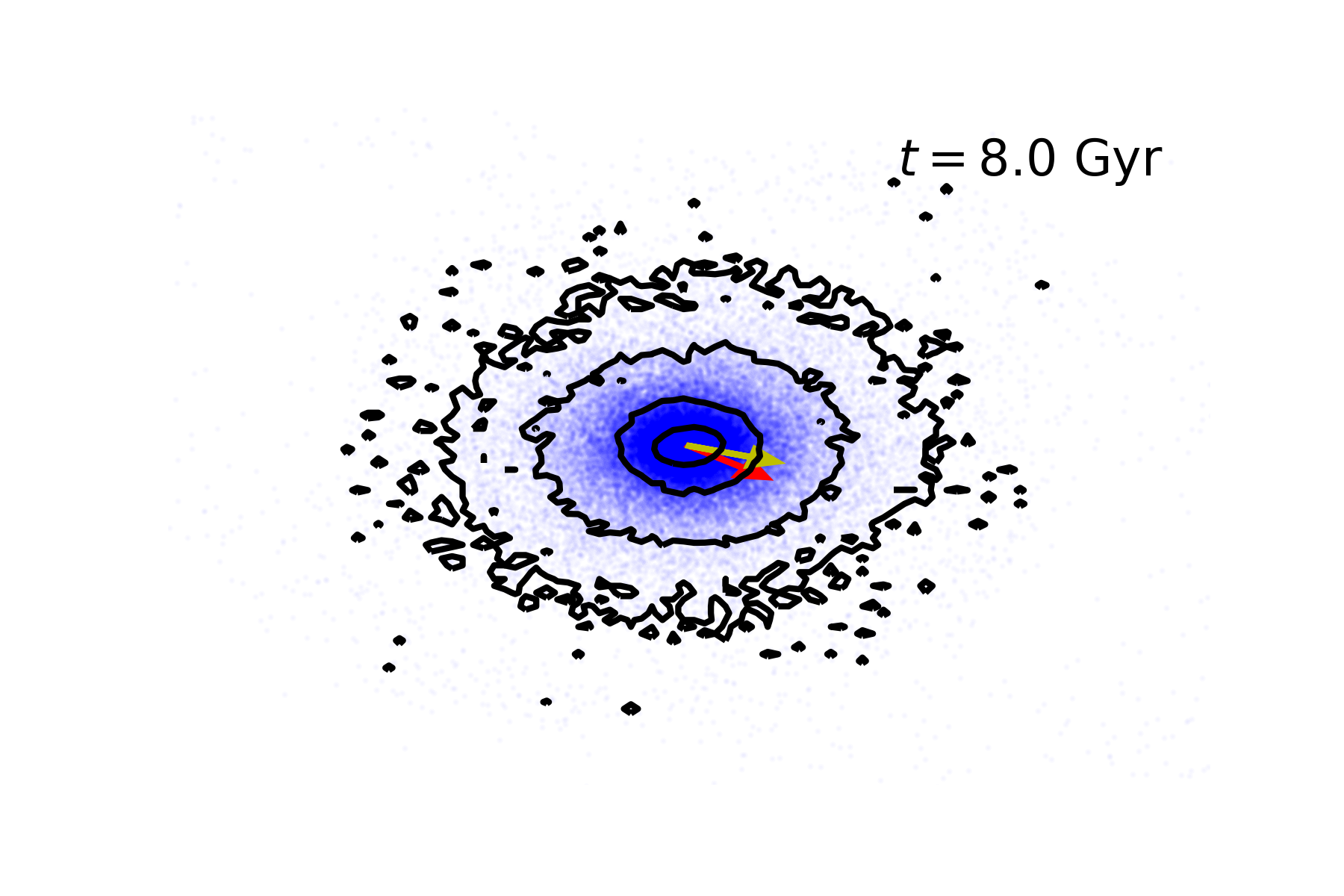} \\
\includegraphics[width=\weeFig\textwidth,trim={1cm 1cm 1cm 1cm},clip]{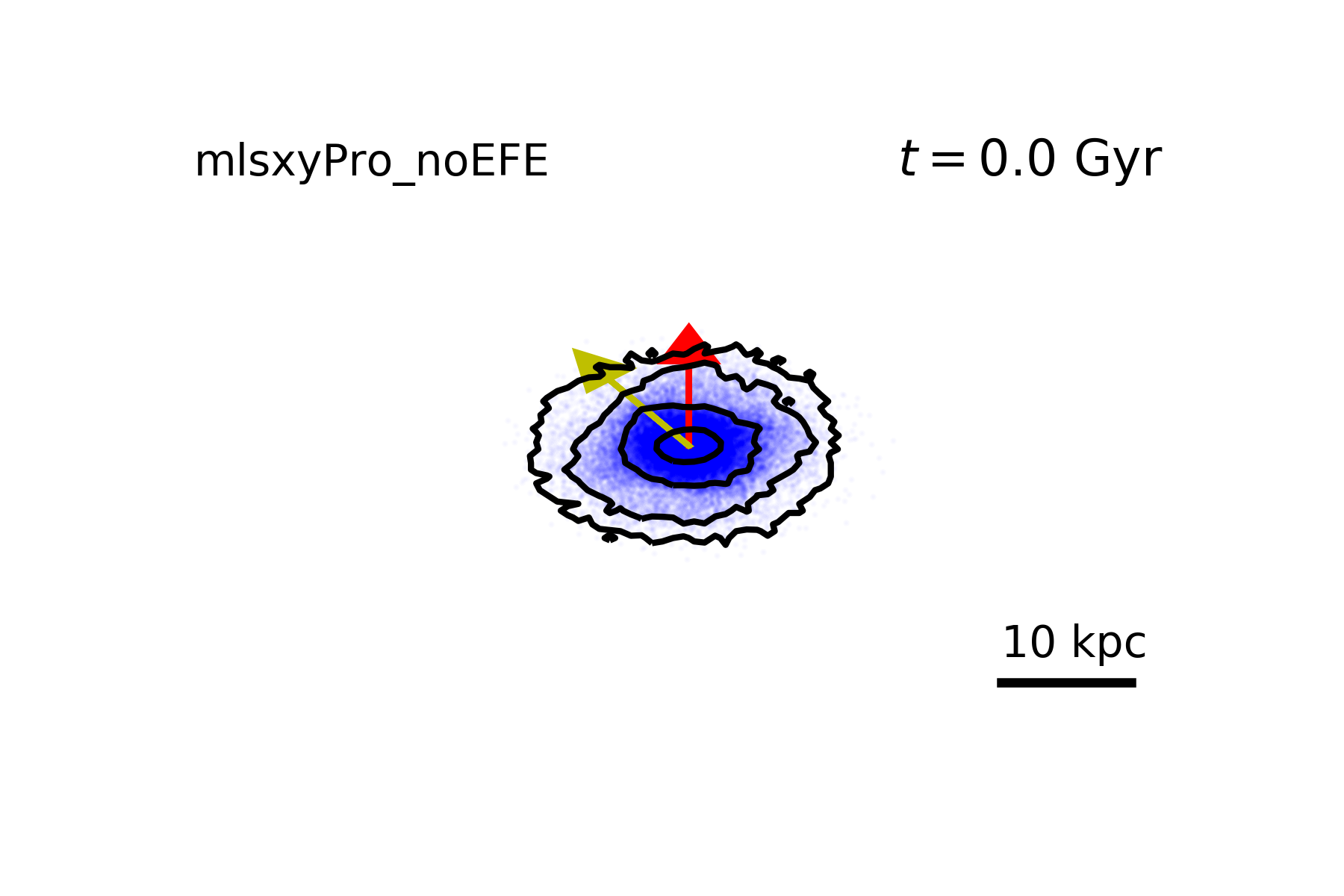} &
\includegraphics[width=\weeFig\textwidth,trim={1cm 1cm 1cm 1cm},clip]{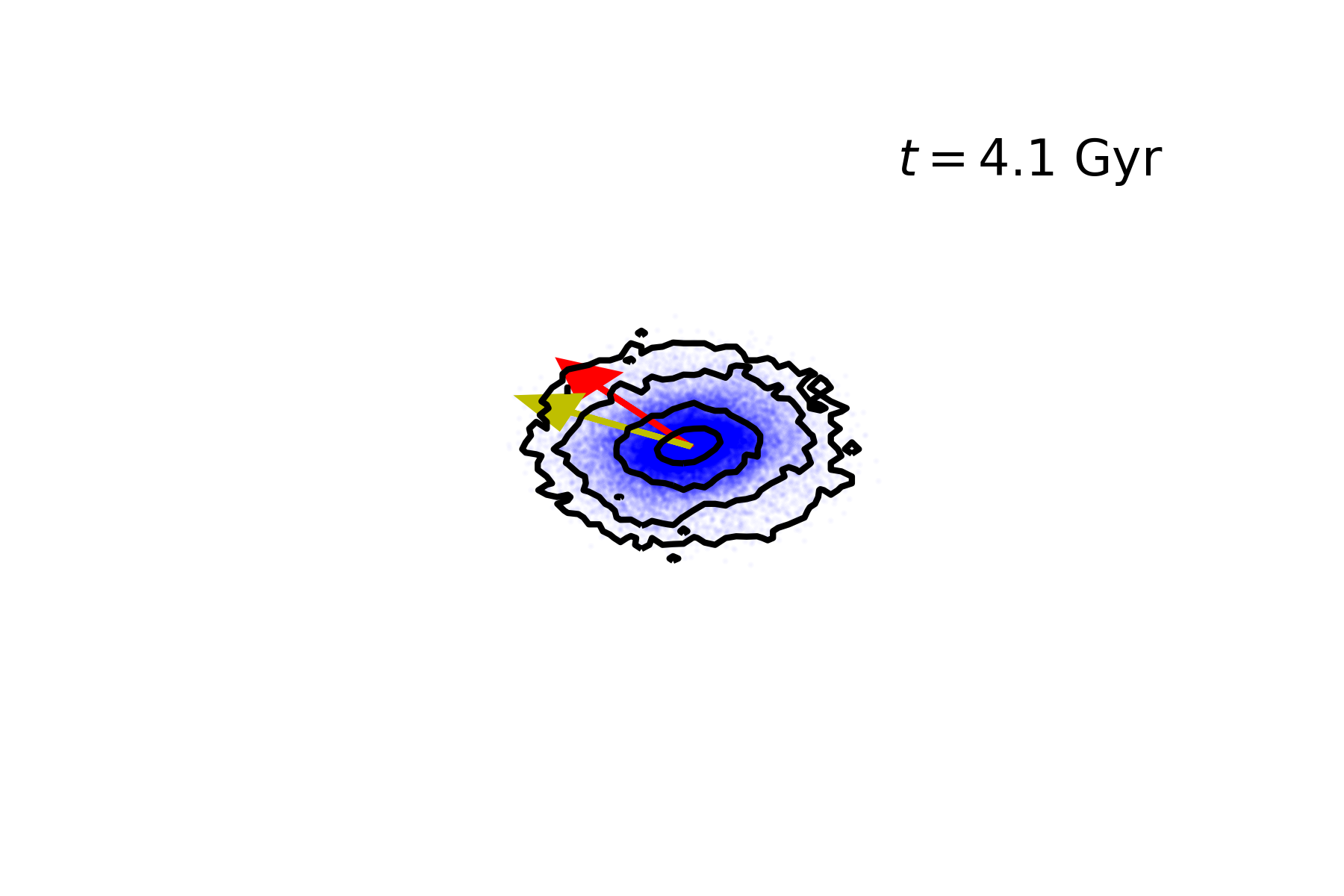} &
\includegraphics[width=\weeFig\textwidth,trim={1cm 1cm 1cm 1cm},clip]{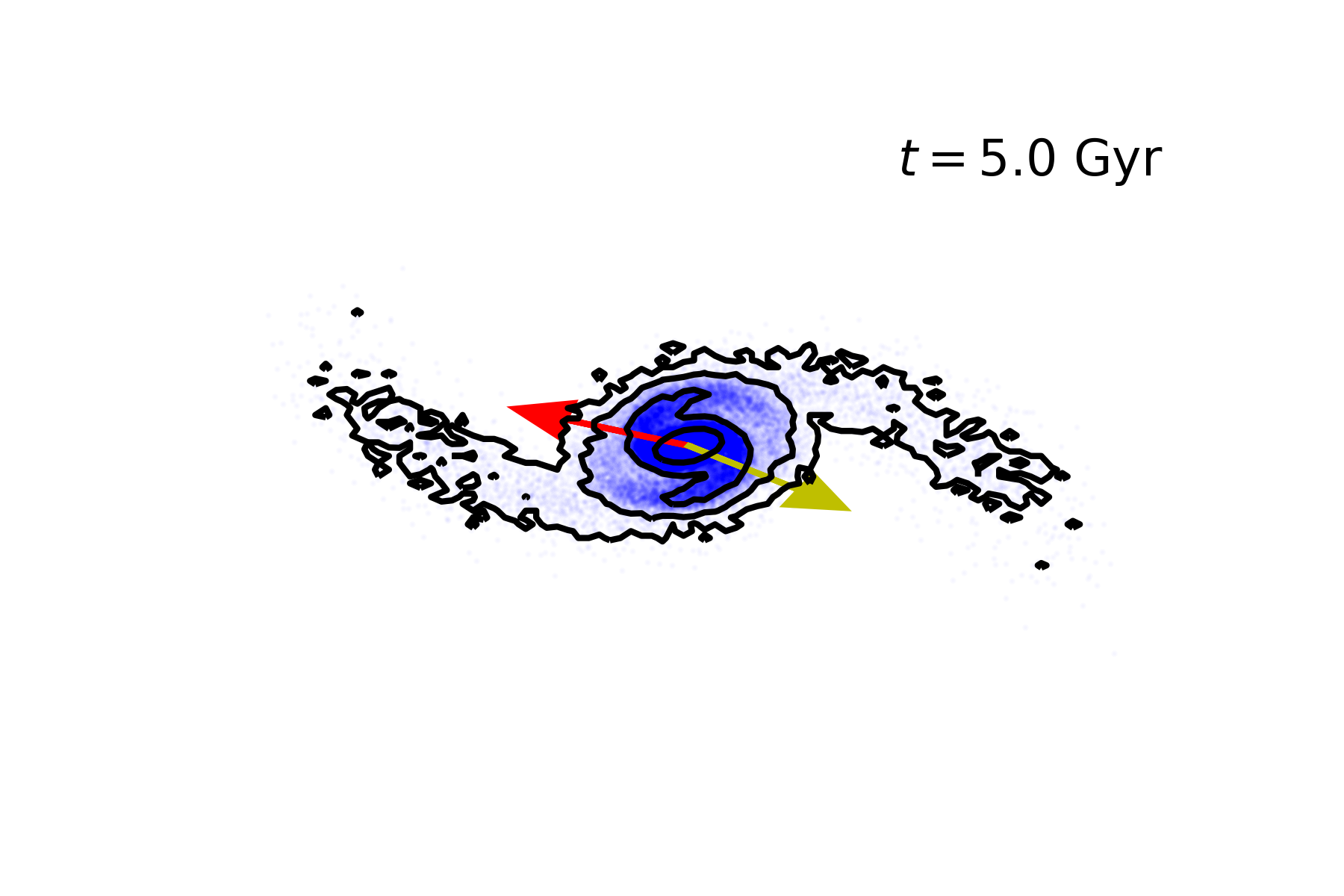} &
\includegraphics[width=\weeFig\textwidth,trim={1cm 1cm 1cm 1cm},clip]{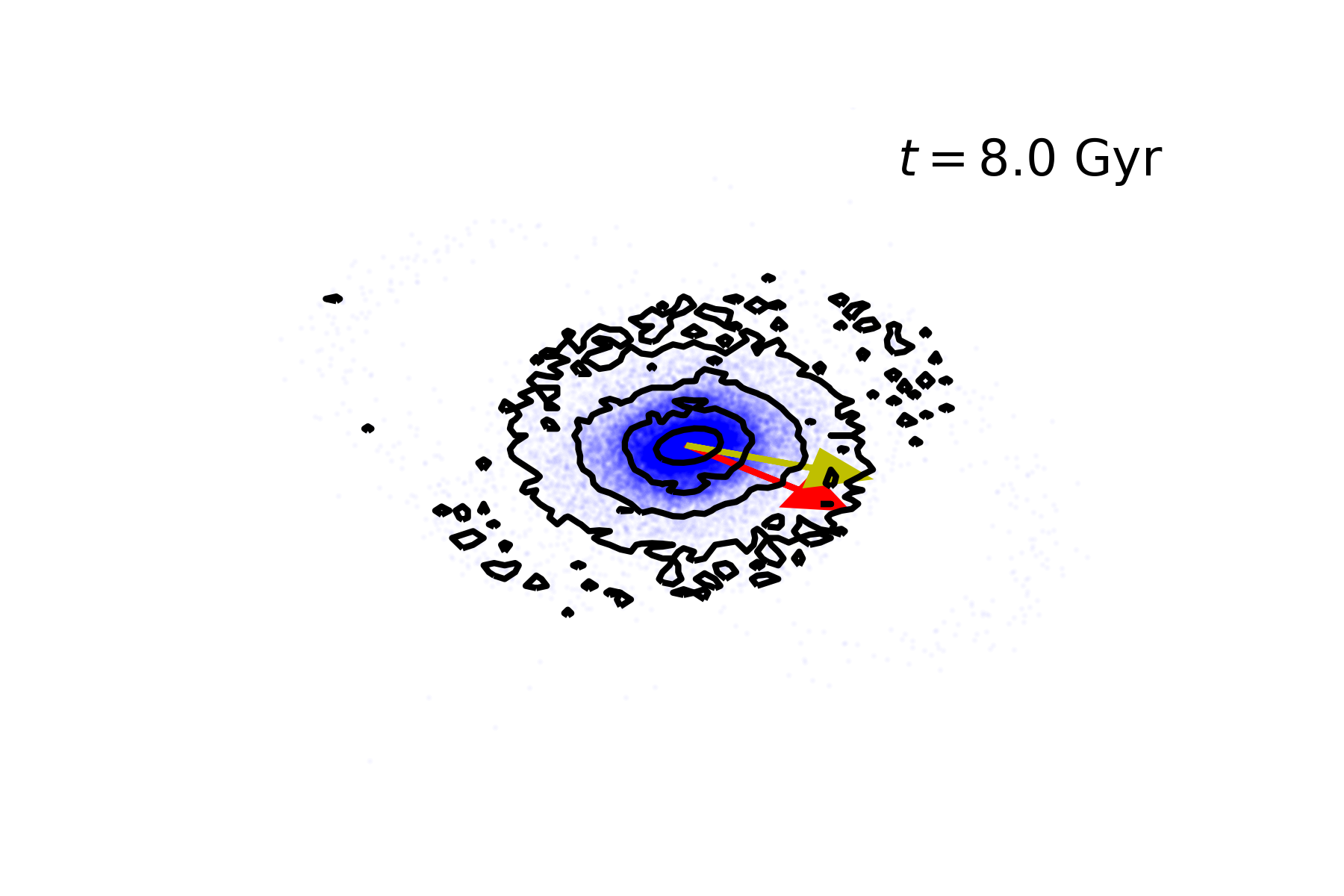}
\end{tabular}
\caption{\textit{Top row:} high mass model without an EFE; \textit{Bottom row:} low mass model without an EFE. In all plots the red arrow indicates the (normalised) centre-of-mass velocity vector, while the yellow arrow points towards the cluster centre. The tidal interaction induces the formation of highly symmetric tidal arms.}
\label{fig:asymmetryNewton}
\end{figure*}

\begin{figure*}
\centering
\begin{tabular}{cccc}
\includegraphics[width=\weeFig\textwidth,trim={1cm 1cm 1cm 1cm},clip]{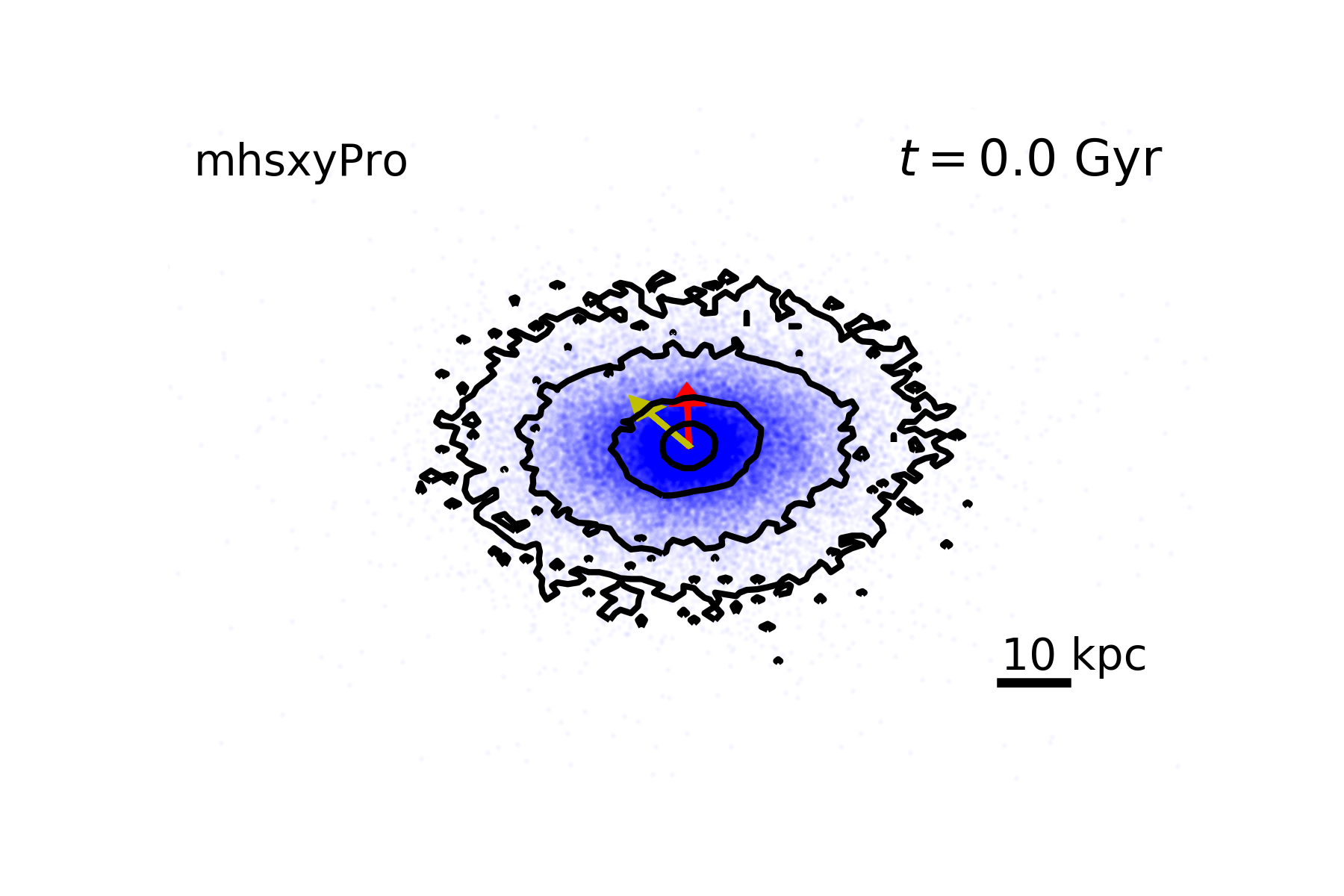} &
\includegraphics[width=\weeFig\textwidth,trim={1cm 1cm 1cm 1cm},clip]{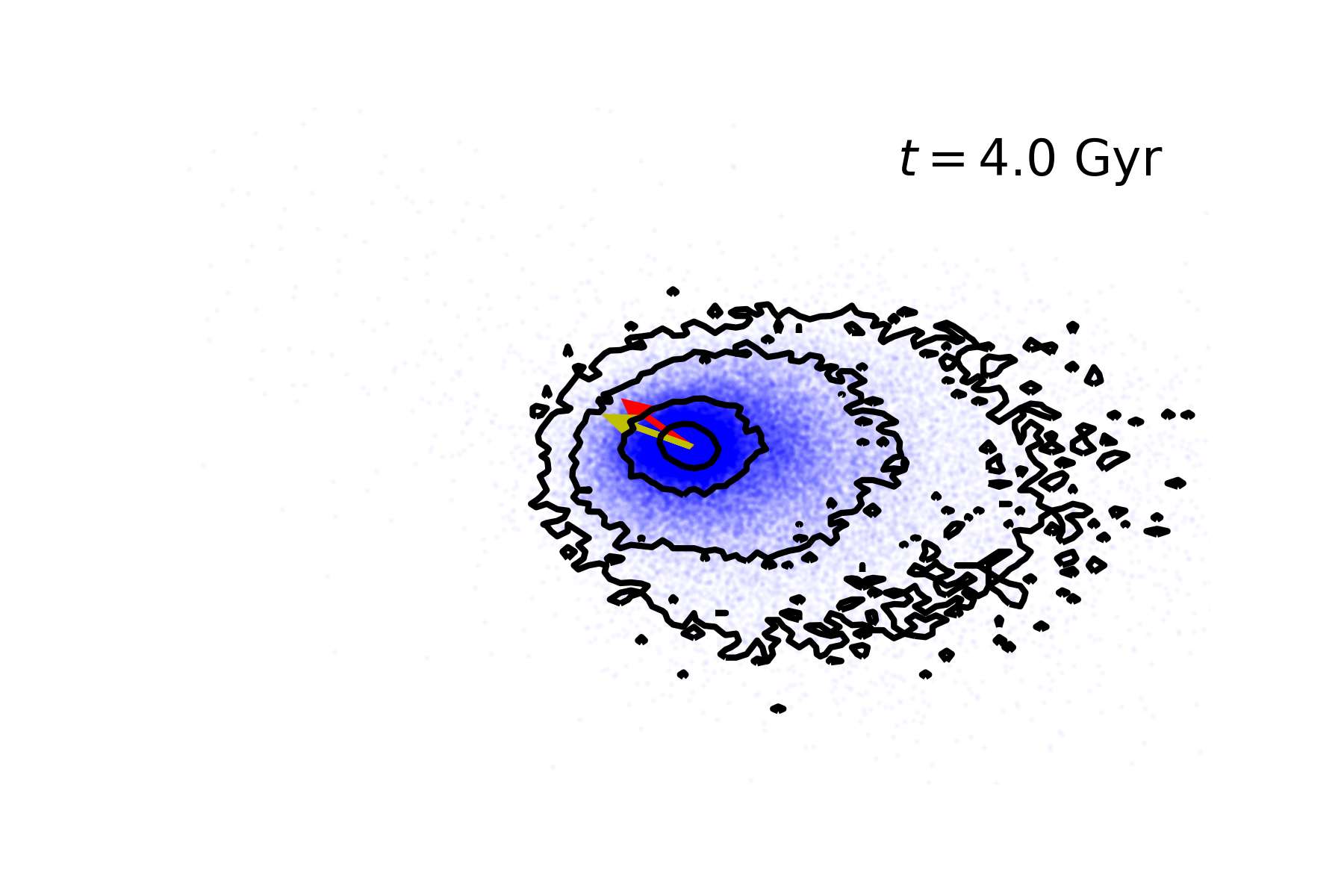} &
\includegraphics[width=\weeFig\textwidth,trim={1cm 1cm 1cm 1cm},clip]{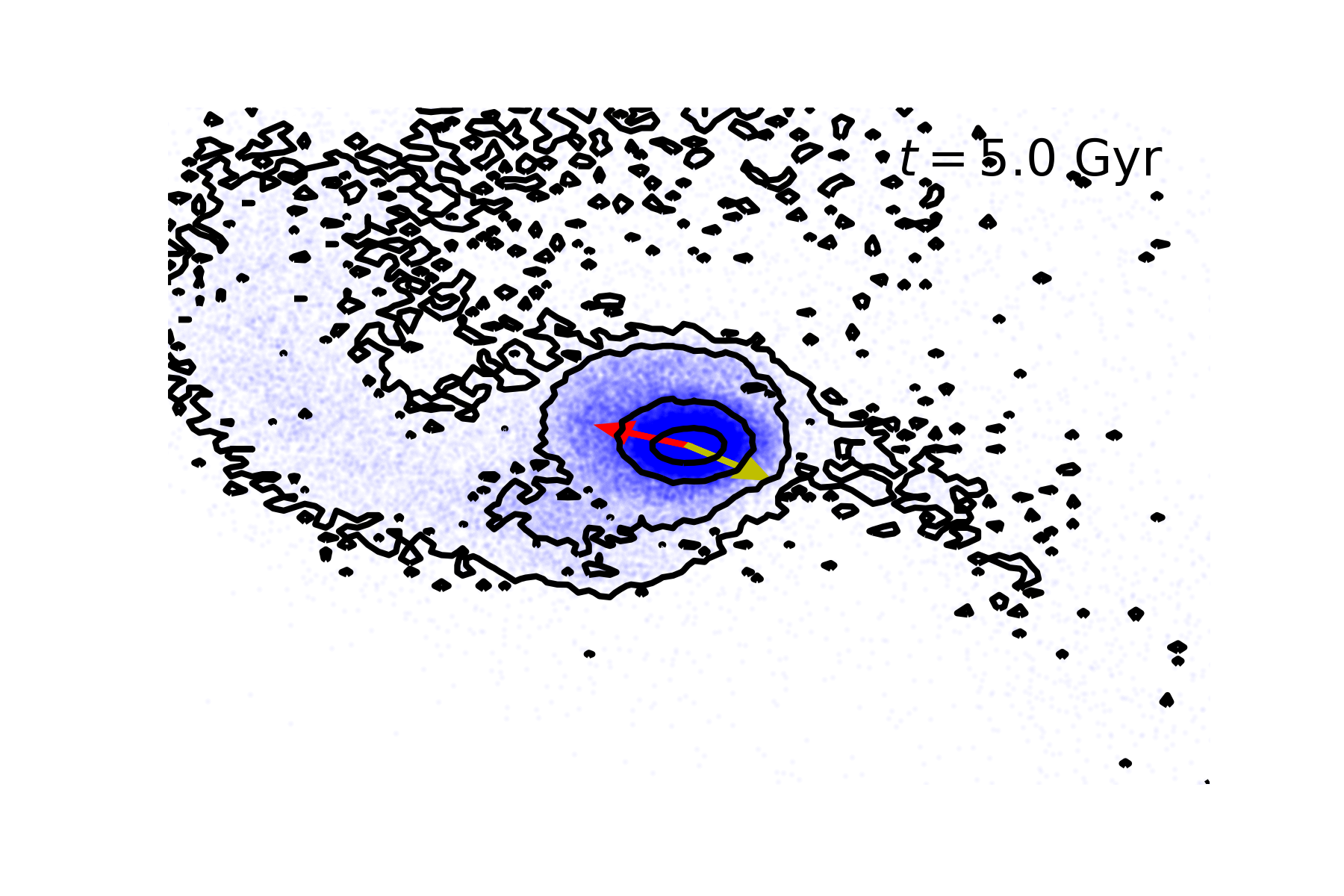} &
\includegraphics[width=\weeFig\textwidth,trim={1cm 1cm 1cm 1cm},clip]{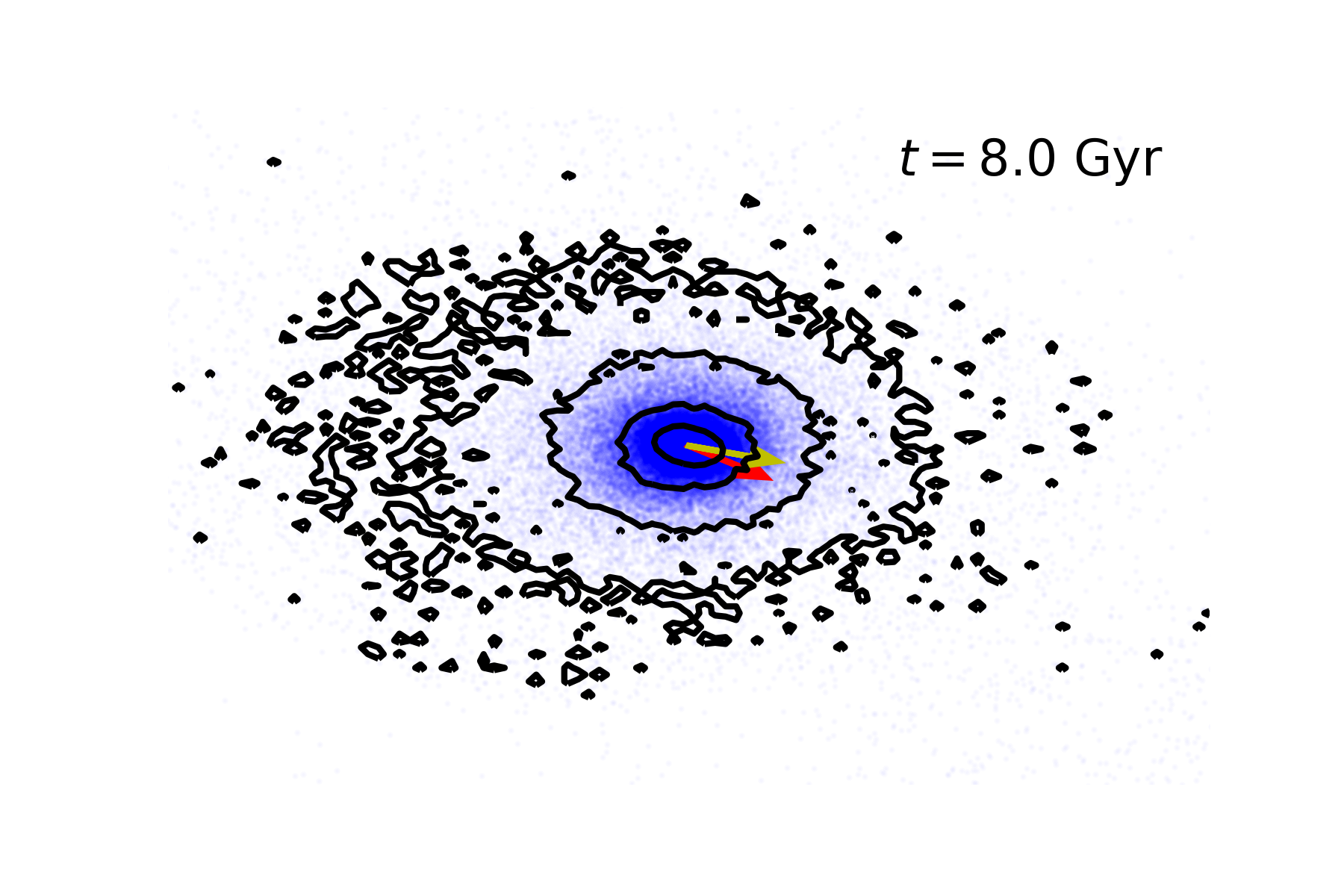} \\
\includegraphics[width=\weeFig\textwidth,trim={1cm 1cm 1cm 1cm},clip]{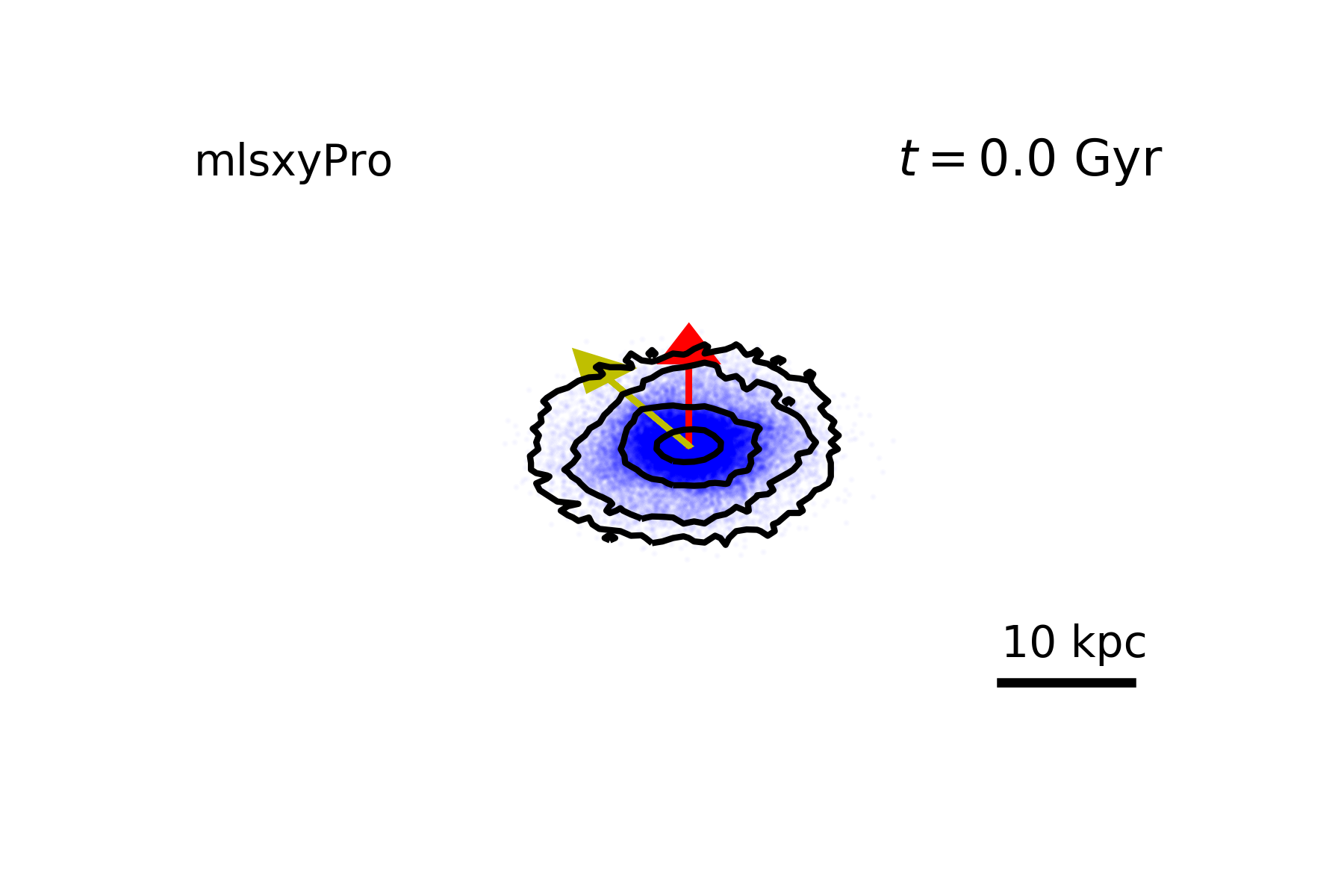} &
\includegraphics[width=\weeFig\textwidth,trim={1cm 1cm 1cm 1cm},clip]{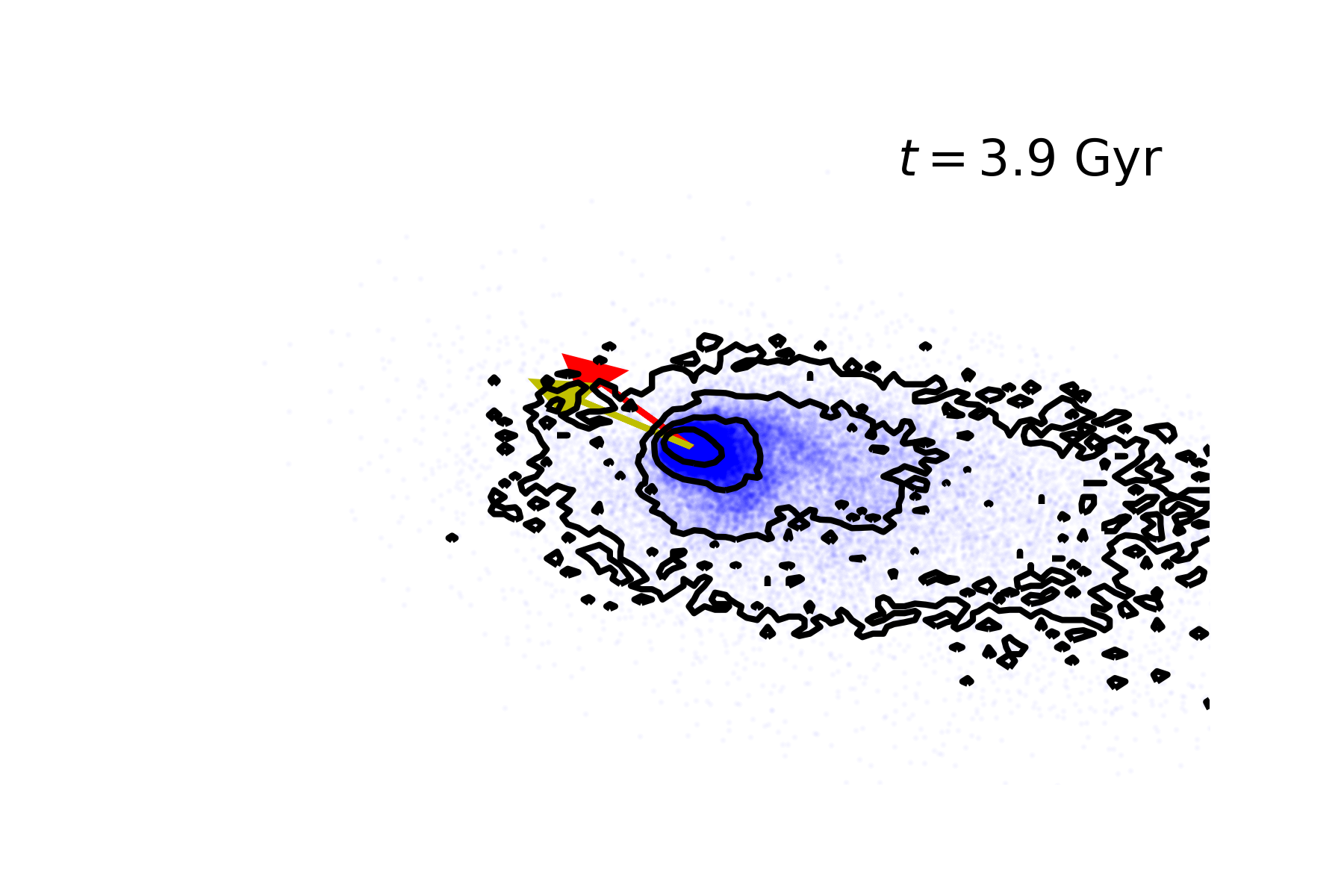} &
\includegraphics[width=\weeFig\textwidth,trim={1cm 1cm 1cm 1cm},clip]{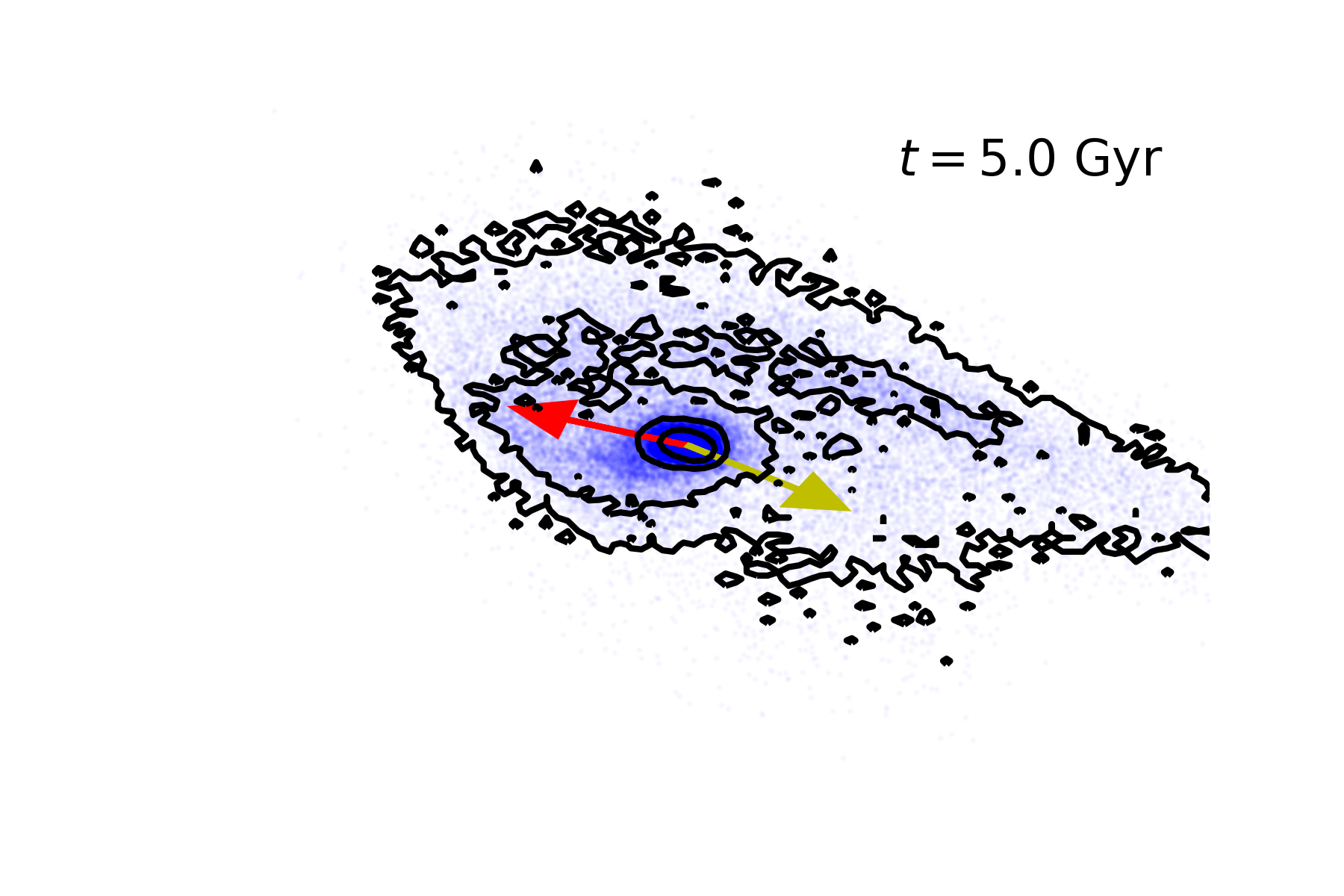} &
\includegraphics[width=\weeFig\textwidth,trim={1cm 1cm 1cm 1cm},clip]{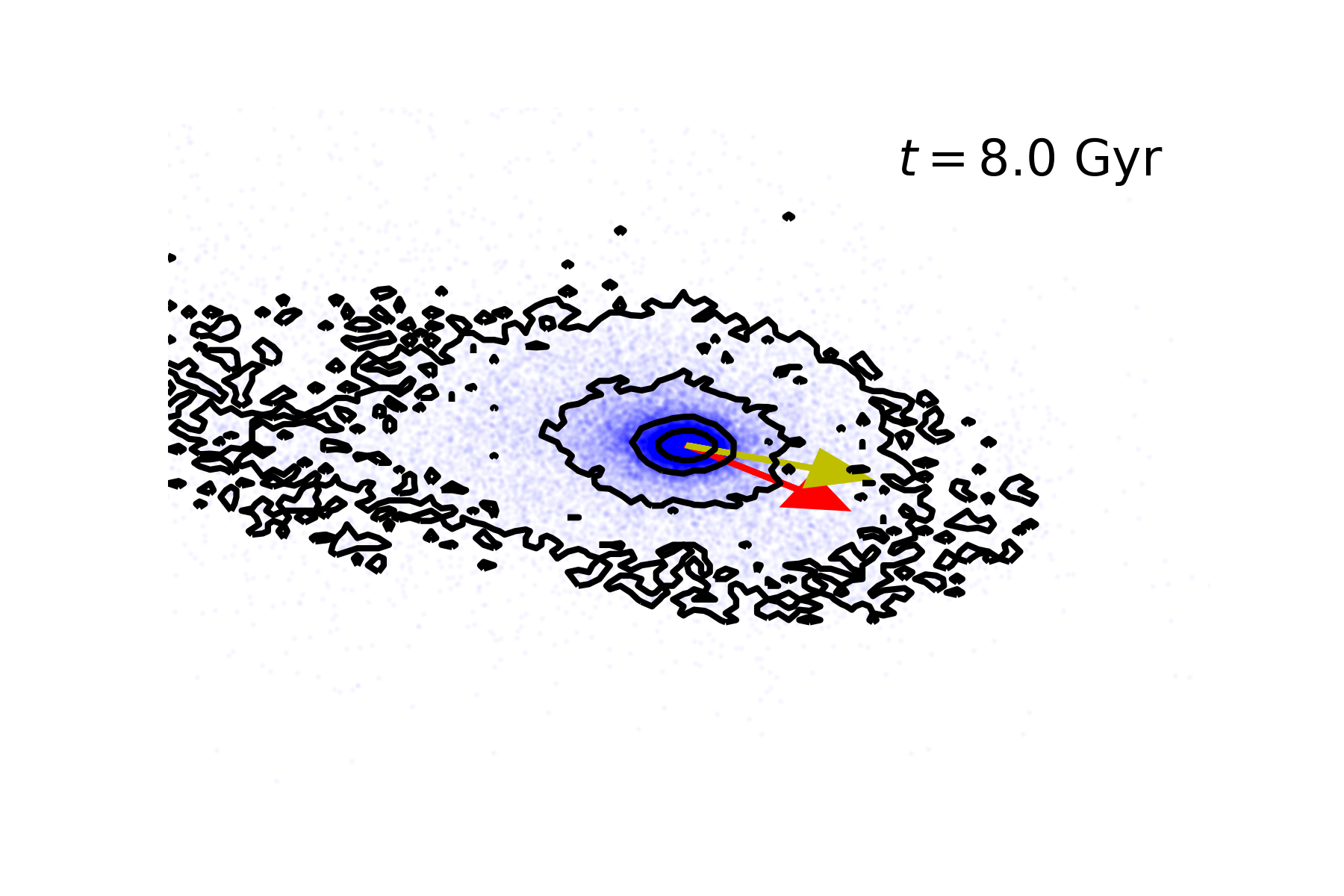} \\
\includegraphics[width=\weeFig\textwidth,trim={1cm 1cm 1cm 1cm},clip]{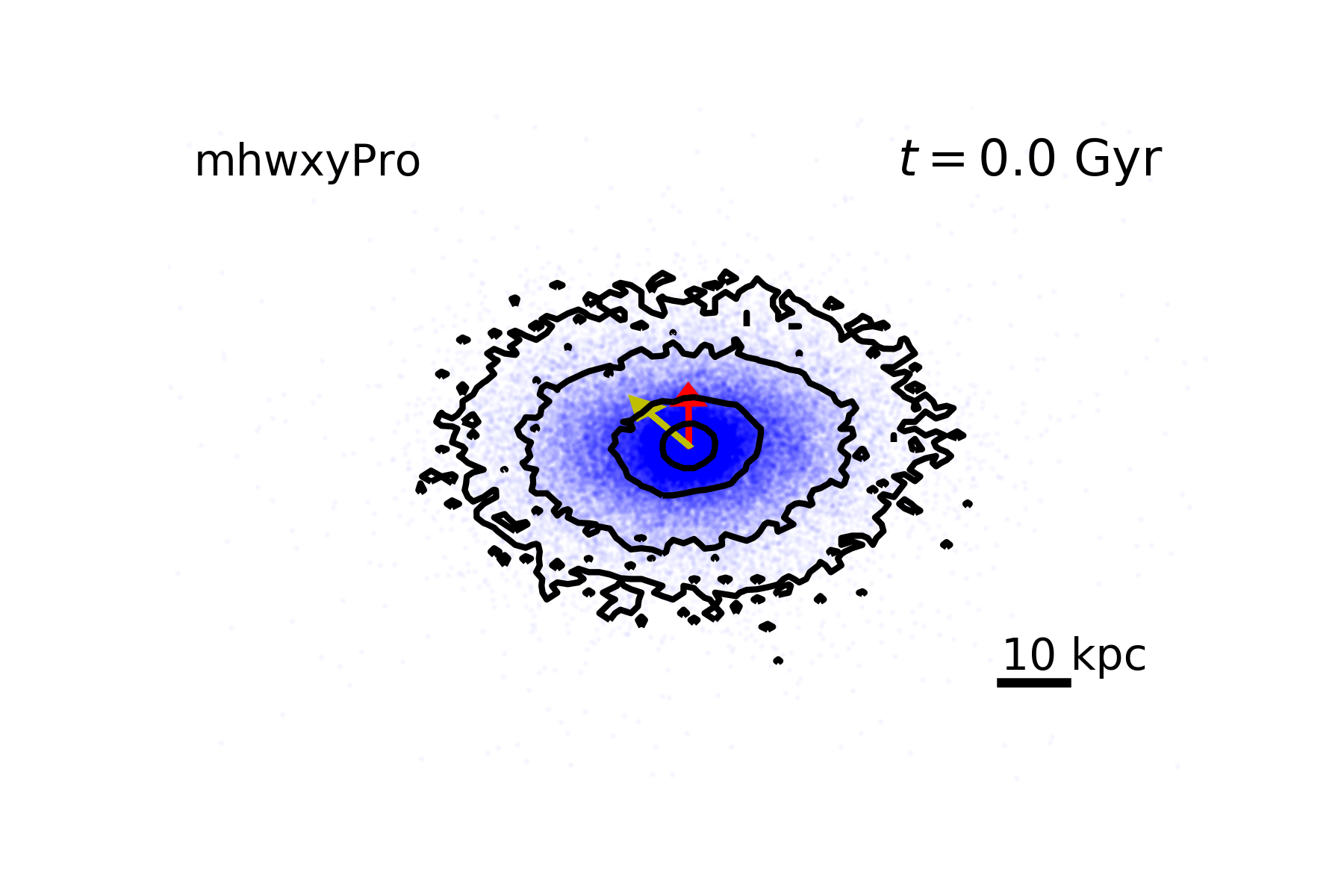} &
\includegraphics[width=\weeFig\textwidth,trim={1cm 1cm 1cm 1cm},clip]{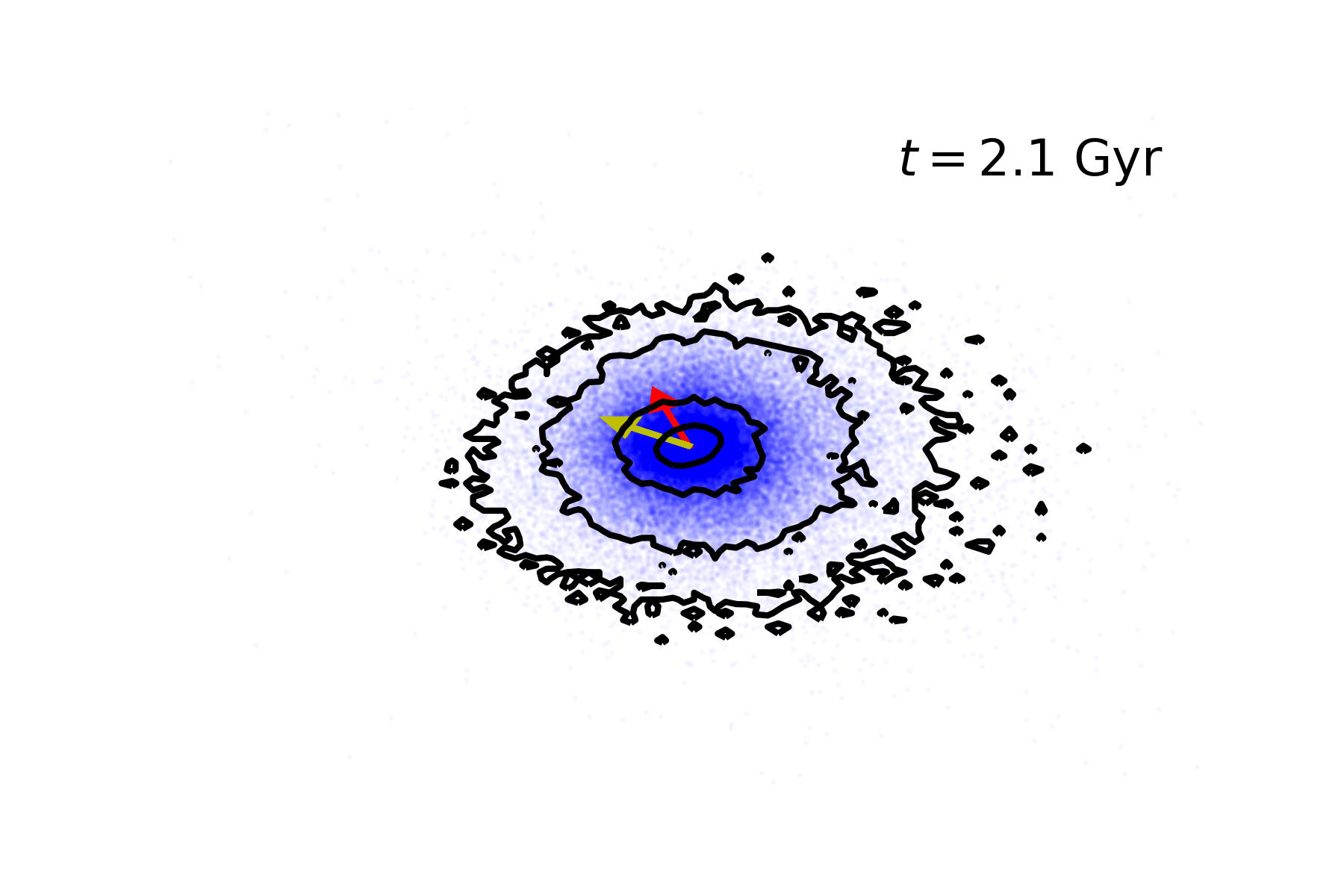} &
\includegraphics[width=\weeFig\textwidth,trim={1cm 1cm 1cm 1cm},clip]{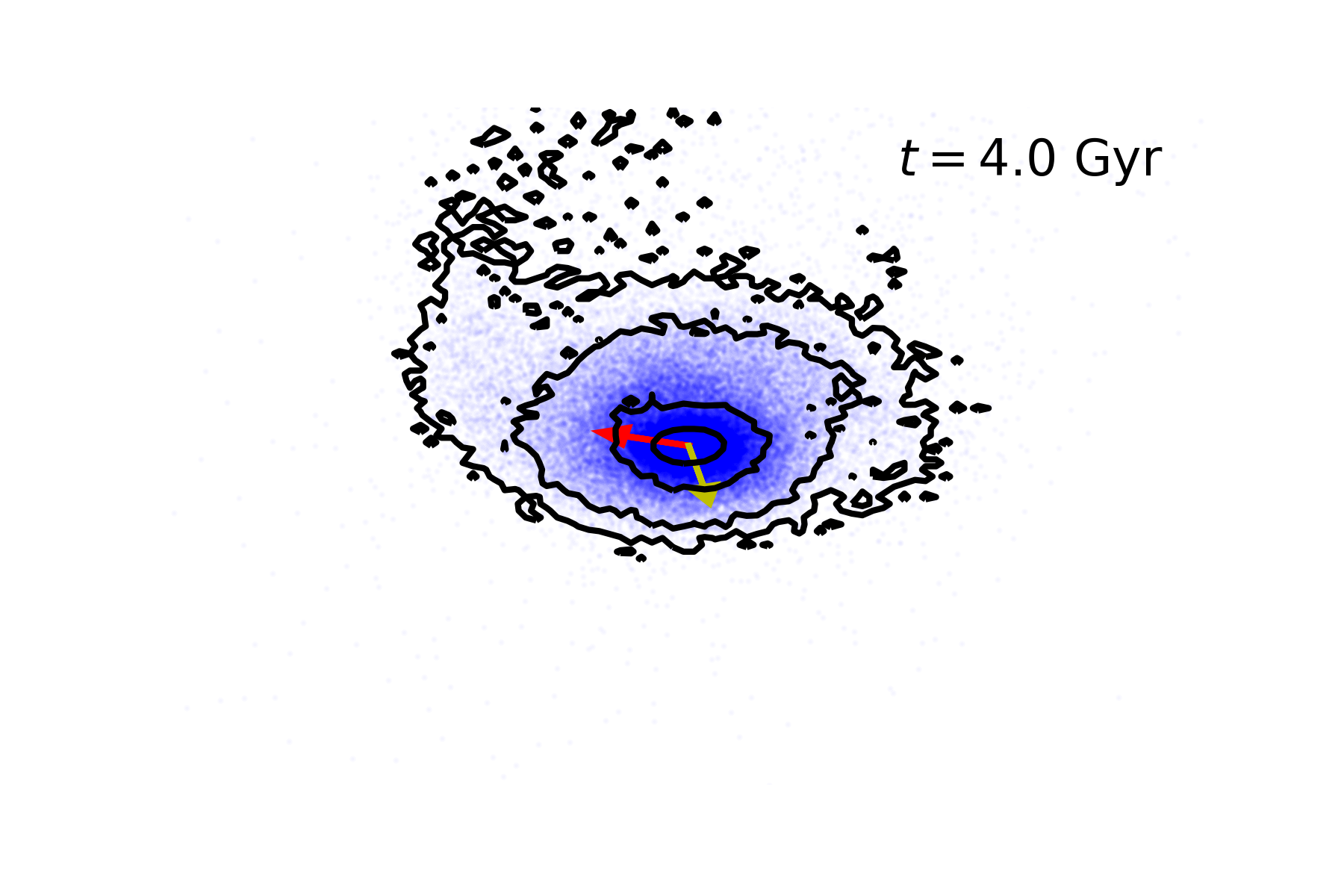} &
\includegraphics[width=\weeFig\textwidth,trim={1cm 1cm 1cm 1cm},clip]{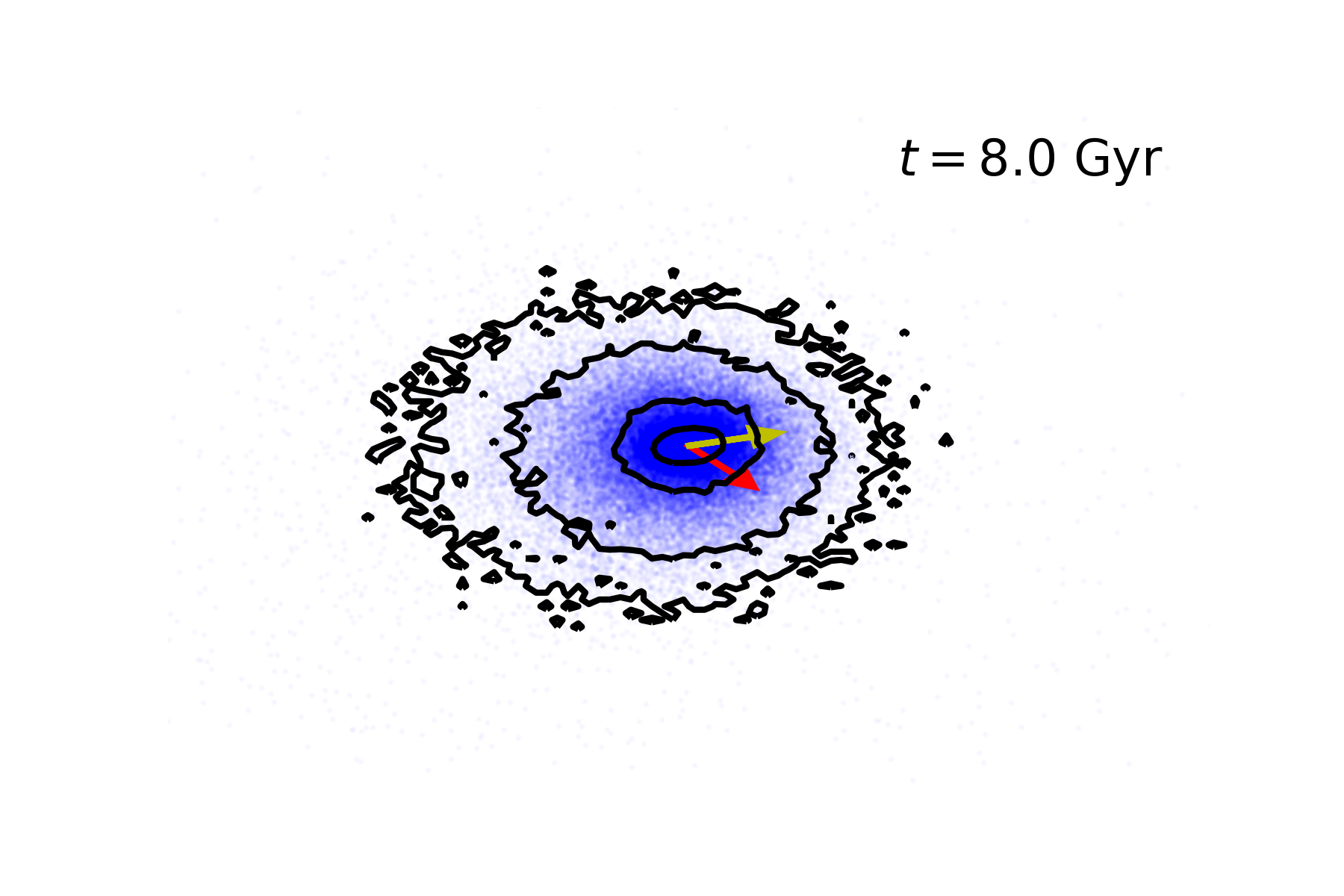} \\
\includegraphics[width=\weeFig\textwidth,trim={1cm 1cm 1cm 1cm},clip]{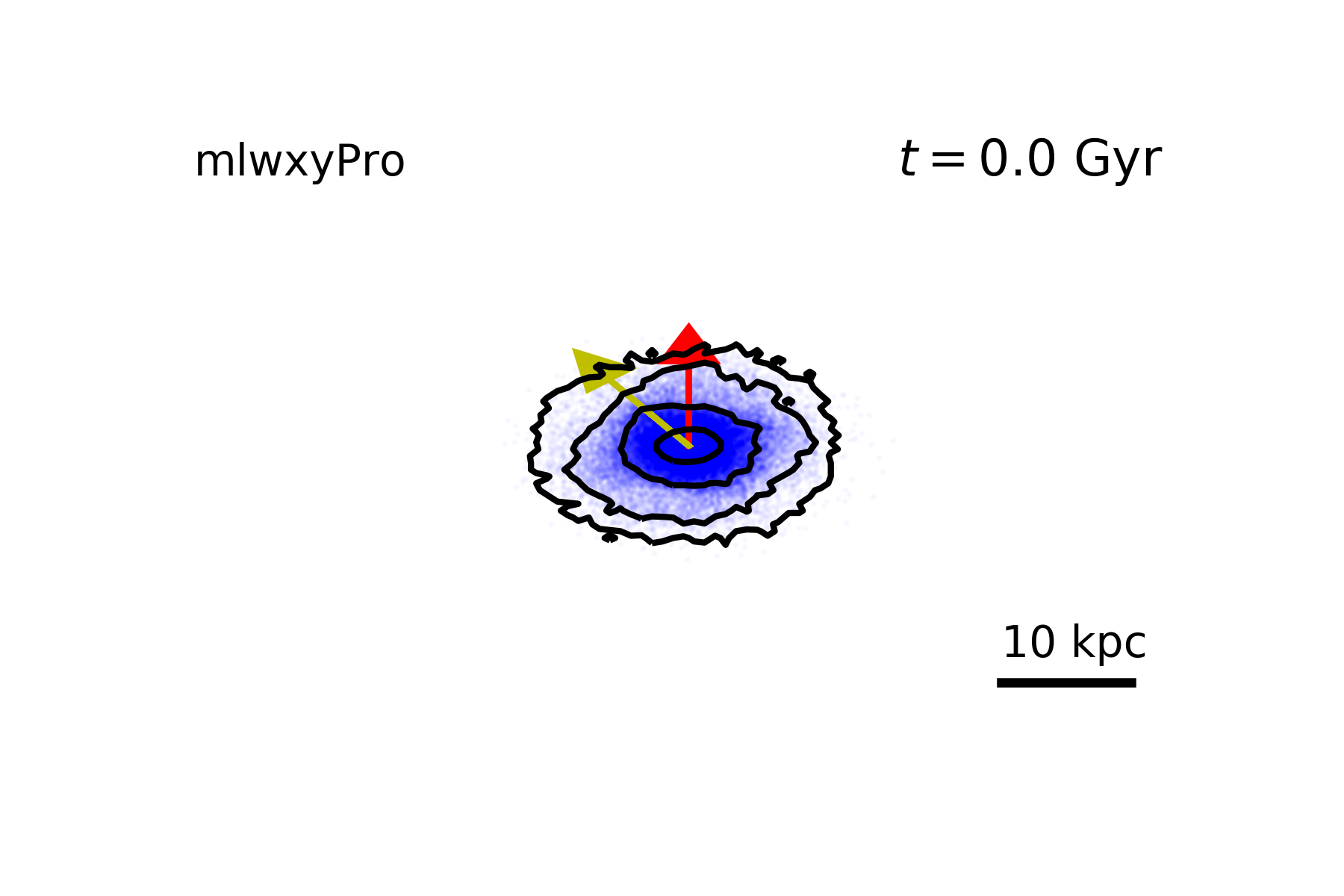} &
\includegraphics[width=\weeFig\textwidth,trim={1cm 1cm 1cm 1cm},clip]{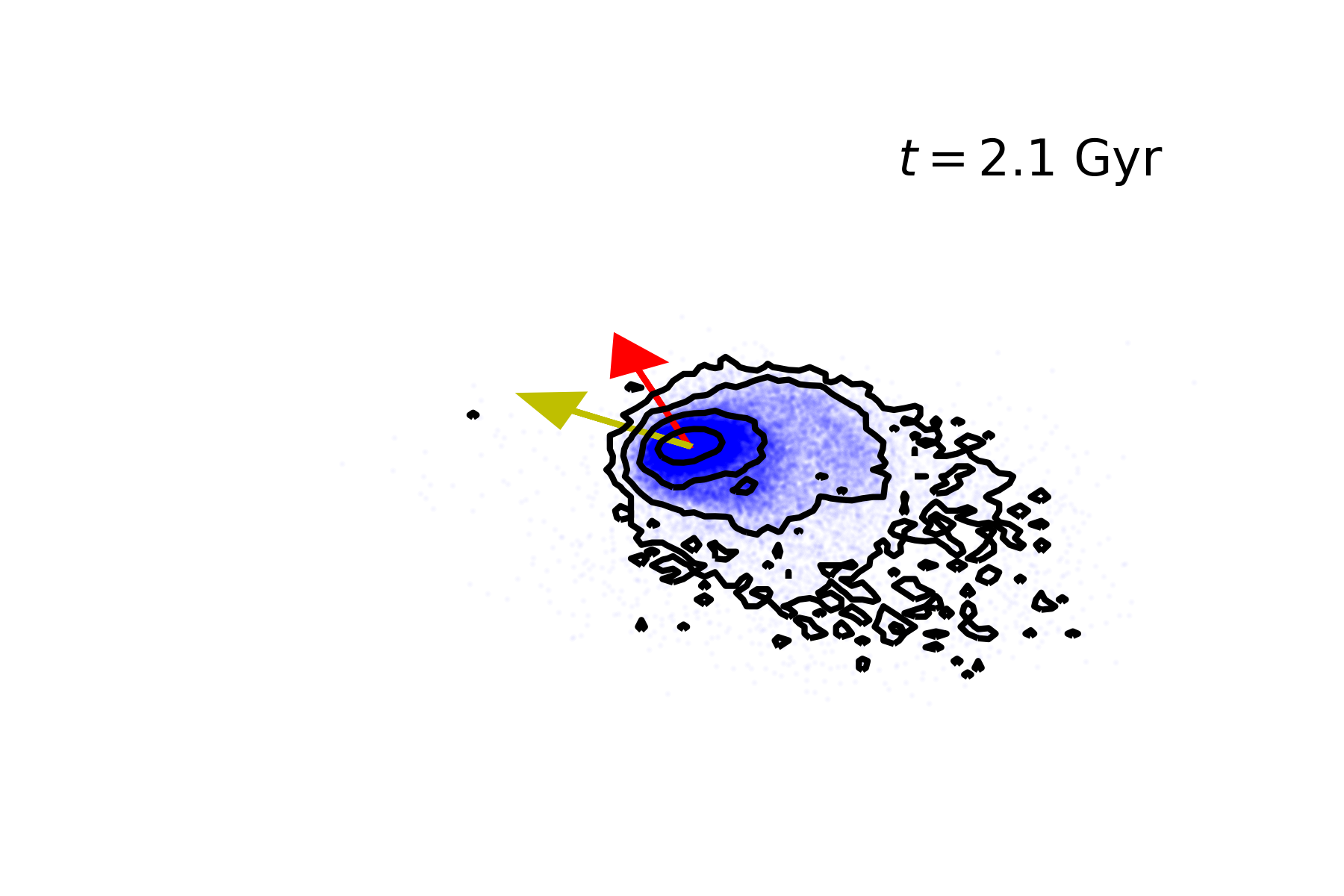} &
\includegraphics[width=\weeFig\textwidth,trim={1cm 1cm 1cm 1cm},clip]{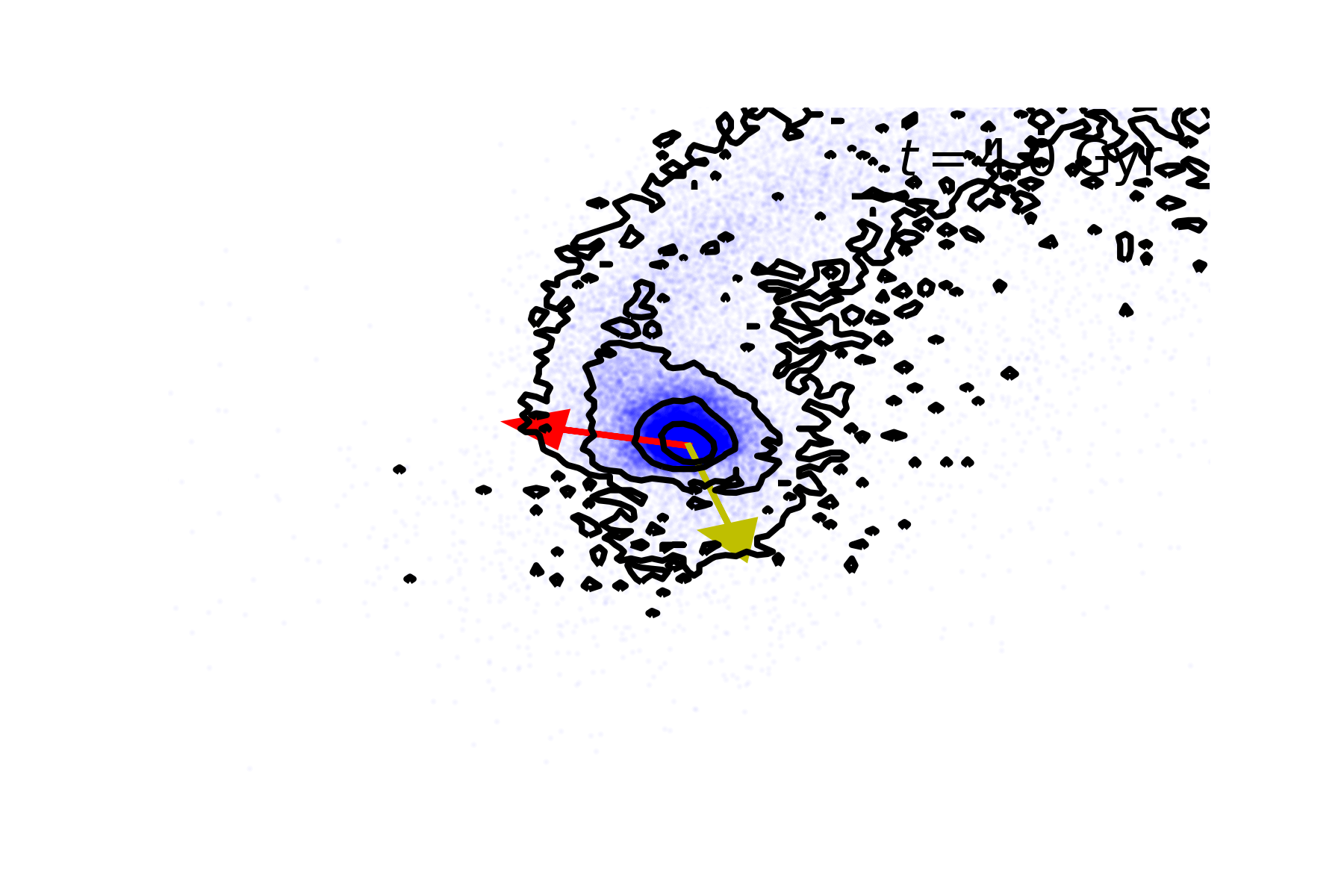} &
\includegraphics[width=\weeFig\textwidth,trim={1cm 1cm 1cm 1cm},clip]{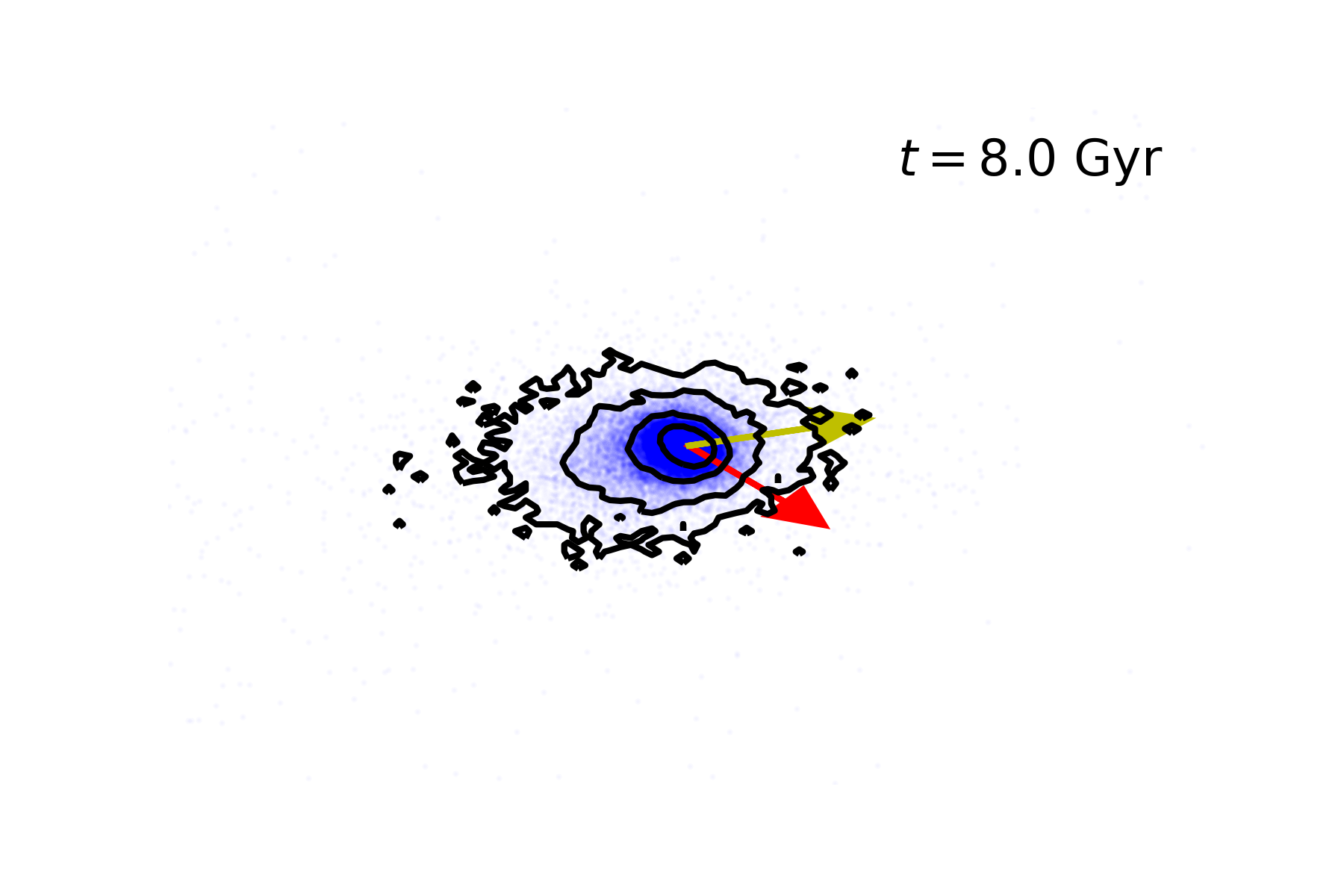} \\
\end{tabular}
\caption{Galaxy models in the presence of an EFE. \textit{Top row}: High mass galaxy in strong tides; \textit{Second row}: low mass galaxy in strong tides; \textit{Third row}: high mass galaxy in weak tides; \textit{Bottom row}: low mass galaxy in weak tides. In all plots the red arrow indicates the (normalised) centre-of-mass velocity vector, while the yellow arrow points towards the cluster centre. The external field effect leads to highly asymmetric disk morphologies, even in the absence of a strong tidal interaction.}
\label{fig:asymmetryMOND}
\end{figure*}

Given these considerations, in Fig.~\ref{fig:asymmetryNewton} we show specific snapshots of the prograde simulations for the low and high mass models without an EFE in strong tides in the Coma-mass cluster. The low and high mass models in weak and strong tides, again in the Coma-mass cluster, in the presence of an EFE, are shown in Fig.~\ref{fig:asymmetryMOND}. All of these plots are face-on views of the galactic disk, i.e. in the $xy$ plane. The contour lines are constant projected number density of $2, 15, 100$ and $600$, using a pixel size of $0.16$~kpc$^2$ for the low mass models, and $0.64$~kpc$^2$ for the high mass models. Recall that the disk particles in the low mass models have a mass of $1.0 \times 10^4$ M$_{\odot}$, while the disk particles in the high mass models have a mass of $9.2 \times 10^5$ M$_{\odot}$, which can be used to determine a surface mass density per pixel. The high mass models also include a bulge component, whose particle mass is $4.6 \times 10^5$ M$_{\odot}$. Using a stellar mass-to-light ratio of $\Upsilon = 1$ we can convert these surface mass densities into surface brightness, which may then be converted to observational units (using $M_g + 21.572 - 2.5 \log_{10}(\mu)$ where $M_g$ is the g-band absolute magnitude of the Sun, \citealp{Conversion}) to give an estimate of the feasibility of observing these features.

Thus the low mass model contours may be interpreted as surface brightness contours of approximately $29, 27, 25$ and $23$~mag/arcsec$^2$. The high mass model contours correspond to $26, 24, 22$ and $20$~mag/arcsec$^2$. If we instead choose a stellar mass-to-light ratio of $10$, these contours increase by approximately $3$~mag/arcsec$^2$. These estimates would suggest that the outermost contour of the low mass models shown in Figs.~\ref{fig:asymmetryNewton} and \ref{fig:asymmetryMOND} are at the limit of observability (or possibly beyond) for a high stellar mass-to-light ratio. All other contours are within observability limits \cite{Trujillo2016}.

In all plots we indicate the direction towards cluster centre by a yellow arrow, and the (normalised) centre-of-mass velocity by a red arrow.

Beginning with the high and low mass models in strong tides without an EFE (Fig.~\ref{fig:asymmetryNewton}), we see clear tidal arms caused by the interaction with the cluster potential, shortly after pericentric pass, as indicated by the antiparallel red and yellow vector arrows. The response to tides is very similar for both models. This symmetric formation of tidal features is a common occurrence in tidal interactions. It should be noted, however, that in our idealised models the symmetry is particularly clear due to the fixed nature of the background potential and the lack of additional substructure in the cluster.

In the top two rows of Fig.~\ref{fig:asymmetryMOND}, we see the effect of the external field added to the tidal disruption. In contrast to the highly symmetric tidal disruption seen in the models without an EFE, we have a highly \textit{asymmetric} disruption in both the low and high mass cases.

Finally, the MOND models in \textit{weak} tides with an EFE are shown in the lower two rows of Fig.~\ref{fig:asymmetryMOND} undergoing a morphological distortion almost \textit{entirely} due to the external field, an effect that does not exist in Newtonian gravity, and is clearly stronger for the low mass galaxy (bottom row) than for the high mass galaxy (second row from bottom), given the stronger external field in that case. This contrasts to the tidal disruption seen for the galaxies without an EFE, where the response was similar for the low and high mass models.

In all cases in which the EFE causes a distortion of the isophotal contours, we can see that the main axis of countour elongation lies roughly opposite to the location of the cluster centre, due to the directional nature of the effect.

It is already known that the asymmetric nature of the external field effect can induce isophotal distortion in galaxies, as studied in \cite{brada_warps, elliptical_efeI}. In these studies, the galaxy was assumed to be in equilibrium with the background potential, which induces a slight distortion in the density profile of elliptical galaxies or warps in disk galaxies.

In our study, however, we see, for the first time, the dynamical consequences of the external field on cluster galaxies. Rather than inducing an asymmetry in an equilibrium model, what we see here are equilibriated galaxies falling into a galaxy cluster and either being substantially distorted by the external field even in the absence of tidal effects, or having their tidal disruption rendered highly asymmetric due to the EFE. It is worth recalling that the EFE becomes stronger closer to the cluster centre, which is also where the tidal disruption becomes more violent, and evolves rapidly over time. Thus these two effects combined for MOND galaxies push the disk far from equilibrium and lead to significantly different morphological changes than would be expected in a Newtonian context.

\subsubsection{Quantifying the asymmetry}
The degree of asymmetry induced in the galaxies by the EFE can be calculated using an asymmetry parameter \citep{schade,conselice} given by
\begin{equation}
A = \frac{\sum |(I_0 - I_{\phi})|}{2\sum |I_0|}.
\end{equation}
Applied to images of galaxies, $I_0$ is the original image (array of pixel intensities) and $I_{\phi}$ is the image after rotation in the image plane by an angle $\phi$.

For our simulated galaxies, we calculate a two-dimensional histogram of particle counts in bins in exactly the same way as done to determine the isophotes in Section~\ref{subsec:EFEasymm}. Thus pixels of size $0.16$~kpc$^2$ (low mass) and $0.64$~kpc$^2$ (high mass) are used in a square region of $40$~kpc $\times$ $40$~kpc (low mass) and $80$~kpc $\times$ $80$~kpc (high mass) centred on the galaxy centre-of-mass. The ``viewing'' plane is taken to be the $xy$ plane, which corresponds to viewing the galaxies exactly face-on. At each snapshot the particle counts are determined for the original orientation (to calculate $I_0$) and then the galaxy is rotated by an angle of $180\degree$ to determine $I_{\phi}$. This choice of rotation angle ensures that we are not sensitive to the tidal features which, from the strong tides models of Fig.~\ref{fig:asymmetryNewton}, can be seen to be approximately invariant under a rotation of $180\degree$.

\begin{figure*}
\centering
\begin{tabular}{cc}
\includegraphics[width=\bigFig\textwidth]{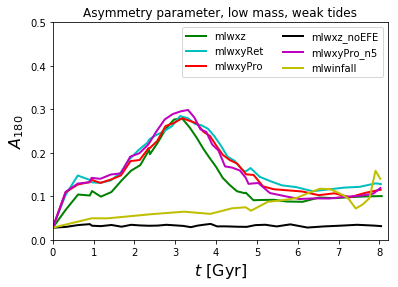} & \includegraphics[width=\bigFig\textwidth]{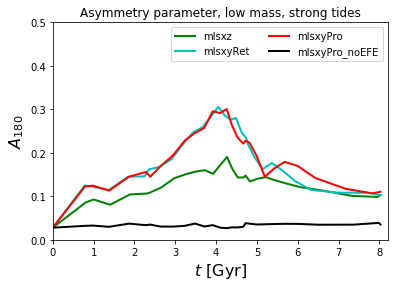} \\
\includegraphics[width=\bigFig\textwidth]{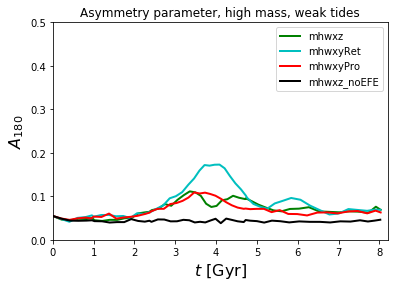} & \includegraphics[width=\bigFig\textwidth]{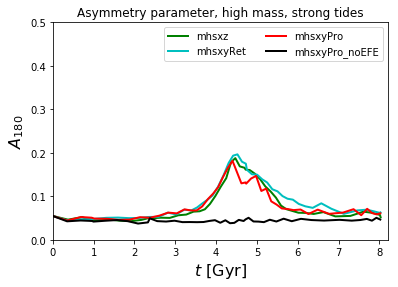}
\end{tabular}
\caption{Time evolution of the asymmetry parameter (using a rotation angle in the viewing plane of $180\degree$ to suppress tidal features) for almost all models (the ``mlsxz\_noEFE'' and ``mhsxz\_noEFE'' models are not shown). \textit{Top row}: low mass models, weak tides (left), strong tides (right). \textit{Bottom row}: high mass models, weak tides (left), strong tides (right). The degree of asymmetry increases substantially for all models during their approach to cluster pericentre. The low mass models, being more affected by the EFE, show particularly large increases in asymmetry, even when their orbits do not bring them very close to cluster centre and they experience only weak tides (upper left panel). In the absence of an EFE (black lines) none of the models show any asymmetry.}
\label{fig:asymmEvo}
\end{figure*}

\begin{figure}
\centering
\includegraphics[width=\bigFig\textwidth]{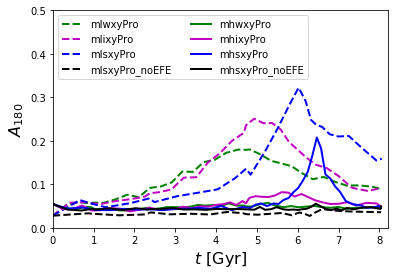}
\caption{Time evolution of the asymmetry parameter for models in the Virgo-mass cluster. Again we see large increases in asymmetry for the low mass models (dashed lines), even in the absence of strong tides. These asymmetries are then exacerbated by a strong tidal interaction (blue dashed line). Although the high mass models are mostly unaffected by the EFE in this cluster background (solid lines) the model in strong tides becomes more asymmetric during the tidal disruption, due to the EFE. In the absence of an EFE (black solid and dashed lines) none of the models show any asymmetry.}
\label{fig:asymmEvoVirgo}
\end{figure}

The time evolution of the asymmetry parameter $A$ is shown for models in the Coma-mass cluster in Fig.~\ref{fig:asymmEvo} and for models in the Virgo-mass cluster in Fig.~\ref{fig:asymmEvoVirgo}. We plot the prograde models in strong tides without an external field rather than the perpendicular models, to match the models shown in Fig.~\ref{fig:asymmetryNewton}.

For all the models without an EFE there is essentially no evolution in the asymmetry (dashed and solid black lines). Although there are clear tidal features (as seen in Fig.~\ref{fig:asymmetryNewton}) the symmetry of these features ensures they are suppressed in the asymmetry parameter by using $\phi = 180\degree$. The main point of interest in Figs.~\ref{fig:asymmEvo} and \ref{fig:asymmEvoVirgo}, in accordance with the features seen in Fig.~\ref{fig:asymmetryMOND}, is that \textit{all} of the MOND models in the presence of the EFE have far higher values of the asymmetry parameter even \textit{before} they pass first pericentre. Note that the amount of asymmetry induced in the high mass models is less than that seen in the low mass models, both because of the weaker EFE for the high mass galaxies and the presence of the highly symmetric bulge component.

In previous observational studies of galaxy asymmetries (e.g. \citealp{conselice}) typical values of $A$ for isolated late-type galaxies are on the order of $0.2-0.3$. The light distribution of observed galaxies may exhibit large asymmetries due to star-forming regions, even while the mass distribution is highly symmetrical. Our simulated galaxies correspond to the mass distribution, and start out with far lower values for $A$ due to lacking spiral arm features, bars, star-forming regions, etc. Thus the absolute values of $A$ given in Figs.~\ref{fig:asymmEvo} and \ref{fig:asymmEvoVirgo} are not expected to be representative of real galaxies. The important point of our results here is that the \emph{increase} in asymmetry in the mass distribution during the evolution of these galaxies within the clusters is large and entirely driven by a MOND effect, and completely dominates over any asymmetry induced by a tidal interaction. Thus an ideal candidate galaxy to investigate this phenomenon observationally would be of sufficiently low mass, without large star-forming regions causing large asymmetries in the light profile, and without any clear companions inducing tidal disruption. It should furthermore be on first infall into the cluster, and still reasonably far from the cluster centre to avoid significant cluster tides.

\begin{figure}
\centering
\includegraphics[width=\bigFig\textwidth]{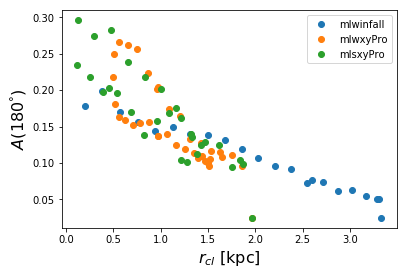}
\caption{Averaged asymmetry parameter, using both $xy$ and $xz$ viewing planes, as a function of clustercentric radius for the distant infall model and the low mass models on prograde orbits in weak and strong tides (all in the Coma-mass cluster). There is a clear trend of increasing asymmetry with decreasing radius, with only a weak dependence on the orbital history of the model.}
\label{fig:asymRad}
\end{figure}

It is important to point out that the low mass models in the Coma-mass cluster, even beginning their evolution at approximately $2R_{200}$, are affected by the external field immediately. This is clear from the initial rapid increase in the asymmetry of these models, which is not evident for the high mass models, or for the low mass models in the Virgo-mass cluster, where the initial EFE is considerably weaker given the lower mass ratio between the galaxies and the cluster. It is instructive to compare the asymmetry seen in the ``mlwinfall'' model (solid yellow line in Fig.~\ref{fig:asymmEvo}), taking into account the orbital trajectory (see Fig.~\ref{setup:orbitVels}). As this model begins at a larger clustercentric radius ($\sim 4R_{200}$) it is not significantly affected by the EFE. Once the model reaches approximately $2R_{200}$ however (at approximately $6.5$ Gyr), it has roughly the same degree of asymmetry as the other low mass models achieve within the first Gyr of evolution. This indicates that the asymmetry arising in our galaxies (due to the distorted potential, as we show in the next Section), is a function of clustercentric radius and does not depend strongly on the galaxy's orbital history. This is made more evident in Fig.~\ref{fig:asymRad} where we plot the averaged asymmetry parameter in both the $xy$ and $xz$ planes as a function of the clustercentric radius. We show only the low mass prograde models in strong and weak tides in the Coma-mass cluster, as well as the distant infall model. It is clear that there is a close correlation between clustercentric radius and amount of asymmetry, with some scatter due to the details of the tidal/EFE interaction during the orbit. Furthermore, this correlation is very similar for all three models, at least for $r_{cl} > 0.5$ where we might expect the tidal distortions to be less pronounced. We use the averaged asymmetry as the amount of distortion visible in each model depends on the orientation of that model with respect to the viewing plane. Note that modifying the MOND interpolation function to have a faster transition from MONDian to Newtonian gravity has only a negligible effect on the degree of asymmetry, as shown by the magenta line in the upper left panel of Fig.~\ref{fig:asymmEvo} which corresponds to the model with a transition function (Eq.~\ref{mondTrans}) where $n=5$.

We estimate the timescale of the asymmetry induced by the external field by determining for how long the galaxy has $A > 0.1$. This choice is, of course, arbitrary, but is at least twice that of the initial asymmetries of the models. For the low mass models, which are more affected by the EFE, on a weak tidal orbit (meaning the galaxy enters the cluster earlier and spends a longer time in the external field) the asymmetry is above $0.1$ for approximately $3-4$ Gyr. The low mass models on a strong tidal orbit enter the cluster more rapidly but have their asymmetries enhanced by the tidal interaction. The timescale for $A > 0.1$ in this case is again of order $3-4$ Gyr. For all low mass models the galaxies remain somewhat asymmetric until the end of the simulation. The high mass models in weak tides are less affected by the EFE, although they still show $A \gtrsim 0.1$ for $\sim 2$ Gyr, particularly the ``xyRet'' model. Finally, the high mass models on the strong tidal orbits again spend less time in the external field before reaching cluster pericentre, and being high mass are less affected by the EFE. Nonetheless, the external field distortion clearly begins about $1$ Gyr before pericentre, and leads to enhanced asymmetries during the tidal disruption. The total timescale is at least of order $1.5$ Gyr.

In summary, the EFE induces a clear asymmetry in both our low and high mass galaxies. The low mass galaxies suffer this effect more strongly due to the larger difference in internal and external accelerations, but the effect is still clear in the high mass galaxies. This asymmetry is then further enhanced by the tidal disruption suffered by the galaxies on strongly tidal orbits. Finally, the timescales over which these asymmetries may be considered to be strong are of the order of $1-2$ Gyr for the high mass galaxies and $3-4$ Gyr for the low mass galaxies. All of this implies that such effects should be clearly observable in cluster galaxies.

\subsubsection{Source of the asymmetry}
\label{subsubsec:asymmSource}

We now demonstrate that the source of this asymmetry is due to the distorted gravitational potential of the galaxy, as the directional dependence of the EFE flattens the isopotential contours of the disk in the radial direction pointing from the cluster centre. This property of the EFE has been well studied in previous works \citep{brada_warps, elliptical_efeI, elliptical_efeII}.

\begin{figure*}
\centering
\begin{tabular}{cc}
\includegraphics[width=\bigFig\textwidth]{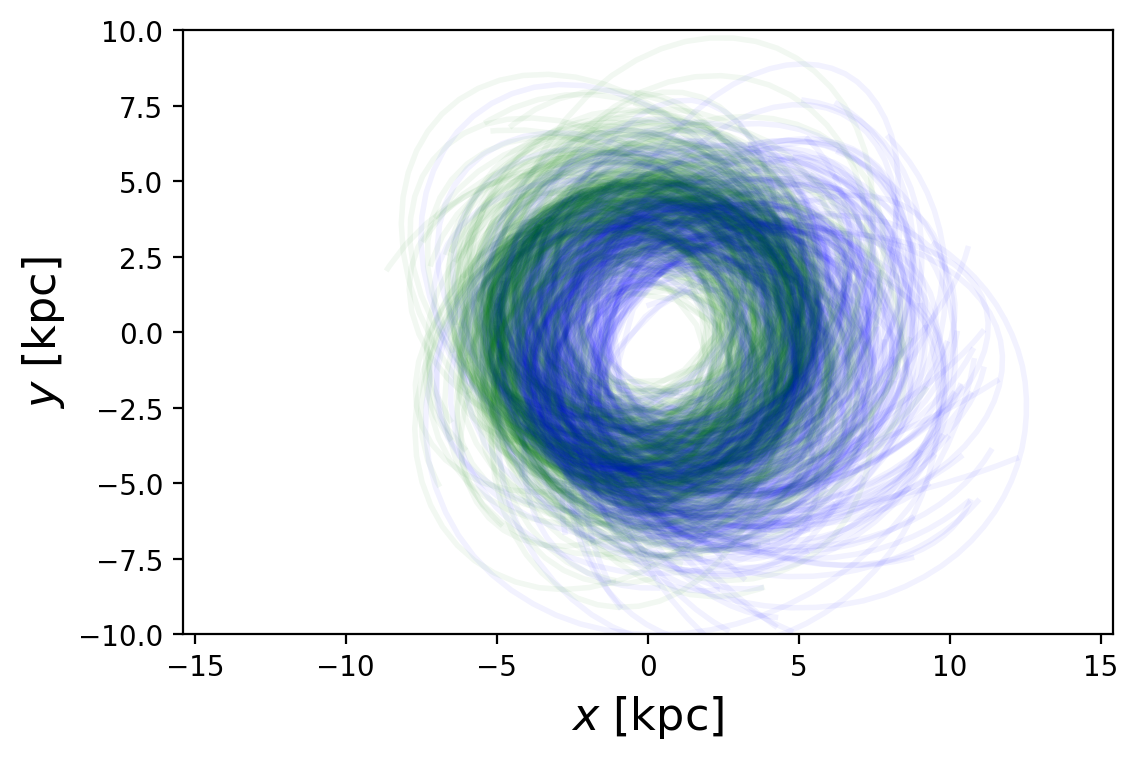} & \includegraphics[width=\bigFig\textwidth]{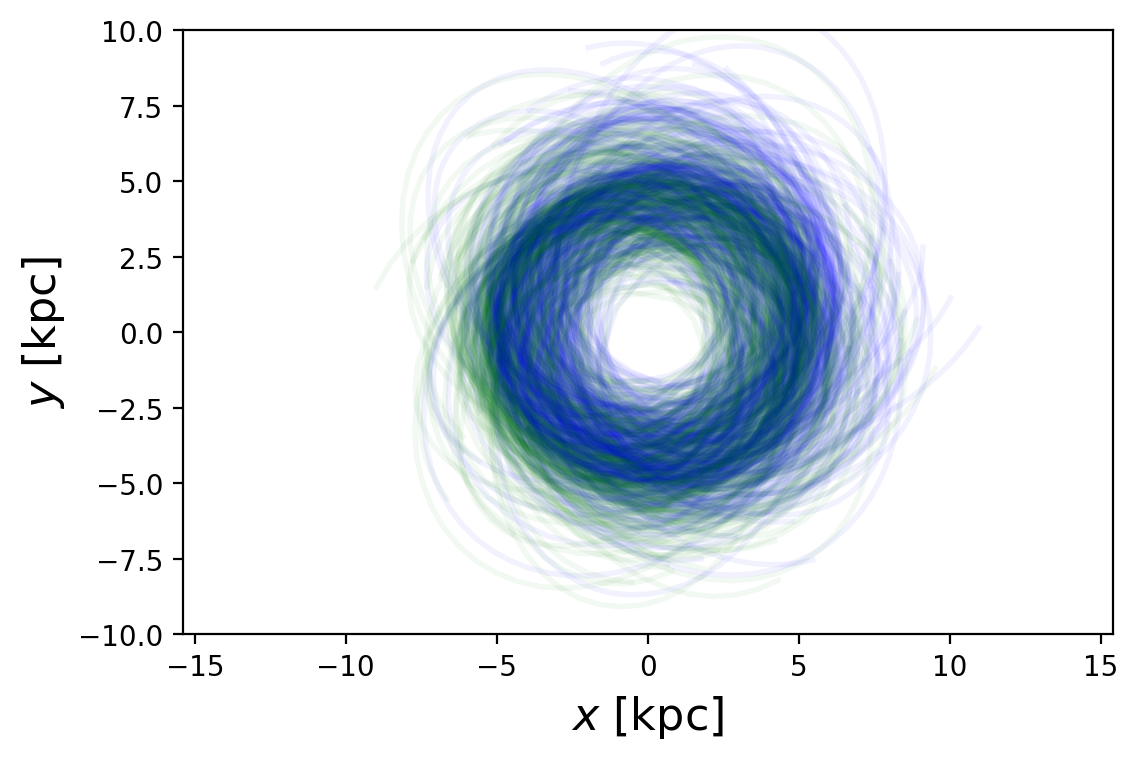}
\end{tabular}
\caption{Orbital trajectories of a subset of particles within the galactic disks, from the beginning of the simulation up to approximately $600$ Myr. The blue lines in both panels correspond to orbits in the models that lie within the background Coma-like cluster, while the green lines are orbits in the same model, but with the external cluster completely removed (i.e. no EFE and no tidal forces). \textit{Left}: ``mlwxyPro'' models; \textit{Right}: ``mlwinfall'' models. In the absence of the external field the average distribution of orbits is mostly circular (green lines), whereas the orbits within the external field (blue lines left panel) become visibly distorted by the modified internal potential of the disk. For the infall model, the initial EFE is weak enough that there is much less orbital distortion (blue lines right panel) compared to the case with the external field removed.}
\label{orbitShifts}
\end{figure*}

To demonstrate this effect, in Fig.~\ref{orbitShifts} we show the orbital trajectories, over the course of $600$ Myr of evolution, of $244$ particles within the galactic disks of the ``mlwinfall'' and ``mlwxyPro'' models, starting at the beginning of the simulation. We have rerun these models for a short period at a much higher time resolution in order to analyse these orbital trajectories, and for this reason we choose a short total time of $600$ Myr, which is approximately $30\%$ of the disk dynamical time. The subset of particles was chosen by restricting to those with an initial radial position of $5 \pm 0.1$~kpc and a vertical position of $\pm 0.1$~kpc from the disk plane.

In Fig.~\ref{orbitShifts}, the right panel shows the orbital trajectories of stars in the ``mlwinfall'' model with the background cluster present (blue lines) and the same model with the background cluster completely removed (green lines). This is easily achieved by simply removing the analytic density in the simulation. This model begins its evolution far from the cluster centre ($r_{cl} \sim 4R_{200}$) where it is not strongly affected by the EFE (we see only a very small increase in the asymmetry of this model in Fig.~\ref{fig:asymmEvo} within the first $600$ Myr). Therefore, the orbital trajectories for the infall models, shown in the right panel of Fig.~\ref{orbitShifts}, are roughly coincident (and on average circular) regardless of whether the background cluster (and therefore the external field) is present or not.

This is not the case for the ``mlwxyPro'' models shown in the left panel of Fig.~\ref{orbitShifts}. In this case the presence of the background cluster (blue lines) clearly induces a shift in the orbital trajectories of the particles, relative to the same model in the absence of the background cluster (green lines). The average distribution of the orbits in the presence of the EFE is noticeably more elliptical, when compared to the roughly circular orbits of the model without a background field. Therefore, the EFE induces an asymmetric distortion of the gravitational potential, and an associated distortion of the particle trajectories. Note that this effect arises \textit{immediately} in our models due to the disruption of the internal gravitational potential of the galaxies. We see distortion in the stellar orbits within just the first $600$ Myr of evolution, during which the galaxy has not fallen far into the cluster.

Finally, as mentioned previously, galaxies infalling into a cluster, due to their typically rapid motion, will experience a rapidly \textit{time-varying} EFE. We thus have two effects: (1) the potentials of the galaxies are distorted due to the external field of the cluster, which depends only on clustercentric radius, and is an ``instaneaous'' effect giving rise to an immediate modification of the stellar orbits within the disks; (2) the degree and direction of the EFE-induced distortion changes very rapidly as the galaxy moves through the cluster, with the rate of change obviously connected to the velocity of the galaxy. Thus the galaxy is both distorted by the EFE, and less able to adjust to its new potential as it gets closer to the cluster centre, placing it far out of dynamical equilibrium.

\subsubsection{Tidal interaction}
\label{subsubsec:tidalAnalytic}
Having quantified the asymmetry induced in these disks, we now quantify the degree of tidal interaction by estimating the tidal radii of our models. The tidal radius in MOND can be estimated with \citep{TidalRadMOND}:
\begin{equation}
\label{eq:tidalRad}
r_{\text{tidal}} = r_{cl} \left( \frac{M_g}{\zeta_1 M_{cl}(r_{cl})} \right)^{1/3}
\end{equation}
where $r_{cl}$ is the clustercentric radius of the galaxy, $M_g$ is the physical galaxy mass and $M_{cl}(r_{cl})$ is the physical cluster mass within the given clustercentric radius of the galaxy. This latter quantity is calculated by a numerical integration of the density distribution given in equation (\ref{density}). The parameter $\zeta_1 = 1 + \zeta$ where
\begin{equation}
\zeta = -\frac{d \log g(r_{cl})}{d \log r_{cl}},
\end{equation}
and $g(r_{cl})$ is the magnitude of the acceleration at the clustercentric radius $r_{cl}$. Note that we can determine this value easily given that our background clusters are NFW potentials.

Our strategy to determine tidal radii for models \textit{without} an EFE is to include an effective dark matter halo for the galaxy assuming a spherically symmetric mass distribution, using the QUMOND formulation. This is because the EFE eliminates (or at least substantially reduces) the effective dark matter halo around our galaxies. We follow the calculation used in \cite{KatzMcGaugh}, based on the work of \cite{AngusDaeferio}. The cumulative equivalent dark matter (eDM) mass is given by 
\begin{equation}
M_{g,eDM} = \frac{1}{2} \left( M_g + \sqrt{M_g^2 + 4M_g\frac{a_0 r^2}{G}} \right).
\end{equation}
We calculate $m_{eDM}$ for the low mass galaxy at $r_c = 8.6$~kpc and $r_c = 20$~kpc for the high mass galaxy, which correspond to approximately $5h_R$ for each disk, thus enclosing $\sim 99\%$ of the disk mass assuming exponential distributions. These values for the galaxy mass are then used in equation \ref{eq:tidalRad} to determine the tidal radii for the models without an EFE.

For all models we use the above values of $r_c$ to compare the calculated tidal radii with the sizes of the disks, to estimate when we would expect tidal stripping to occur.

\begin{figure*}
\centering
\begin{tabular}{cc}
\includegraphics[width=\bigFig\textwidth]{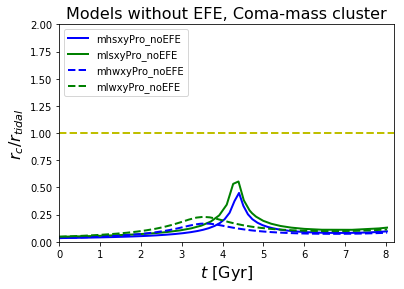} & \includegraphics[width=\bigFig\textwidth]{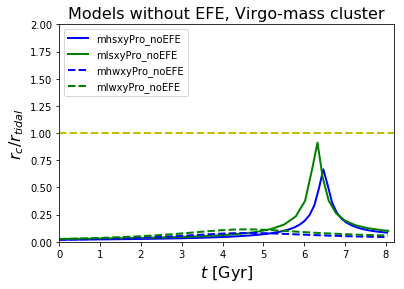} \\
\includegraphics[width=\bigFig\textwidth]{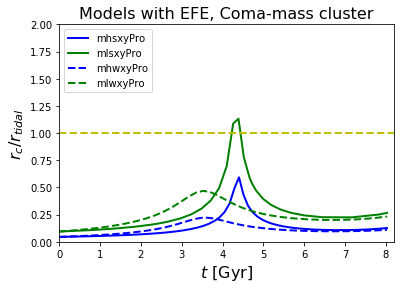} & \includegraphics[width=\bigFig\textwidth]{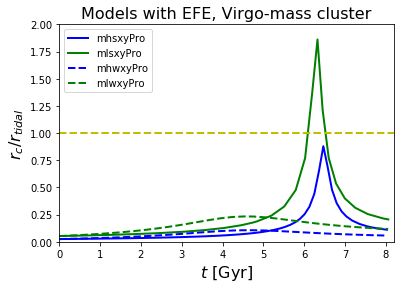}
\end{tabular} 
\caption{Ratio of $r_c$ to the tidal radius for the low and high mass models during orbits of both weak and strong tides, \textit{Left:} Coma-mass cluster; \textit{Right:} Virgo-mass cluster; \textit{Top row:} without an external field effect; \textit{Bottom row:} including the external field. For values of $r_c/r_{tidal} \sim 1$ tidal disruption would be expected within the disk. The higher concentration of the Virgo-mass cluster leads to a larger mass within the orbital radius, especially close to the cluster centre, and a stronger tidal field. When the EFE is active, the galaxies become more susceptible to tidal disruption, particularly the low mass models (green lines).} 
\label{tidalRadius}
\end{figure*}

In Fig.~\ref{tidalRadius} we plot the ratio $r_c/r_{\text{tidal}}$ as a function of time for the low and high mass galaxies on tidally weak and strong orbits in both background cluster models, in the presence and the absence of an EFE. Whenever this ratio becomes greater than $1$ the tidal radius lies within the stellar disk (or, more specifically, within $\sim 5h_R$) and we would expect some tidal disruption. This is the case only for the models on strongly tidal orbits (solid lines), in both cluster backgrounds. In the presence of the EFE, however, we can see that the tidal radius decreases (the peaks in the figure increase in height) with the low mass models on orbits that include strong tides expected to experience significant tidal disruption, as $r_c/r_{tidal} > 1$. Furthermore, the difference between the solid green and blue lines is larger when the EFE is included, indicating that the low mass disks become more affected by tides as they are more affected by the external field. We would thus expect that, for the models in strong tides and with an EFE, the low mass models will be more tidally disrupted than the high mass models. In the absence of an external field effect the models in strong tides will suffer only marginal tidal mass loss. Finally, for models in weak tides, we would not expect to see any tidal mass loss. The tidal disruption is expected to be somewhat stronger in the Virgo-mass cluster due to its higher concentration.

These estimates of the tidal disruption will be compared with our mass loss analysis in the next Section.

\subsection{Mass loss}
\label{subsec:massLoss}

\begin{figure*}
\centering
\begin{tabular}{cc}
\includegraphics[width=\bigFig\textwidth]{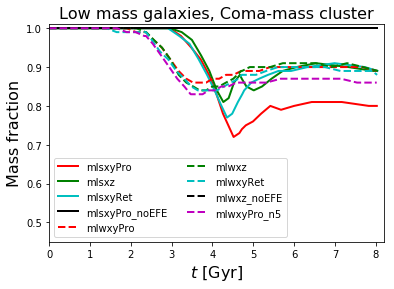} & \includegraphics[width=\bigFig\textwidth]{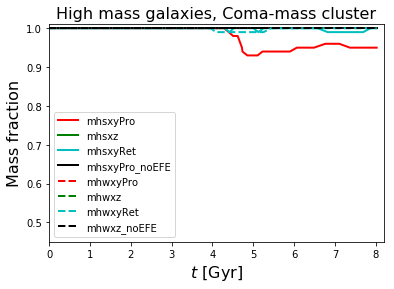}
\end{tabular}
\caption{Bound mass fractions for the galaxy models in the Coma-like cluster using the method described in the text. \textit{Left}: low mass galaxies; \textit{Right}: high mass galaxies. The low mass galaxies, being more affected by the EFE, suffer more mass loss. For the models in weak tides (dashed lines) this mass loss is not dominated by tidal stripping, but is instead due to the EFE. For the strong tides orbits, the prograde model is more stripped due to resonance effects.}
\label{results:boundMass}
\end{figure*}

Following on from the main results reported in the previous subsection, we now analyse the rate of mass loss of our models. Given the non-linear nature of MOND gravity, it is a non-trivial task to accurately determine if a particle is bound to another. Furthermore, the external field complicates matters by modifying the internal potential of each galaxy. Therefore, we use the following procedure to estimate the bound mass fraction of each galaxy model throughout its evolution within the cluster. 

For all snapshots of each model we determine all the Lagrange radii corresponding to between $50\%$ and $100\%$ of the total mass of the galaxy (including the bulge in the case of the high mass disks) in increments of $1\%$. We then determine, at each snapshot, the innermost Lagrange radius that has grown to at least twice its initial value. We assume that all material outside this radius is stripped mass. If none of the Lagrange radii satisfy this criterion we assume that there is no stripping. In this way we can estimate what percentage of the galaxy mass is ``stripped'' at each snapshot. Clearly our decision to label a doubling of a Lagrange radius as indicating ``mass loss'' is arbitrary, but we consider this to be sufficiently large to be insensitive to strong distortions of the disk morphology, yet sufficiently small to capture significant changes to the mass distribution that are likely to be associated to mass loss. We can check the suitability of our criterion by seeing how often this ``mass loss'' is recovered, as this would indicate merely a morphological disruption.

\begin{figure}
\centering
\includegraphics[width=\bigFig\textwidth]{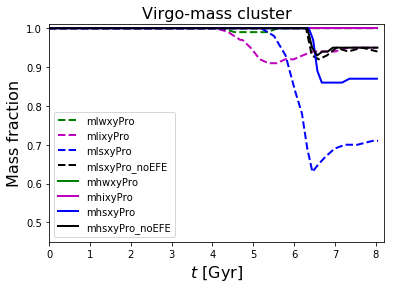}
\caption{Bound mass fractions for the galaxy models in the Virgo-mass cluster using the method described in the text. The lower mass of this cluster reduces the strength of the EFE, but the low mass models that have closer pericentric passages in the cluster (blue and magenta dashed lines) experience both EFE and tidal mass loss. The tidal stripping induced in the high mass model in strong tides (blue solid line) is enhanced by the EFE (compare with the same model without an EFE, solid black line).}
\label{results:boundMassVirgo}
\end{figure}

The results of our analysis for the low and high mass galaxy models in the Coma-like cluster are shown in Fig.~\ref{results:boundMass} and for the Virgo-mass cluster in Fig.~\ref{results:boundMassVirgo}. We can see several instances of small ``dips'' in the mass loss evolution, where the Lagrange radius criterion is detecting significant morphological disruption during the tidal interaction  (much more than occurs due to the EFE-induced asymmetry) rather than actual mass loss. After the tidal interaction, however, the mass fractions become approximately constant, likely indicating genuine mass loss.

In general we see a mixture of mass loss induced by the EFE and the tidal interaction in our plots. When the EFE is not active, almost none of the models lose any mass, except for those on strong tides in the Virgo-mass cluster, due to the higher concentration of this cluster potential. Furthermore, the final mass fraction of the low and high mass galaxies in this case are very similar. This is consistent with our tidal radius analysis in Fig.~\ref{tidalRadius} (upper row), which predicts negligible tidal stripping for all non-EFE models in the Coma-mass cluster, and only very minor stripping of the non-EFE models in the Virgo-mass cluster. Note that our results for the models without an EFE are very similar to those seen in our comparison Newtonian models with galaxies embedded within dark matter halos (at least after one pericentric passage).

For the high mass galaxies, which are less affected by the external field, we see essentially \textit{only} tidal mass loss, and only when there are strong tides. In the Coma-mass cluster we see additional evidence that this mass loss is truly tidal because it is only the prograde orbit model that suffers any significant stripping, due to the well-known resonance effect (the Virgo-mass cluster models are all prograde). Again, this is consistent with our tidal radius analysis for the high mass galaxies in Fig.~\ref{tidalRadius} (lower row, blue lines).

In the case of the low mass galaxies, the situation becomes more complicated. For those on orbits with strong tides, we expect from our tidal radius analysis that they should suffer significant stripping (especially in the Virgo-mass cluster), which we do indeed see in Figs.~\ref{results:boundMass} and \ref{results:boundMassVirgo}. Furthermore, the resonance effect is again visible: the low mass prograde model on strong tides in the Coma-mass cluster suffers greater mass loss than the others. In the Virgo-mass cluster, the dependence on the amount of mass loss with tidal strength is made very clear: the strong tides orbit suffers more mass loss than the intermediate tides orbit, and the weak tides orbit indicates none at all.

The low mass galaxies in the Coma-mass cluster on \textit{weak} tides also show substantial mass loss, which is not expected to be tidal, given our tidal radius analysis (none of the weak tides models had tidal radii anywhere near the chosen disk edge). Thus here we see that the external field effect \textit{alone} causes mass loss in these models. This mass loss begins as early as $2$ Gyr into their orbits and is more gradual than the mass loss evolution that we can confidently attribute to tides. Given this, it is important to note that the models on strong tides must also suffer mass loss due to the EFE, as they move through a much stronger external field than those on weak tides. Indeed, these models begin to lose mass at about $3$ Gyr into their orbits. Comparing the clustercentric radii of the weak and strong tides models (Fig.~\ref{setup:orbitVels}) we see that they reach $R_{200}$ at approximately $2$ and $3$ Gyr after the beginning of the simulation, matching the onset of the mass loss. We thus conclude that, for low mass galaxies in the presence of an external field, the mass loss is a result of a double whammy: the EFE itself and the tidal disruption.

\subsection{Kinematical and morphological evolution}
\label{subsec:rotCurves}

We have seen that the external field causes an asymmetric redistribution of the stellar density, mass loss due to the modified internal potential of the galaxy and an increased susceptibility to tidal mass loss, particularly for low mass galaxies in high mass clusters, where the external field is strongest.

Now we wish to examine the evolution of the kinematics of the disks, as well as their general morphology (i.e. disk-like or spherical) to determine if the external field effect combined with tides results in dispersion dominated objects. The kinematical evolution is also a means of checking how the internal potential of the galaxies evolves, although we must bear in mind that due to the rapidly-changing EFE, especially close to the cluster centre, these galaxies will be far from equilibrium throughout much of their evolution. In this section we will focus exclusively on the models in the Coma-mass cluster, where the external field is strongest.

\begin{figure*}
\centering
\begin{tabular}{cc}
\includegraphics[width=\bigFig\textwidth]{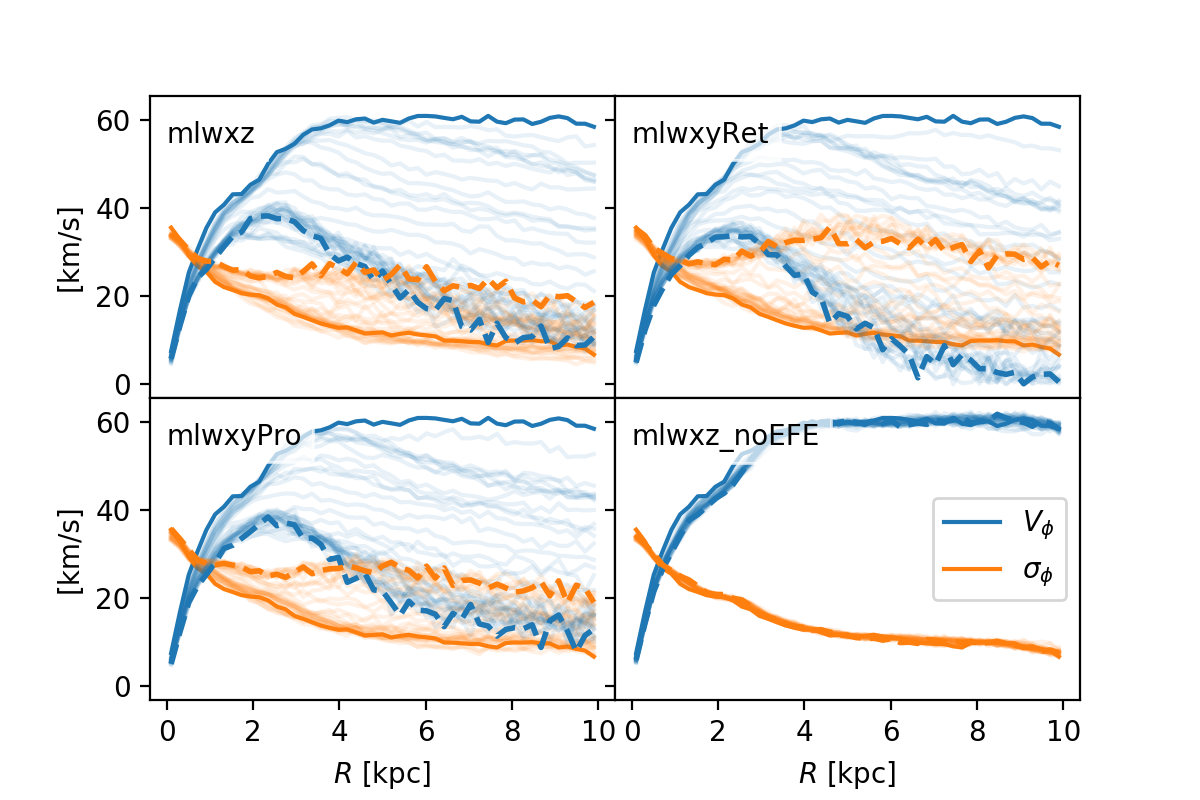} & \includegraphics[width=\bigFig\textwidth]{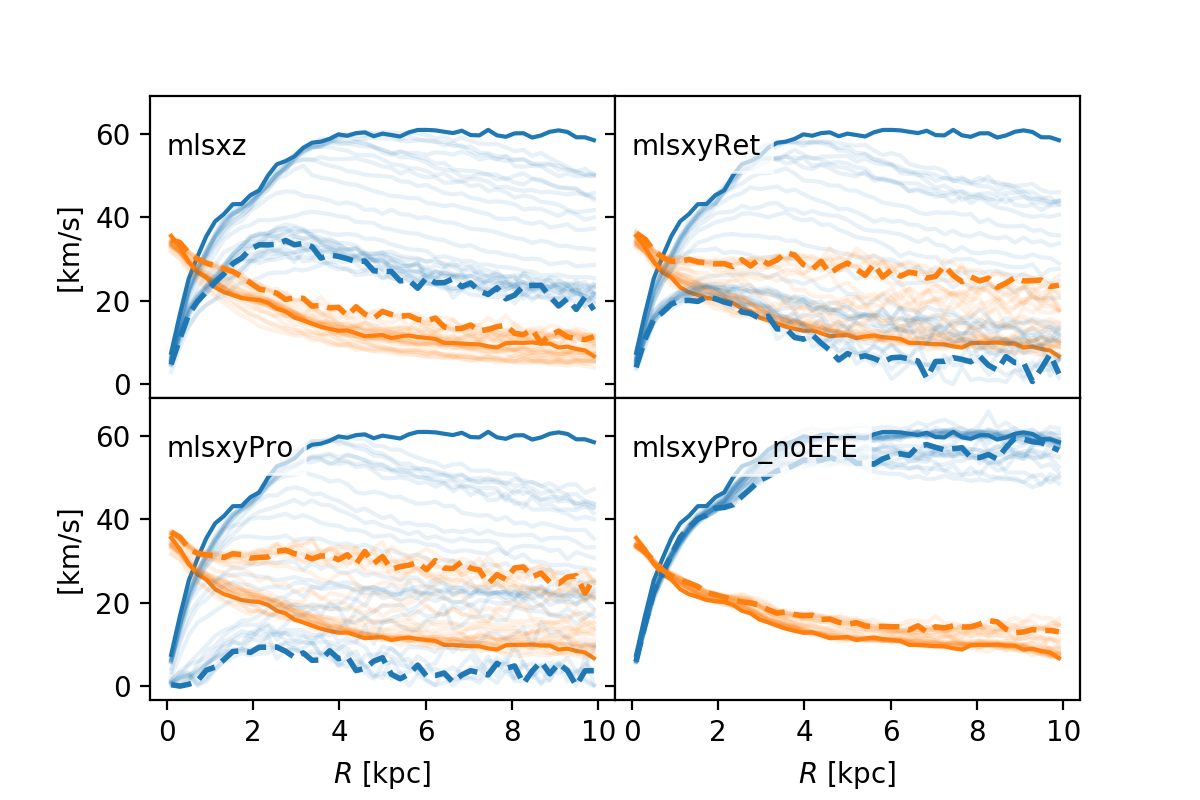} \\
\includegraphics[width=\bigFig\textwidth]{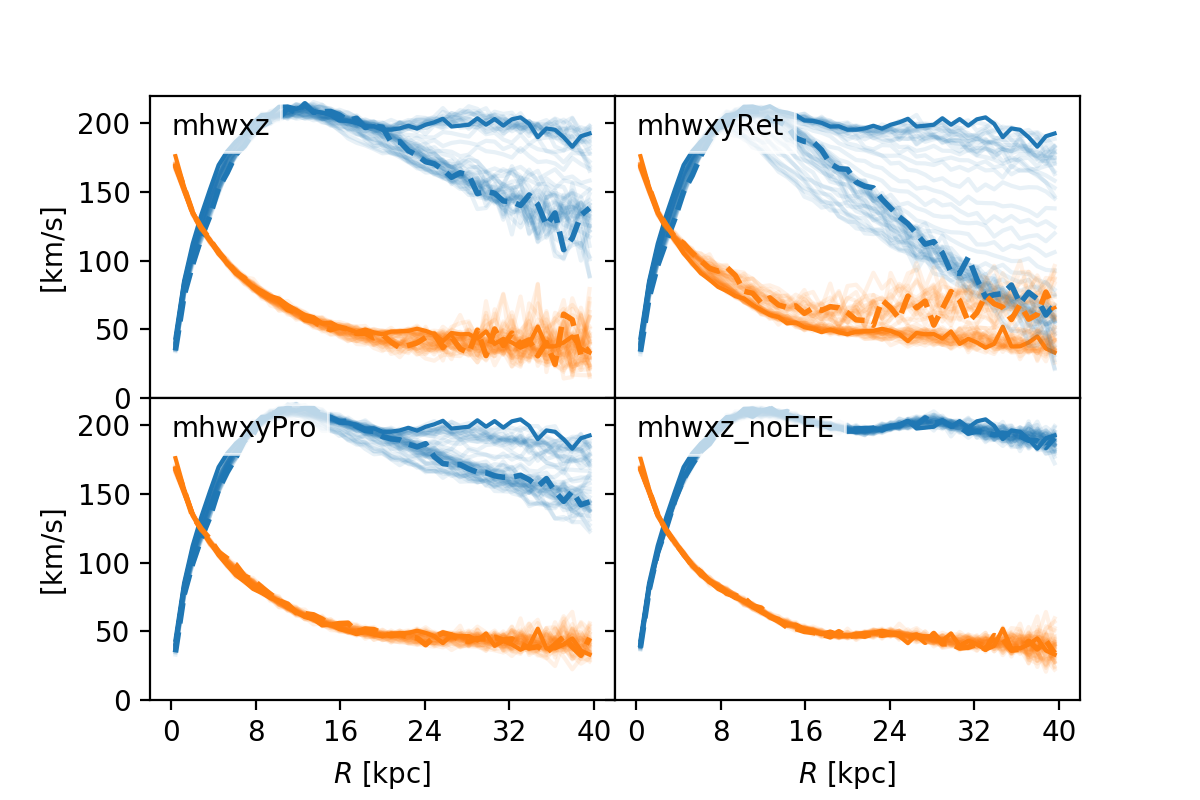} & \includegraphics[width=\bigFig\textwidth]{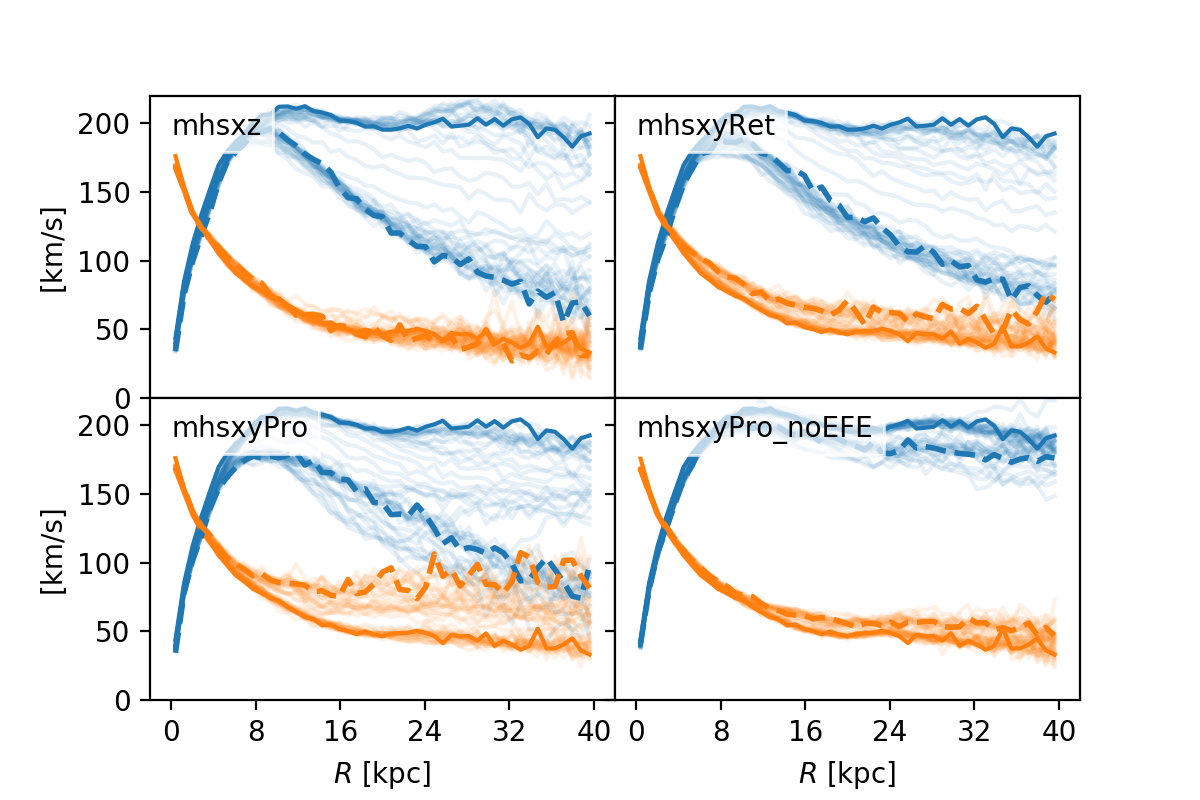}
\end{tabular}
\caption{Evolution of the rotation curves for the galaxies in the Coma-mass background cluster. \textit{Upper left}: low mass models in weak tides; \textit{Upper right}: low mass models in strong tides; \textit{Lower left}: high mass galaxies in weak tides; \textit{Lower right}: high mass galaxies in strong tides. We include the orbit that lies perpendicular to the galactic plane (`xz'), the retrograde and prograde orbits (`xyRet' and `xyPro') and the models without an EFE. The blue lines indicate the azimuthal velocities and the orange lines give the dispersion in these values. The initial profiles are given by solid lines, and the final profiles by dashed lines. The intermediate profiles at each snapshot are shown with lightly shaded lines. In the presence of the EFE all models show a significant reduction of their rotation curves, particularly the low mass galaxies, while in the absence of the EFE there is essentially no evolution at all.}
\label{results:rotComa}
\end{figure*}

\subsubsection{Kinematical evolution}
\label{subsubsec:kine}
To analyse the kinematical evolution of the disks, we look at how the rotation curve evolves over time. We first determine the principal axes of our models using an iterative calculation of the inertia tensor, as described in the next Section. This allows us to rotate our coordinate system until it is coincident with these principal axes, ensuring we are correclty calculating the rotation curve within the disk. We then determine the average rotation velocities $V_{\phi}$ in $50$ radial bins, and the dispersion in these velocities $\sigma_{\phi}$, out to a maximum radius $R_{max}$, only considering particles within $\pm 3$ kpc from the disk plane. For the low mass galaxies $R_{max} = 10$ kpc, while for the high mass galaxies $R_{max} = 40$ kpc. This procedure is performed for each snapshot of the simulations to obtain the time evolution of the outer rotation curve.

It is worth making the point at this stage that the evolution of the rotation curves is the result of a complex interplay between both the tidal interaction and the external field effect. The former can strip mass from the disk, or redistribute mass within the disk (thereby affecting the gravitational potential and by extension the circular velocity) as well as perturb the stellar orbits from mostly circular to mostly radial. The latter distorts the internal gravitational isopotential lines in an asymmetric manner (as seen in Section~\ref{subsubsec:asymmSource}), thus perturbing the stellar orbits, and also reduces the overall depth of the potential. All these effects are combined in our galaxies, as the closer the galaxy is to cluster centre, the stronger the tides and the stronger the external field effect. Furthermore, the susceptibility of the galaxy to tidal disruption is enhanced in the presence of a strong external field. Thus, although we can use the models without an EFE to attempt to separate out tides from the EFE, we must remember that the tidal disruption will be made more pronounced in the presence of a strong EFE.

Given these considerations, in Fig.~\ref{results:rotComa} we examine the galaxies in the high mass cluster. The blue lines show the binned average azimuthal velocities while the orange lines show the dispersion in these values. The solid lines indicate the initial state of the model, and the dashed lines are the state at the end of the simulation. The intermediate states are shown with the lightly shaded lines. We can see that in the absence of an external field, there is essentially no dynamical evolution, regardless of galaxy mass or tides. Thus the tides alone do not significantly affect the rotation curve, consistent with our results regarding mass loss.

With the inclusion of the external field, however, the rotation curves of \textit{all} the models reduce significantly over time, particularly in the low mass galaxies where the EFE is strongest, independently of tides or orbital orientation. The velocity dispersions are also generally enhanced, particularly in those models on perpendicular orbits where the tidal heating of the disk is expected to be strongest. The low mass retrograde and prograde models in strong tides are dispersion-dominated at all radii by the end of the simulation. This is partly explained by the more substantial mass loss for the prograde model, as seen in Fig.~\ref{results:boundMass}.

In summary, the low mass models in a high mass background cluster undergo significant reduction of their rotation curves, \textit{even on an orbit with weak tidal forces}. A strong external field both modifies the internal potential, reducing the rotation curve, and exacerbates the tidal disruption of all galaxies (mass stripping and/or orbit perturbation), leading to a far more substantial reduction of the rotational component than would take place in the absence of the external field.

\subsubsection{Morphological evolution}
\label{subsubsec:morphEvo}
The final diagnostic that we apply to our models, is an analysis of their morphological evolution, for which we will follow a modified version of the approach used in \cite{elliptical_efeII}, where the axis ratios of a bounding ellipsoid were used to analyse elliptical galaxy models. Similarly to the approach in that work, we use an iterative algorithm to find the principal axes of our models using the moment of inertia tensor.

Firstly, a cut is applied to the model (at the initial time) to select a spherical region around the centre-of-mass at a chosen radius. For the low mass galaxies we choose $r = 5h_R \approx 8.6$ kpc, while for the high mass galaxies we choose $r = 5h_R \approx 20$ kpc. Here we refer to the initial scale radii of these models. Then the components of the inertia tensor are calculated using the positions of the selected particles, relative to the centre-of-mass of the galaxy:
\begin{equation}
\begin{split}
I_{xx} &= \sum_i m_i (y_i^2 + z_i^2) \\
I_{xy} &= -\sum_i m_i x_i y_i
\end{split}
\end{equation}
and similarly for the other 4 independent components of this symmetric tensor. This tensor is then diagonalised, and the eigenvalues $\lambda_1$, $\lambda_2$ and $\lambda_3$ are then sorted into ascending order and used to determine the principal axes:
\begin{equation}
\begin{split}
a &= \sqrt{\lambda_2 + \lambda_3 - \lambda_1} \\
b &= \sqrt{\lambda_1 + \lambda_3 - \lambda_2} \\
c &= \sqrt{\lambda_1 + \lambda_2 - \lambda_3}.
\end{split}
\end{equation}
We then calculate the $L_2$ norm of the off-diagonal components of the inertia tensor, which we refer to as $I'$:
\begin{equation}
I' = \sqrt{I_{xy}^2 + I_{xz}^2 + I_{yz}^2}.
\end{equation}
The value of $I'$ is then compared to a user-defined threshold (we choose $10^{-13}$) to determine if the iterative calculation can be terminated. If not, the diagonlisation matrix is used to rotate the coordinates, the inertia tensor is recalculated, diagonalised again, and a new value of $I'$ is determined. The process repeats until $I' < 10^{-13}$) or a maximum of $200$ iterations is reached. We carry out this procedure for each snapshot of each simulation in order to determine the time evolution of the axis ratios, $p = b/a$ and $q = c/a$. Note that this procedure to align the coordinate system with the principal axes is also used in the analysis of the rotation curves.

\begin{figure*}
\centering
\begin{tabular}{cc}
\includegraphics[width=\bigFig\textwidth]{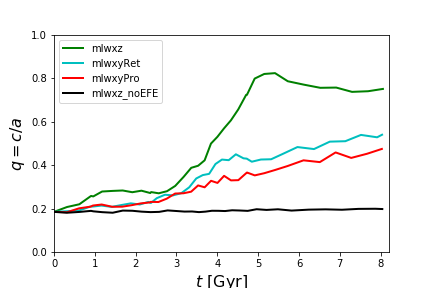} & \includegraphics[width=\bigFig\textwidth]{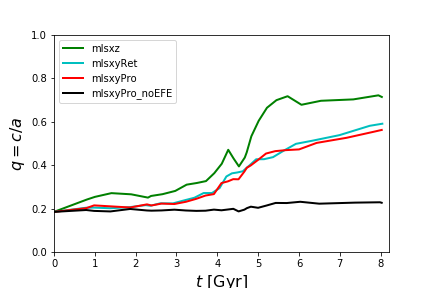} \\
\includegraphics[width=\bigFig\textwidth]{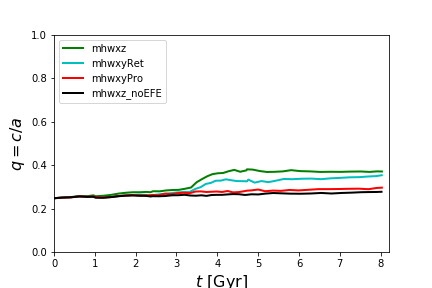} & \includegraphics[width=\bigFig\textwidth]{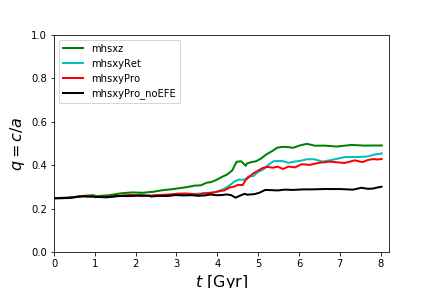}
\end{tabular}
\caption{Evolution of the axis ratio $q = c/a$ for galaxy models in the Coma-mass background cluster. The low mass models exhibit strong morphological changes from disks towards a spheroidal morphology, particularly when the disk plane lies perpendicular to the EFE direction (the `xz' models, green lines). Note that the low mass prograde and retrograde models (cyan and red lines) are still not in equilibrium by the end of the simulation: their axis ratio continues to grow.}
\label{results:morphComa}
\end{figure*}

The results for the models in the Coma-mass cluster are shown in Fig.~\ref{results:morphComa}. We concentrate on the $q$ ratio, rather than the $p$ ratio, as these are disk models and the ratio of the axes within the disk shows little evolution, apart from some tidal squeezing. Referring to the solid black lines in all panels of this figure, we can see that there is no appreciable morphological evolution in the absence of an external field, regardless of galaxy mass or orbital trajectory, consistent with the mass loss and kinematic evolution results.

For the low mass galaxy models (upper panels in Fig.~\ref{results:morphComa}) we can see clear morphological evolution, regardless of the tidal strength. Thus the EFE alone is responsible for a major morphological change for the models in weak tides, and a combination of tides and EFE leads to a similar result for the models in strong tides. Interestingly, there is a very similar separation in morphological evolution between orbit types in the weak and strong tides, connected to the directional nature of \textit{both} effects. This connection between the degree of disk thickening and the orientation of the disk during a tidal interaction has also been explored in the Newtonian context \citep{bialas}. For the high mass galaxy models (lower panels in Fig.~\ref{results:morphComa}) these effects are reduced given the weaker EFE, although we still see a slight increase in the axis ratio, even in weak tides, and a larger increase in strong tides.

\begin{figure}
\centering
\includegraphics[width=\bigFig\textwidth]{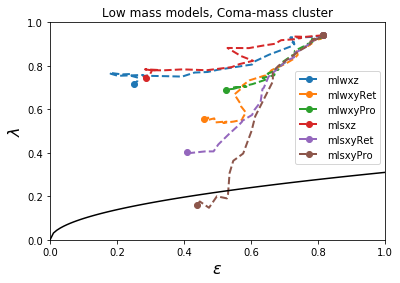}
\caption{Evolution over the full timescale of our simulations ($\sim 8$ Gyr) of the spin parameter and ellipticity of the low mass models in the Coma-mass cluster (excluding models without an EFE). The points indicate the initial and final states. All models begin at the upper right of the diagram. The solid black line separates fast and slow rotators, as in \protect\cite{emsellem2011}. The models without an EFE begin at the same point, but there is essentially no evolution in their spin or ellipticity.}
\label{spinEllip}
\end{figure}

We can combine the results of the analysis of the kinematical evolution with the morphological evolution using a plot of the spin parameter with ellipticity. Such plots have been used to classify early-type galaxies as slow or fast rotators \citep{emsellem2007,emsellem2011,scott2014}. Our definition of the spin parameter is that of \cite{emsellem2007}:
\begin{equation}
\lambda = \frac{\sum_i F_i R_i |V_{\phi,i}|}{\sum_i F_i R_i \sqrt{V_{\phi,i}^2 + \sigma_{\phi,i}^2}}
\end{equation}
where $R_i$ is the radius of the $i$-th bin within the disk, $V_{\phi,i}$ and $\sigma_{\phi,i}$ are the mean azimuthal velocity and azimuthal velocity dispersion in the bin, and $F_i$ is the flux. For the latter value we simply use the projected particle count. We define the ellipticity $\epsilon = 1 - (c/a)$. As the morphological evolution is most pronounced for the low mass galaxies in the Coma-mass cluster, we show only these models in Fig.~\ref{spinEllip}. The initial and final values are plotted as points, with dashed lines indicating their evolutionary trajectory in the $(\lambda,\epsilon)$ space. All models begin in the far upper right of the diagram, with high ellipticity and high spin. The perpendicular models (``mlwxz'' and ``mlsxz'') evolve in a very similar manner, regardless of tides, with a large reduction in their ellipticity due to the EFE and tides acting vertically across their disks. The prograde and retrograde models all show a strong reduction in their spin parameters, with a relatively minor change in their ellipticities. This is due to the EFE and tides acting within the plane of the disks in these models. Note that, in weak tides $\lambda_{\text{Ret}} < \lambda_{\text{Pro}}$, as the EFE induces violent relaxation effects in the disk, but there is little tidal stripping. Conversely, in strong tides $\lambda_{\text{Pro}} < \lambda_{\text{Ret}}$ as the mass is resonantly stripped by the strong tides, significantly reducing the rotation curve.

In summary, when the combined effects of the EFE and the cluster tides are strong enough, we see a morphological evolution towards a spherical distribution. It is worth noting that the rotation curve evolution for the low mass prograde and retrograde models in the Coma-mass cluster (in strong tides) would suggest a fully spherical distribution, which we do not see here. It is clear from Fig.~\ref{results:morphComa}, however, that these models are not yet in equilibrium, regardless of the tides: their axis ratios continue to increase right to the end of our simulation. Thus one would expect these models to evolve further over to the lower left corner of Fig.~\ref{spinEllip}. Indeed it is in $(\lambda,\epsilon)$ space where we can see the combined influences of tides, EFE and orbital orientation all playing a role.

\section{Discussion and conclusions}
\label{sec:conclusions}
We have studied idealised models of low and high mass MOND disk galaxies falling into galaxy clusters of different masses. For more control over the orbital trajectories, we have used an analytic NFW model for the background cluster, with a time varying mass in order to approximate the mass growth of the clusters and ensure bound infall trajectories. Our models do not incorporate dynamical friction, although this is only significant for the most massive galaxies in galaxy clusters. The orbits of the galaxy models are either perpendicular to the disk, or parallel, with either a retrograde or prograde motion around the cluster centre in the latter case. Finally, we have analysed the consequences of removing the MOND external field effect.

The main conclusion of our study is that both the low and high mass MOND galaxies are substantially distorted by the EFE as they fall into the cluster, leading to significant asymmetries in the galactic disks, even in the absence of any tidal interaction. In the presence of such an interaction, however, these asymmetries are further enhanced. The weakened potentials of the MOND galaxies within the cluster also leads to mass loss in the absence of tides and increased susceptibility to mass loss due to tidal stripping, loss of their rotational components and a morphological evolution towards a more spherical distribution.

More specifically, our study has found that:
\begin{itemize}
\item MOND disk galaxies in clusters have distorted isophotal contours due to the EFE, with low mass galaxies in a high mass cluster the most severely affected. When further combined with a tidal interaction, the distortion is even more enhanced, producing highly asymmetric forms. The effect is weaker for high mass galaxies, but still long-lived (for all models the asymmetries last at least $1$ Gyr, and up to $4$ Gyr for the low mass models) and likely to be clearly detectable.
\item All of our model galaxies that are subjected to an EFE suffer a gradual mass loss even before reaching pericentre on their orbits and are more easily tidally stripped, both due to the EFE. Again, as the EFE is stronger for low mass galaxies (given a fixed background field), these results are very clear for those models. Thus our galaxies suffer more mass loss than would be expected in the absence of the EFE or for a Newtonian galaxy embedded in a dark matter halo.
\item Given the modifed internal potential of the galaxies, as well as the induced disequilibrium due to the rapidly-evolving external field (effectively a form of violent relaxation), they undergo a kinematical evolution within the cluster towards being more dispersion dominated objects. Several of the low mass galaxies in the high mass cluster evolve to dispersion-supported systems in the Coma-mass cluster, regardless of tides, although they remain out of equilibrium by the end of our simulation, as indicated by their evolving morphology.
\item Finally, this loss of rotational dominance is reflected in a morphological change from disks to more spherical distributions. This morphological change has a clear dependency on the orientation of the galactic disk relative to the orbital trajectory within the cluster, given the directional nature of the orbital perturbations induced by the tidal disruption and the EFE.
\end{itemize}

Furthermore, from our results we can infer that, in the absence of an external field effect, MOND galaxies within galaxy clusters would behave similarly to Newtonian galaxies within dark matter halos, at least on first infall within the cluster. Given that our models do not undergo more than one pericentric passage, it is not clear if the tidal stripping of a MOND galaxy in the absence of an EFE would be comparable to that of a Newtonian model that had already lost a significant fraction of its halo after the first pericentric passage.

Given our asymmetry results, it would be of great interest to explore the prevalence of galactic asymmetries in massive nearby clusters and to compare this with values in the field. Typically this parameter is taken to be indicative of galaxy mergers in the standard context of galaxy evolution, yet in our study we show large increases in asymmetry as galaxies move within the cluster, \textit{in the absence of any galaxy-galaxy interaction}. In \cite{homeier} the asymmetry of cluster and field galaxies is compared for late-type galaxies at $z \sim 1$, where the galaxies in lower mass (X-ray faint) clusters were found to be more asymmetric than those in higher mass (X-ray luminous) clusters, a property attributed to the likelihood of more mergers in low mass clusters. In our study we suggest that \textit{high mass} clusters should have a larger population of asymmetric galaxies. Our simulations do not include the possibility of galaxy-galaxy interactions, but it is worth mentioning that mergers are expected to be much rarer in MOND \cite{kroupa} due to the lack of dynamical friction. Thus it seems likely that in MOND galaxy asymmetries are indeed more common in high mass clusters, whereas in $\Lambda$CDM they would be more common in low mass clusters/groups.

A natural extension of this study, therefore, would be to test this hypothesis by considering a realistic live cluster evolution of multiple infalling galaxies and groups, allowing for possible galaxy-galaxy interactions. This would allow us to investigate the effect of dynamical friction on the galaxy interactions and on their orbital trajectories, given the clear differences expected between $\Lambda$CDM and MOND in this area \cite{SanchezSalcedoDF,NipotiDF,HongShengDF}. In addition, such a study would allow a phase-space analysis of the EFE as an environmental effect \citep{oman,jaffe1,jaffe2,rhee}.

All known MOND theories based on the modification of the Newtonian Poisson equation include an external field effect. More specifically, the relativistic extensions of MOND that invoke additional degrees of freedom in the gravitational sector give rise to violations of the Strong Equivalence Principle and an external field effect. Furthermore, the recent discovery of an electromagnetic counterpart to a gravitational wave detection \citep{GravWaveDiscovery} has placed very stringent bounds on the (low redshift) propagation velocity of gravitational waves, ruling out large classes of relativistic modified gravity theories that predict gravitational wave propagation at velocities other than the speed of light \citep{GWchallenge}. If observational data of galaxies within clusters consistently contradicts the results of this study, this may present a significant challenge to the presence of an external field effect, putting pressure on the standard non-relativistic formulations of MOND. Other formulations that do not predict an external field effect (e.g. modified inertia) are known, but are much less studied, and have highly unusual theoretical properties, such as extreme non-locality \citep{modInertia1,modInertia2}.

The consequences for the dynamics of galaxies and galaxy clusters of modifying gravity are profound and varied. Further investigations along these lines, using numerical simulations within the MOND context, may allow us to find further interesting and potentially observable signatures.

\section*{Acknowledgements}
Powered@NLHPC: This research was partially supported by the supercomputing infrastructure of the NLHPC (ECM-02). GNC acknowledges support from CONICYT PAI No. 79150053 and FONDECYT Regular 1181708. Y. J. acknowledges support from CONICYT PAI No. 79170132. AC would like to thank CNPq for the fellowship 4150977/2017-4.

%\begin{thebibliography}{99}


\begin{thebibliography}{}
\bibitem[\protect\citeauthoryear{Abbott et al.}{2017}]{GravWaveDiscovery} Abbott B.~P., Abbott R., Abbott T.~D., Acernese F., Ackley K., Adams C., Adams T., Addesso P., Adhikari R.~X., Adya V.~B. et al. (LIGO Scientific Collaboration, Virgo Collaboration, Fermi Gamma-Ray Burst Monitor, INTEGRAL), 2017, ApJL, 848, L13
\bibitem[\protect\citeauthoryear{Alzain}{2017}]{modInertia2} Alzain M., 2017, JApA, 38, 59
\bibitem[\protect\citeauthoryear{Angus \& Diaferio}{2011}]{AngusDaeferio} Angus G.~W., Diaferio A., 2011, MNRAS, 417, 941
\bibitem[\protect\citeauthoryear{Bekenstein}{2004}]{teves} Bekenstein J.D., 2004, Phys. Rev. D, 70, 083509
\bibitem[\protect\citeauthoryear{Bekenstein \& Milgrom}{1984}]{milgrom2} Bekenstein J., Milgrom M., 1984, ApJ, 286, 7
\bibitem[\protect\citeauthoryear{Berezhiani \& Khoury}{2015}]{superfluidDM} Berezhiani L., Khoury J., 2015, Phys Rev D, 92, 103510
\bibitem[\protect\citeauthoryear{Bialas et al.}{2015}]{bialas} Bialas D., Lisker T., Olczak C., Spurzem R., Kotulla R., 2015, A\&A, 576, 103
\bibitem[\protect\citeauthoryear{Binney \& Merrifield}{1998}]{Conversion} Binney J., Merrifield M., 1998, Princeton University Press
\bibitem[\protect\citeauthoryear{Blanchet \& Le Tiec}{2008}]{dipolarDM1} Blanchet L., Le Tiec A., 2008, Phys Rev D, 78, 024031
\bibitem[\protect\citeauthoryear{Blanchet \& Heisenberg}{2017}]{dipolarDM2} Blanchet L., Heisenberg L., 2017, Phys Rev D, 96, 083512
\bibitem[\protect\citeauthoryear{Boran et al.}{2018}]{GWchallenge} Boran S., Desai S., Kahya E.~O., Woodard R.~P., Phys Rev D, 97, 041501
\bibitem[\protect\citeauthoryear{Brada \& Milgrom}{2000}]{brada_warps} Brada R., Milgrom M., 2000, ApJL, 531, L21
\bibitem[\protect\citeauthoryear{Bullock \& Boylan-Kolchin}{2017}]{smallscale} Bullock J., Boylan-Kolchin M., 2017, ARAA, 55, 343
\bibitem[\protect\citeauthoryear{Candlish et al.}{2015}]{raymond} Candlish G.~N., Smith R., Fellhauer M., 2015, MNRAS, 446, 1060
\bibitem[\protect\citeauthoryear{Candlish}{2016}]{raymondCosmo} Candlish G.~N., 2016, MNRAS, 460, 2571
\bibitem[\protect\citeauthoryear{Conselice et al.}{1999}]{conselice} Conselice C.~J., Bershady M.~A., Jangren A., 2000, ApJ, 529, 886
\bibitem[\protect\citeauthoryear{DeMaio et al.}{2015}]{ICLbuildup} DeMaio T., Gonzalez A.~H., Zabludoff A., Zaritsky D., Brada\u{c}, Maru\u{s}a, 2015, MNRAS, 448, 1162
\bibitem[\protect\citeauthoryear{D'Onghia et al.}{2009}]{resonant_strip} D'Onghia E., Besla G., Cox T.~J., Hernquist L., 2009, Nature, 460, 605
\bibitem[\protect\citeauthoryear{Emsellem et al.}{2007}]{emsellem2007} Emsellem E., Cappellari M., Krajnovi{\'c} D., van de Ven G., Bacon R., Bureau M., Davies R.~L., de Zeeuw P.~T., Falcón-Barroso J., Kuntschner H., McDermid R., Peletier R.~F., Sarzi M., 2007, MNRAS, 379, 401
\bibitem[\protect\citeauthoryear{Emsellem et al.}{2011}]{emsellem2011} Emsellem E., Cappellari M., Krajnovi{\'c} D., Alatalo K., Blitz L., Bois M., Bournaud F., Bureau M., Davies R.~L., Davis T.~A., de Zeeuw P.~T., Khochfar S., Kuntschner H., Lablanche P.-Y. McDermid R.~M., Morganti R., Naab T., Oosterloo T., Sarzi M., Scott N., Serra P., van de Ven G., Weijmans A.-M., Young L.~M., 2011, MNRAS, 414, 888
\bibitem[\protect\citeauthoryear{Hees et al.}{2014}]{MONDsolar} Hees A., Folkner W.M., Jacobson R.A., Park R.S., 2014, Phys. Rev. D, 89, 102002
\bibitem[\protect\citeauthoryear{Famaey \& McGaugh}{2012}]{mondreview} Famaey B., McGaugh S.~S., 2012, Living Reviews in Relativity, 15
\bibitem[\protect\citeauthoryear{Fan et al.}{2013}]{dissipativeDM} Fan J., Katz A., Randall L., Reece M., 2013, Phys Rev Lett, 110, id:211302
\bibitem[\protect\citeauthoryear{Gill et al.}{2004}]{gill2004} Gill S.~P.~D., Knebe A., Gibson B.~K., 2004, MNRAS, 351, 399
\bibitem[\protect\citeauthoryear{Homeier et al.}{2006}]{homeier} Homeier N.~L., Mei S., Blakeslee J.~P., Postman M., Holden B., Ford H. C., Bradley L.~D., Demarco R., Franx M., Illingworth G.~D., Jee M.~J., Menanteau F., Rosati P., van der Wel A., Zirm A., 2006, ApJ, 647, 256
\bibitem[\protect\citeauthoryear{Jaff\'e et al.}{2015}]{jaffe1} Jaff{\'e}, Y.~L., Smith R., Candlish G.~N., Poggianti B.~M., Sheen Y.-K., Verheijen M.~A.~W., 2015, MNRAS, 448, 1715
\bibitem[\protect\citeauthoryear{Jaff\'e et al.}{2018}]{jaffe2} Jaff{\'e}, Y.~L., Poggianti B.~M., Moretti A., Gullieuszik M., Smith R., Vulcani B., Fasano G., Fritz J., Tonnesen S., Bettoni D., Hau G., Biviano A., Bellhouse C., McGee S., 2018, MNRAS, \textit{in press}
\bibitem[\protect\citeauthoryear{Katz et al.}{2013}]{KatzMcGaugh} Katz H., McGaugh S., Teuben T., Angus G.~W., 2013, ApJ, 772, 10
\bibitem[\protect\citeauthoryear{Klasen et al.}{2015}]{DMpaper} Klasen M., Pohl M., Sigl G., 2015, PrPNP, 85, 1
\bibitem[\protect\citeauthoryear{Kroupa}{2015}]{kroupa} Kroupa P., 2015, CaJPh, 93, 169
\bibitem[\protect\citeauthoryear{Laporte et al.}{2013}]{laporte} Laporte C.~F.~P., White S.~D.~M., Naab T., Gao L., 2013, MNRAS, 435, 901
\bibitem[\protect\citeauthoryear{Llinares et al.}{2008}]{amiga_mond} Llinares C., Knebe A., Zhao H., 2008, MNRAS, 391, 1778
\bibitem[\protect\citeauthoryear{Londrillo and Nipoti}{2009}]{nmody} Londrillo P. and Nipoti C., 2009, Memorie della Societ\`{a} Astronomica Italiana Supplement, v.13, p.89
\bibitem[\protect\citeauthoryear{Ludlow et al.}{2013}]{massConc} Ludlow A.~D., Navarro J.~F., Angulo R.~E., Boylan-Kolchin M., Springel V., Frenk C., White S.~D.~M., 2014, MNRAS, 441, 378
\bibitem[\protect\citeauthoryear{L{\"u}ghausen et al.}{2014}]{efe_mw} L{\"u}ghausen F., Famaey B., Kroupa P., 2014, MNRAS, 441, 2497
\bibitem[\protect\citeauthoryear{L{\"u}ghausen et al.}{2015}]{por} L{\"u}ghausen F., Famaey B., Kroupa P., 2015, Canadian Journal of Physics, 93, 232
\bibitem[\protect\citeauthoryear{Milgrom}{1983}]{milgrom1} Milgrom M., 1983, ApJ, 270, 365
\bibitem[\protect\citeauthoryear{Milgrom}{2006}]{modInertia1} Milgrom M., 2006, EAS Publications Series, 20, 217
\bibitem[\protect\citeauthoryear{Milgrom}{2009}]{bimetricmond} Milgrom M., 2009, Phys. Rev. D, 80, 123536
\bibitem[\protect\citeauthoryear{Milgrom}{2010}]{milgrom3} Milgrom M., 2010, MNRAS, 403, 886
\bibitem[\protect\citeauthoryear{Milgrom}{2015}]{milgromUDG} Milgrom M., 2015, MNRAS, 454, 3810
\bibitem[\protect\citeauthoryear{M{\"u}ller et al.}{2018}]{centA} M{\"u}ller O., Pawlowski M.~S., Jerjen H., Lelli F., 2018, Science, 359, 534
\bibitem[\protect\citeauthoryear{Nipoti et al.}{2008}]{NipotiDF} Nipoti C., Ciotti L., Binney J., Londrillo P., 2008, MNRAS, 386, 2194
\bibitem[\protect\citeauthoryear{Oman et al.}{2013}]{oman} Oman K.~A., Hudson M.~J., Behroozi P.~S., 2013, MNRAS, 431, 2307
\bibitem[\protect\citeauthoryear{Pawlowski}{2018}]{pawlowskiPlanes} Pawlowski M.~S., 2018, MPLA, 33, id:1830004
\bibitem[\protect\citeauthoryear{Perret}{2014}]{DICE} Perret V., Renaud F., Epinat B., Amram P., Bournaud F., Contini T., Teyssier R., Lambert J.-C., 2014, A\&A, 562, A1
\bibitem[\protect\citeauthoryear{Rhee et al.}{2017}]{rhee} Rhee J., Smith R., Choi H., Yi S.~K., Jaff{\'e} Y., Candlish G., S{\'a}nchez-J{\'a}nssen R., 2017, ApJ, 843, 128
\bibitem[\protect\citeauthoryear{Sanchez-Salcedo et al.}{2006}]{SanchezSalcedoDF} S{\'a}nchez-Salcedo F. J., Reyes-Iturbide J., Hernandez X., 2006, MNRAS, 370, 1829
\bibitem[\protect\citeauthoryear{Sanders}{2003}]{mondgalcl} Sanders R.~H., 2003, MNRAS, 342, 901
\bibitem[\protect\citeauthoryear{Sawala et al.}{2016}]{baryonicPhysSim} Sawala T., Frenk C.~S., Fattahi A., Navarro J.~F., Bower R.~G., Crain R.~A., Dalla Vecchia C., Furlong M., Helly J.~C., Jenkins A., Oman K.~A., Schaller M., Schaye J., Theuns T., Trayford J., White S.~D.~M., 2016, MNRAS, 457, 1931
\bibitem[\protect\citeauthoryear{Schade et al.}{1995}]{schade} Schade D., Lilly S.~J., Crampton D., Hammer F., Le Fevre O., Tresse L., 1995, ApJ, 451, L1
\bibitem[\protect\citeauthoryear{Scott et al.}{2014}]{scott2014} Scott N., Davies R.~L., Houghton R.~C.~W., Cappellari M., Graham A.~W., Pimbblet K.~A., 2014, MNRAS, 441, 274
\bibitem[\protect\citeauthoryear{Smith et al.}{2013}]{roryMassLoss} Smith R., S{\'a}nchez-Janssen R., Fellhauer M., Puzia T.~H., Aguerri J.~A.~L., Farias J.~P., 2013, 429, 1066
\bibitem[\protect\citeauthoryear{Teyssier}{2002}]{ramses} Teyssier R., 2002, A\&A, 385, 337
\bibitem[\protect\citeauthoryear{Thomas et al.}{2017a}]{tidal_streamsI} Thomas G.~F., Famaey B., Ibata R., L{\"u}ghausen F., Kroupa P., 2017, A\&A, 603, A65
\bibitem[\protect\citeauthoryear{Thomas et al.}{2017b}]{tidal_streamsII} Thomas G.~F., Famaey B., Ibata R., Renaud F., Martin N.~F., Kroupa P., 2017, arXiv:1709.01934
\bibitem[\protect\citeauthoryear{Trujillo \& Fliri}{2016}]{Trujillo2016} Trujillo I., Fliri J., 2016, ApJ, 823, 123
\bibitem[\protect\citeauthoryear{Wu et al.}{2010}]{elliptical_efeI} Wu X., Zhao H., Famaey B., 2010, JCAP, 6, 010
\bibitem[\protect\citeauthoryear{Wu et al.}{2017}]{elliptical_efeII} Wu X., Wang Y., Feix M., Zhao H., 2017, ApJ, 844, 130
\bibitem[\protect\citeauthoryear{Zhao \& Tian}{2006}]{TidalRadMOND} Zhao H., Tian L., 2006, A\&A, 450, 1005
\bibitem[\protect\citeauthoryear{Zhao et al.}{2013}]{HongShengDF} Zhao H., Famaey B., L{\"u}ghausen F., Kroupa P., 2013, A\&A, 557, 3
\bibitem[\protect\citeauthoryear{Zhao \& Li}{2010}]{darkfluid} Zhao H., Li B., 2010, ApJ, 712, 130
\bibitem[\protect\citeauthoryear{Zlosnik et al.}{2006}]{GEA} Zlosnik T.~G., Ferreira P.~G., Starkman G.~D., 2006, Phys. Rev. D, 74, 044037
\end{thebibliography}
\end{document}